\documentclass[11pt]{article}
\usepackage{jheppub}
\usepackage{graphicx} 
\usepackage{graphbox}
\usepackage[usenames,dvipsnames]{xcolor} 
\usepackage{tikz}	
\usepackage{dirtytalk}
\usepackage{dsfont}
\usepackage{amsthm}
\usepackage{float}
\usepackage{bbm, dsfont}
\usepackage[normalem]{ulem}
\usepackage{slashed}
\usepackage{mathtools}
\usepackage{cancel}
\usepackage{cleveref}
\usepackage{orcidlink}
\usepackage{enumerate}
\usepackage{fontawesome}
\usepackage{adjustbox}
\usepackage{tcolorbox}

\renewcommand{\th}{\th}
\newcommand{\vep}{\varepsilon}

\renewcommand{\d}{\text{d}}
\newcommand{\rd}{\mathrm{d}}

\newcommand{\mbf}[1]{\mathbf{#1}}
\newcommand{\bs}[1]{\boldsymbol{#1}}
\newcommand{\gem}{ {\gamma_{ {}_\text{EM}}} }

\newcommand{\e}{ {{\color{BrickRed}(123)}} }
\newcommand{\ee}{ {{\color{Blue}(345)}} }
\newcommand{\rb}[1]{{(#1)}}
\newcommand{\nn}{\nonumber}

\newcommand\scalemath[2]{\scalebox{#1}{\mbox{\ensuremath{\displaystyle #2}}}}
\tikzset {_hq8agl9h9/.code = {\pgfsetadditionalshadetransform{ \pgftransformshift{\pgfpoint{-19.5 bp } { -37.5 bp }  }  \pgftransformrotate{0 }  \pgftransformscale{2 }  }}}
\pgfdeclarehorizontalshading{_4sn6rfw6e}{150bp}{rgb(0bp)=(1,1,1);
rgb(50.5bp)=(1,1,1);
rgb(72.5bp)=(0.92,0.01,0.11);
rgb(100bp)=(1.3,0.01,0.11)}
\tikzset{every picture/.style={line width=0.75pt}}
\tikzset {_bhh1yolhd/.code = {\pgfsetadditionalshadetransform{ \pgftransformshift{\pgfpoint{11.5 bp } { -8 bp }  }  \pgftransformrotate{-90 }  \pgftransformscale{2 }  }}}
\pgfdeclarehorizontalshading{_90qpx1llj}{150bp}{rgb(0bp)=(0.82,0.01,0.11);
rgb(10.5bp)=(0.82,0.01,0.11);
rgb(52.5bp)=(1,1,1);
rgb(100bp)=(1.3,1,1)}
\tikzset{every picture/.style={line width=0.75pt}}
\tikzset {_h90ekjryl/.code = {\pgfsetadditionalshadetransform{ \pgftransformshift{\pgfpoint{11.5 bp } { -8 bp }  }  \pgftransformrotate{-90 }  \pgftransformscale{2 }  }}}
\pgfdeclarehorizontalshading{_k8eguni3i}{150bp}{rgb(0bp)=(0,0.11,1);
rgb(10.5bp)=(0,0.11,1);
rgb(52.5bp)=(1,1,1);
rgb(100bp)=(1.3,1,1)}
\tikzset{every picture/.style={line width=0.75pt}}
\tikzset {_xqcddvd9p/.code = {\pgfsetadditionalshadetransform{ \pgftransformshift{\pgfpoint{-4 bp } { -27.5 bp }  }  \pgftransformrotate{0 }  \pgftransformscale{2 }  }}}
\pgfdeclarehorizontalshading{_tl6yhambn}{150bp}{rgb(0bp)=(1,1,1);
rgb(50.5bp)=(1.3,1,1);
rgb(72.5bp)=(0,0.11,1);
rgb(100bp)=(0,0.11,1)}
\tikzset{every picture/.style={line width=0.75pt}}
\newcommand\wedgedot[1][1]{
\,
\tikzset{every picture/.style={line width=0.75pt}}      
\begin{tikzpicture}[x=0.75pt,y=0.75pt,yscale=-1,xscale=1]
\draw   (100,130.12) -- (105.06,120) -- (110.12,130.12) ;
\draw  [fill={rgb, 255:red, 0; green, 0; blue, 0 }  ,fill opacity=1 ] (104,127) .. controls (104,126.45) and (104.45,126) .. (105,126) .. controls (105.56,126) and (106.01,126.45) .. (106.01,127) .. controls (106.01,127.56) and (105.56,128.01) .. (105,128.01) .. controls (104.45,128.01) and (104,127.56) .. (104,127) -- cycle ;
\end{tikzpicture}
\,
}
\makeatletter
\newenvironment{sqcases}{
  \matrix@check\sqcases\env@sqcases
}{
  \endarray\right.
}
\def\env@sqcases{
  \let\@ifnextchar\new@ifnextchar
  \left\lbrack
  \def\arraystretch{1.2}
  \array{@{}l@{\quad}l@{}}
}
\makeatother
\definecolor{darkGreen}{cmyk}{1,.2,1,0.2}

\title{The soaring kite: a tale of two punctured tori}

\abstract{We consider the 5-mass kite family of self-energy Feynman integrals and present a systematic approach for constructing an $\varepsilon$-form basis, along with its differential equation pulled back onto the moduli space of two tori. Each torus is associated with one of the two distinct elliptic curves this family depends on. We demonstrate how the locations of relevant punctures, which are required to parametrize the full image of the kinematic space onto this moduli space, can be extracted from integrals over maximal cuts. A boundary value is provided such that the differential equation is systematically solved in terms of iterated integrals over $g$-kernels and modular forms. Then, the numerical evaluation of the master integrals is discussed, and important challenges in that regard are emphasized. In an appendix, we introduce new relations between $g$-kernels.}

\author[a,\orcidlink{0000-0002-2672-634X}]{Mathieu Giroux,}
\emailAdd{mathieu.giroux2@mail.mcgill.ca}
\affiliation[a]{
    Department of Physics, 
    McGill University, 
    3600 Rue University, 
    Montr\'eal, 
    QC 
    Canada
    H3A 2T8
}

\author[b,\orcidlink{0000-0003-1186-4624}]{Andrzej Pokraka,}
\emailAdd{andrzej\_pokraka@brown.edu}
\affiliation[b]{
	Department of Physics, 	
        Brown University, 	
	Providence, 	
	RI 02912, 
	USA
}

\author[c,\orcidlink{0000-0002-3328-499X}]{Franziska Porkert,}
\emailAdd{fporkert@uni-bonn.de}
\affiliation[c]{Bethe Center for Theoretical Physics, Universität Bonn, D-53115, Germany
}

\author[d,\orcidlink{0009-0007-0651-0676}]{Yoann Sohnle}
\emailAdd{yoann.sohnle@physics.uu.se}
\affiliation[d]{
	Department of Physics and Astronomy, Uppsala University, Box 516, 75120 Uppsala, Sweden
}

\allowdisplaybreaks
\begin{document}
\preprint{\begin{tabular}{r}
UUITP–03/24 \\
\vspace{-25pt} BONN-TH-2024-01
\end{tabular}}

\maketitle
\normalem
\allowdisplaybreaks
\raggedbottom

\newpage
\section{Introduction}
With the lack of conclusive evidence for new particles in the Large Hadron Collider (LHC) data (other than the Higgs boson), recent phenomenological interests and efforts have shifted to understanding known standard model couplings, especially those that involve the Higgs. 

A pressing concern is that the current theoretical and statistical uncertainties for Yukawa couplings between matter and the Higgs boson (e.g., the $t\bar{t}H$ coupling \cite{CMS:2018uxb}) are approximately equal.
With the prospect of the LHC high-luminosity phase and of higher-energy FCC-$ee$ colliders, it is therefore essential to improve theoretical predictions as much as possible to match (and possibly go beyond) the expected statistical uncertainties of upcoming measurements, which will naturally decrease as more data become available. To meet this demand, it is necessary to have 2-loop processes with massive particles in the loops (e.g., $gg\to t\bar{t}H$) under good computational control \cite{Huss:2022ful,Delto:2023kqv}. Furthermore, the ability to perform precision measurements of various properties of electroweak bosons (including their mass and decay widths) at the LHC (TeV) scale provides a strong motivation for new perturbative calculations of scattering amplitudes with \emph{unequal masses}.

However, massive 2-loop diagrams often give rise to special functions beyond conventional multiple polylogarithms (MPLs): iterated integrals with rational kernels that have at most simple poles \cite{Bourjaily:2022bwx}. 
Like MPLs, these special functions have a geometric interpretation connected to complex Riemann surfaces. 
Although the simplest class of such functions evaluates to elliptic multiple polylogarithms (eMPLs) \cite{Brown:2011wfj, Broedel:2016wiz, Broedel:2018qkq, Broedel:2017kkb, Broedel:2017siw, weinzierl2022feynman}, there is currently no upper bound on the geometric complexity of Feynman integrals (i.e., higher genus surfaces \cite{Huang:2013kh, Georgoudis:2015hca, Doran:2023yzu, Marzucca:2023gto}, Calabi-Yaus \cite{Bourjaily:2018ycu, Bourjaily:2018yfy, Bourjaily:2019hmc, Pogel:2022vat, Pogel:2022ken, Broedel:2021zij, Bonisch:2021yfw} and potentially beyond). 
To have control over these functions, the underlying function spaces must be understood. In particular, it is crucial to better understand the space of differential forms that generate these integrals, the analytic structure of the resulting integrals, the relations they satisfy and their
numerical evaluation.
Our hope is that improving our understanding of these problems streamlines the computation of new phenomenologically relevant processes that are at the moment out of reach, just as an improved mathematical understanding of MPLs led to the NLO revolution in the 2010's \cite{Remiddi:1999ew,Aglietti:2008fe,Goncharov:2010jf,Duhr:2012fh,Gehrmann-DeRidder:2012too,Dixon:2014voa,Caron-Huot:2016owq,Badger:2021nhg,Badger:2019djh,Abreu:2019odu,Hartanto:2019uvl,Duhr:2019tlz,Duhr:2019wtr}. Let us also mention that the toolbox developed for the evaluation of Feynman integrals in particle physics is slowly extending to other fields, such as gravitational wave physics \cite{Jakobsen:2023hig,Jakobsen:2023pvx,Frellesvig:2023bbf,Klemm:2024wtd} and cosmology \cite{anastasiou2023efficiently,De:2023xue,Arkani-Hamed:2023kig}, where, in particular, elliptic 2-loop integrals with many masses, like the ones discussed in this paper, are starting to appear (see \cite{anastasiou2023efficiently}).
 
As a result of the significant efforts put into studying elliptic Feynman integrals over the past decade, the function space associated with these families is increasingly well understood. Key developments include the generation of functional relations between eMPLs \cite{Broedel:2019tlz, Wilhelm:2022wow, Bhardwaj:2023vvm}, a good notion of transcendentality \cite{Frellesvig:2023iwr}, systematic ways of constructing $\varepsilon$-forms basis \cite{G_rges_2023, Frellesvig:2023iwr, giroux_loop-by-loop_2022}, and the introduction of the first publicly available tools for the numerical evaluation of eMPLs \cite{Walden:2020odh}.
In addition, amplitudes in integer dimension have been bootstrapped from a massive Schubert analysis and the Symbol Prime \cite{Wilhelm:2022wow, Morales:2022csr, McLeod:2023qdf,Cao:2023tpx,He:2023umf}. 
Less understood aspects include defining a consistent coaction \cite{tapušković2023cosmic}, the image of the Landau locus on the torus, and families that depend on multiple elliptic curves \cite{Muller:2022aba, Bourjaily:2021vyj}.

\begin{figure}
    \centering
\includegraphics[width=.45\textwidth]{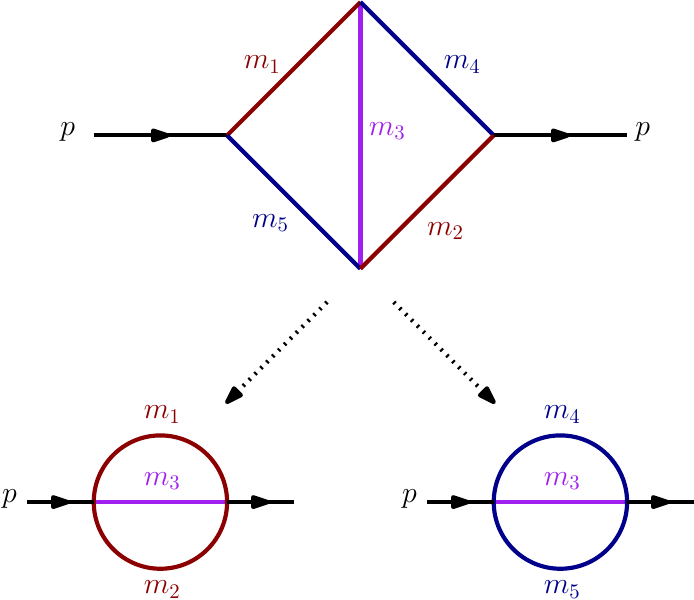}
    \caption{The 5-mass kite integral. All other integrals belonging to the family are obtained by pinching the 5-mass kite. The internal edges are colored according to which sunrise topology they belong to: {\color{BrickRed}red} for the $\e$-sunrise, {\color{Blue}blue} for the $\ee$-sunrise and {\color{Plum}purple} for both sunrises. This color scheme will be used throughout this work. }
    \label{fig:5mkite}
\end{figure}

The main goal of this paper is to further elucidate the properties of elliptic Feynman integrals by studying the 5-mass kite family of integrals (see figure \ref{fig:5mkite}) in dimensional regularization (see \cite{Broadhurst:2022bkw} for the strict 2-dimensional limit). 
This family of integrals is particularly interesting because it contains two separate elliptic subsectors corresponding to two sunrise subtopologies.
Of particular interest is the interplay between these two distinct elliptic sectors, since eMPLs are only defined with respect to a \emph{single} modulus.

\paragraph{Outline} This paper is organized as follows. In section \ref{sec_kitefam}, we introduce the 5-mass kite family of integrals, sketch our calculational method, and review the necessary background on elliptic curves.
In particular, we introduce a non $\varepsilon$-form basis whose differential equation is computed using the public IBP software \texttt{LiteRed} \cite{lee_presenting_2012}.
Then, we systematically construct the gauge transformation that brings our naive starting basis into $\varepsilon$-form in section \ref{epssec}.
We illustrate our method in detail for both the eyeball diagram (a 4-propagator subtopology of the kite; see figure \ref{fig.3}) and the kite itself (see figure \ref{fig:5mkite}). 

In section \ref{sec_punctureslab},  we show that the new elliptic functions (beyond those of the standard sunrise) in the gauge transformation correspond to the positions of additional punctures needed to embed the kinematic space into the moduli space of the tori. 
We also demonstrate how the location of these extra punctures are directly related to the integration of various maximal cuts, thereby simplifying their systematic extraction and rooting their definition on more physical grounds. 

The $\varepsilon$-form differential equation is then expressed in terms of Kronecker-Eisenstein (KE) forms ($g$-kernels) and modular forms in section \ref{pullback}. 
To construct the appropriate ansatz using such forms, we infer all relevant arguments (linear combinations of these punctures) 
from the diagonal dlog terms by examining the leading-order term of their $q$-expansion. 
Additionally, we perform a (partial) mapping of the Landau variety to the moduli space of the tori. 

In section \ref{bdry}, we provide a boundary value for the differential equation, discuss our integration technique, outline the checks performed, and emphasize pressing challenges regarding high-precision numerical evaluation of multiscale elliptic Feynman integrals through $\varepsilon$-form differential equations. 

Finally, while not central to this work, we also present interesting new identities satisfied by $g$-functions in appendix \ref{sec_newrel} and explain how they can be useful in the context of eMPLs.

\begin{tcolorbox}[width=15.5cm, colframe=black!15, colback=white]
The main results of this paper are available in \textsc{Mathematica} format at the following GitHub repository: \href{https://github.com/StrangeQuark007/kite_ancillary}{\faGithub}.
\end{tcolorbox}

\section{The 5-mass kite family}
\label{sec_kitefam}

In this section, we spell out our conventions for the 5-mass kite family of integrals. 
Then, in section \ref{subsec.twoell}, we describe our starting ``Laporta-like'' basis and the associated noncanonical differential equations. 
In section \ref{subsec_ellipticint}, we provide a primer on the fundamentals of elliptic Feynman integrals with a focus on the now well studied sunrise family, which plays a pivotal role in the kite family.

In the mostly-minus signature, integrals belonging to the kite family take the form
\begin{equation} \label{eq:5mFam}
    I^{\text{5m-kite}}_{\bs{\nu}}
    \equiv I_{\bs{\nu}}
    = (-1)^{\vert\nu\vert}e^{2\gem\vep} (\mu^2)^{|\nu|-d} 
    \int \frac{\d^d \ell_1}{i\pi^{d/2}} \frac{\d^d \ell_2}{i\pi^{d/2}} 
    \frac{1}{ {\mbf{D}}^{\bs{\nu}} } 
    \, , 
\end{equation}
where $\bs{\nu}\in\mathbbm{Z}^5$, $\mbf{D}$ is a vector of the propagators
\begin{equation} \label{eq:5mProps}
    \hspace{-0.3cm}\begin{aligned}
    D_1 &= - \ell_1^2 + m_1^2-i\epsilon\,, 
    \quad&
    D_2 &= - (\ell_2 - p)^2 + m_2^2-i\epsilon\,, 
    \quad&
    D_3 &= - (\ell_1 - \ell_2)^2 + m_3^2-i\epsilon\,,\\
    D_4 &= - \ell_2^2 + m_4^2-i\epsilon\,, \quad&
    D_5 &= - (\ell_1 - p)^2 + m_5^2-i\epsilon\,,
    \quad&
    \end{aligned}
\end{equation}
and $i\epsilon$ is the Feynman $i\epsilon$ that enforces causal propagation. 
Here, we use the compact notation $\mbf{x}^\mbf{y} = \prod_{i=1}^{|\mbf{y}|=|\mbf{x}|} x_i^{y_i}$ (common in the mathematics literature).
Throughout this paper, the spacetime dimension is set to $d=2-2\vep$ since
the integral family has uniform transcendentality
and all integrals (besides the trivial double-tadpoles) are finite in two dimensions.
The corresponding integrals in higher dimensions (i.e., $d \to d + n$ for $n\in\mathbbm{N}_{>0}$) are obtained by the dimension-shift relations \cite{LEE2010474,PhysRevD.54.6479,Tarasov:1997kx}. 
We also fix $\mu=m_3$ to match the normalization choice for the known canonical basis for the 3-mass sunrise integrals \cite{bogner_unequal_2020}. 
As a consequence, our integrals become dimensionless functions of the following dimensionless parameters
\begin{align}
     X_0 = p^2/m_3^2\,, \ \, \,
     X_1 = m_1^2/m_3^2\,, \ \, \,
     X_2 = m_2^2/m_3^2\,, \ \, \,
     X_4 = m_4^2/m_3^2\,, \ \, \,
     X_5 = m_5^2/m_3^2\,.
\end{align}

In this paper, the \emph{kinematic space} parameterized by these variables is denoted by 
\begin{align*}
    \mathcal{K} = \{ \mbf{X}=(X_0,X_1,X_2,X_4,X_5) \in \mathbbm{C}^5 \setminus \text{Landau loci} \}\,.
\end{align*} 
Perturbatively, the singularity structure of QFT scattering amplitudes is governed by nonlinear polynomial systems known as \emph{Landau equations} \cite{Landau:1959fi} (see \cite{Mizera:2021icv,Gardi:2022khw,Fevola:2023kaw,Fevola:2023fzn} for recent developments). The solutions of the Landau equations (Landau loci) for the kite integral can be found at this \href{https://mathrepo.mis.mpg.de/PLD/}{URL} in the text file \texttt{kite\_generic\_generic.txt}.

\subsection{Starting basis and differential equations}
\label{subsec.twoell}

Our algorithm for constructing an $\varepsilon$-form basis (section \ref{epssec}) requires an initial/starting choice of basis and the associated differential equations as input. 

\begin{figure}
    \centering
    \includegraphics[align=c, scale=.75]{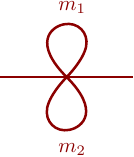}
    \qquad
    \includegraphics[align=c, scale=.75]{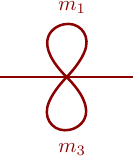}
    \qquad
    \includegraphics[align=c, scale=.75]{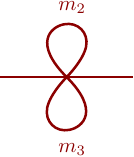}
    \qquad
    \includegraphics[align=c, scale=.75]{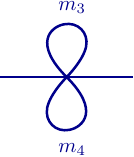}
    \qquad
    \includegraphics[align=c, scale=.75]{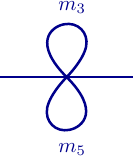}
    \qquad
    \includegraphics[align=c, scale=.75]{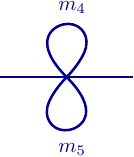}
    \\[1em]
    \includegraphics[align=c, scale=.75]{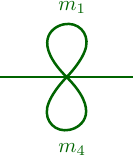}
    \qquad
    \includegraphics[align=c, scale=.75]{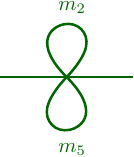}
    \caption{2-propagator subtopologies (all double tadpoles). The double tadpoles can be partitioned into three sets: those that can be obtained from pinching the {\color{BrickRed}$(123)$-sunrise} (top left triple), those that can be obtained from pinching the {\color{Blue}$(345)$-sunrise} (top right triple) and those that are {\color{darkGreen}not pinches of either sunrise} (bottom).}
    \label{fig:2propTop}
\end{figure}

\begin{figure}
    \centering
    \includegraphics[align=c, scale=.6]{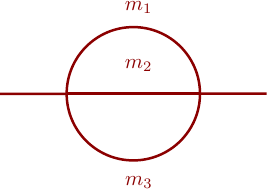}
    \qquad
    \includegraphics[align=c, scale=.6]{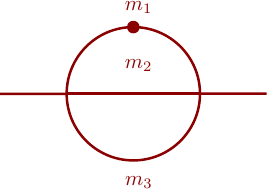}
    \qquad 
    \includegraphics[align=c, scale=.6]{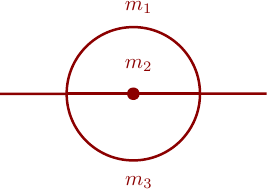}
    \qquad
    \includegraphics[align=c, scale=.6]{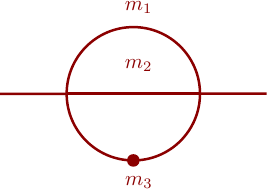}
    \\[2em]
    \includegraphics[align=c, scale=.6]{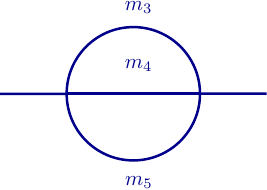}
    \qquad 
    \includegraphics[align=c, scale=.6]{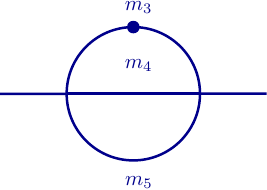}
    \qquad
    \includegraphics[align=c, scale=.6]{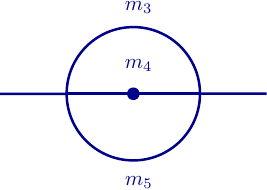}
    \qquad
    \includegraphics[align=c, scale=.6]{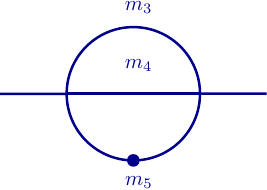}
    \\[2em] 
    \includegraphics[align=c, scale=.6]{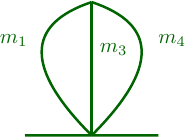}
    \qquad
    \includegraphics[align=c, scale=.6]{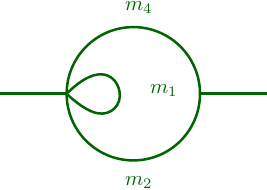}
    \qquad
    \includegraphics[align=c, scale=.6]{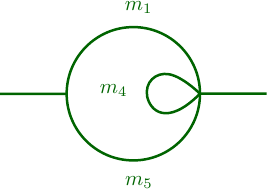}
    \qquad
    \includegraphics[align=c, scale=.6]{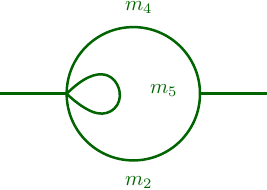}
    \\[2em]
    \includegraphics[align=c, scale=.6]{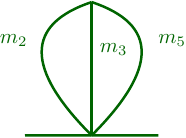}
    \qquad
    \includegraphics[align=c, scale=.6]{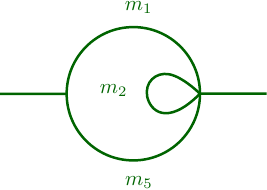}
    \qquad
    \includegraphics[align=c, scale=.6]{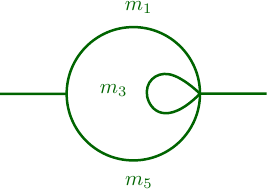}
    \qquad
    \includegraphics[align=c, scale=.6]{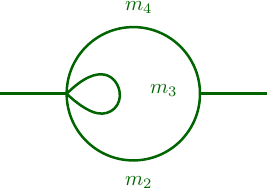}
    \caption{3-propagator subtopologies. Again, we can partition these subtopologies into three sets: the {\color{BrickRed}$(123)$-sunrise} sector (first row),  the {\color{Blue}$(345)$-sunrise} sector (second row) and the {\color{darkGreen} non-sunrise} sector (remaining rows). The two distinct sunrise sectors are associated to \emph{distinct} elliptic curves, while the non-sunrise sector is polylogarithmic.}
    \label{fig:3propTop}
\end{figure}

\begin{figure}
    \centering
    \includegraphics[align=c, scale=.6]{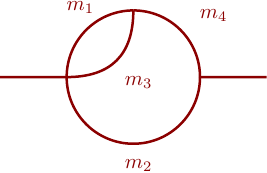}
    \qquad
    \includegraphics[align=c, scale=.6]{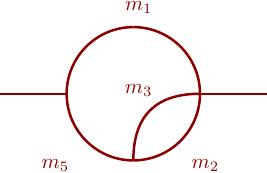}
    \qquad
    \includegraphics[align=c, scale=.6]{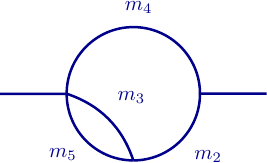}
    \qquad
    \includegraphics[align=c, scale=.6]{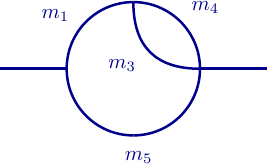}
    \\[2em]
    \includegraphics[align=c, scale=.65]{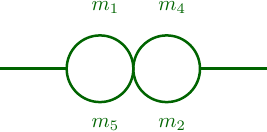}
    \qquad
    \includegraphics[align=c, scale=.6]{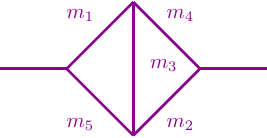}
    \caption{The 4-propagator subtopologies and the 5-propagator top sector. Clearly, all 4- and 5-propagator topologies are polylogarithmic according to the maximal cut classification. However, only the {\color{darkGreen}double bubble} (bottom left) is truly polylogarithmic. 
    The {\color{BrickRed}red eyeball} and {\color{Blue}blue eyeball} (top left and right pairs, respectively) are elliptic since they contain the {\color{BrickRed}$(123)$-sunrise} and {\color{Blue}$(345)$-sunrise} as subtopologies.
    There is only one diagram in the top-sector: the {\color{Plum} kite}. Since it contains both sunrise sectors as subtopologies, it knows about two elliptic curves. }
    \label{fig:45propTop}
\end{figure}

We choose a naive ``Laporta-like'' basis of master integrals and derive the associated differential equations using the publicly available software for IBP reduction \texttt{LiteRed} \cite{lee_presenting_2012} (see \cite{Smirnov:2008iw,smirnov_fire6_2020,Klappert_2021,Maierh_fer_2018} for alternatives). 
Explicitly, we choose the following basis for the kite family \cite{Tarasov:1997kx}
\begin{align} \label{eq:5mbasis}
\begin{aligned}
    \mbf{I}^\top
    =(&
    {\color{BrickRed} I_{1,1,0,0,0}}
    ,
    {\color{BrickRed} I_{1,0,1,0,0} }
    ,
    {\color{BrickRed} I_{0,1,1,0,0} }
    , 
    {\color{Blue} I_{0,0,1,1,0} }
    ,
    {\color{Blue} I_{0,0,1,0,1} }
    ,
    {\color{Blue} I_{0,0,0,1,1} }
    ,
    {\color{darkGreen} I_{1,0,0,1,0} }
    ,
   {\color{darkGreen} I_{0,1,0,0,1} }
    ,
    \\&
    {\color{BrickRed} I_{1,1,1,0,0} }
    ,
    {\color{BrickRed} I_{2,1,1,0,0} }
    ,
    {\color{BrickRed} I_{1,2,1,0,0} }
    ,
    {\color{BrickRed} I_{1,1,2,0,0} }
    ,
    {\color{Blue} I_{0,0,1,1,1} }
    ,
    {\color{Blue} I_{0,0,2,1,1} }
    ,
    {\color{Blue} I_{0,0,1,2,1} }
    ,
    {\color{Blue} I_{0,0,1,1,2} }
    ,
    \\&
    {\color{darkGreen} I_{1,0,1,1,0} }
    ,
    {\color{darkGreen} I_{1,1,0,1,0} }
    ,
    {\color{darkGreen} I_{1,0,0,1,1} }
    ,
    {\color{darkGreen} I_{0,1,0,1,1} }
    ,
    {\color{darkGreen} I_{0,1,1,0,1} }
    ,
    {\color{darkGreen} I_{1,1,0,0,1} }
    ,
    {\color{darkGreen}    I_{1,0,1,0,1}}
    ,    
     {\color{darkGreen}  I_{0,1,1,1,0}}
    ,
    \\&
    {\color{BrickRed} I_{1,1,1,1,0} }
    ,
    {\color{BrickRed} I_{1,1,1,0,1} }
    ,
    {\color{Blue} I_{0,1,1,1,1} }
    ,
    {\color{Blue} I_{1,0,1,1,1} }
    ,
    {\color{darkGreen}   I_{1,1,0,1,1}}
    ,
    \\&
    {\color{Plum}   I_{1,1,1,1,1} }
    )\,.
\end{aligned}
\end{align}
A diagrammatic representation of this basis is presented in figures \ref{fig:2propTop}, \ref{fig:3propTop} and \ref{fig:45propTop}.

Two distinct sunrise subtopologies appear in this basis: one with propagators $D_1,D_2$ and $D_3$ (the {\textcolor{BrickRed}{(123)}-sunrise} family) and one with propagators $D_3,D_4$ and $D_5$ (the {\textcolor{Blue}{(345)}-sunrise} family). 
Since it is well known that the sunrise integral is associated with an elliptic curve \cite{bogner_unequal_2020},
two distinct elliptic curves appear in the kite family: a {\textcolor{BrickRed}{(123)}-sunrise} and a {\textcolor{Blue}{(345)}-sunrise} elliptic curve. 
As a simple organizing principle, we use the following color code: diagrams in {\color{BrickRed}red} ({\color{Blue}blue}) either contain the {\textcolor{BrickRed}{(123)}-sunrise} ({\textcolor{Blue}{(345)}-sunrise}) as a subtopology or are a subtopology of the {\textcolor{BrickRed}{(123)}-sunrise} ({\textcolor{Blue}{(345)}-sunrise}). 
The only exception to this color coding is the {\color{Plum} kite} (denoted in {\color{Plum}purple}) since it contains both the {\textcolor{BrickRed}{(123)}-sunrise} and {\textcolor{Blue}{(345)}-sunrise} as subtopologies. 
All other topologies that  \textit{do not} contain a sunrise as a subtopology or \textit{ are not} a subtopology of a sunrise are colored {\color{darkGreen}green}.
These are not associated with any of the two elliptic curves and are polylogarithmic.

The basis \eqref{eq:5mbasis} satisfies the following system of differential equations
\begin{align} \label{eq:5mNCanDEQ}
    \d\mbf{I} = \mbf{A}\cdot\mbf{I}\,,
\end{align}
where ``$\d$'' stands for the exterior derivative on $\mathcal{K}$ and each component $A_{ij}$ is a differential 1-form on $\mathcal{K}$. By construction, the matrix $\mbf{A}$ in \eqref{eq:5mNCanDEQ} is \emph{integrable}, meaning that it is subject to the constraint 
\begin{equation}\label{eq:integrable}
    \d\mbf{A}+\mbf{A}\wedge\mbf{A}=0\,.
\end{equation}
Note that \eqref{eq:integrable} is just a compact way of saying that partial derivatives commute. 

Although the physically relevant integrals are contained in $\mbf{I}$, it is mathematically convenient to work with a so-called $\vep$-factorized basis of integrals $\mbf{J}= \mbf{U}\cdot\mbf{I}$. In dimensional regularization, such a basis satisfies (by definition) a differential equation of the form 
\begin{align} \label{eq:5mCanDEQ}
    \d\mbf{J}=\vep\,\mbf{B}\cdot\mbf{J}\,,
\end{align}
where $\mbf{B}$ is independent of $\vep$ and contains at most simple poles at the Landau loci. The differential equation of $\mbf{J}$ is related to that of $\mbf{I}$ by the gauge transformation
\begin{align}
    \label{trafogen}
    \varepsilon \mbf{B} = \mbf{U}\cdot\mbf{A}\cdot\mbf{U}^{-1} + \rd \mbf{U}\cdot\mbf{U}^{-1}\,. 
\end{align}
The differential equation \eqref{eq:5mCanDEQ} (and even \eqref{eq:5mNCanDEQ}) are formally solved by the \emph{path ordered exponential}
\begin{align}\label{eq:POE}
    \mbf{J}(\mbf{X})
    =\mathbbm{P} \exp\Big(\vep \int_\gamma \mbf{B}\Big) \cdot \mbf{J}_0\,,
\end{align}
where $\mbf{J}_0$ denotes a boundary value at some $\mbf{X}_0\in\mathcal{K}$ that needs to be computed separately (see section \ref{bdry}). The advantage of a canonical basis is that the path ordered exponential \eqref{eq:POE} can be expanded in small $\varepsilon$. That is, it admits a perturbative series expansion in terms of iterated integrals
 \begin{equation}\label{eq:ittSol}
 \begin{split}
    \mbf{J}(\mbf{X})
    =~&\left(
        \mathbbm{1}
        + \varepsilon \int_\gamma \mbf{B}
        + \varepsilon^2 \int_\gamma \mbf{B} \cdot \mbf{B} 
        + \mathcal{O}(\varepsilon^3)
    \right) \cdot \mbf{J}_0\, . 
\end{split}
\end{equation}

The fact that $\mathbf{B}$ is integrable (c.f., \eqref{eq:integrable}) guarantees that \eqref{eq:POE} depends only on the homotopy class $[\gamma]$; once the endpoints of $\gamma$ are fixed, the value of \eqref{eq:POE} does not change under \emph{smooth} deformations of the contour $\gamma$. Moreover, in the case where the $B_{ij}$'s are differential 1-forms over the punctured Riemann sphere (i.e., dlog forms), the iterated integrals in \eqref{eq:POE} evaluate to multiple polylogarithms (MPLs) \cite{kummerpolylogs,REMIDDI_2000,goncharov2001multiple, goncharov2011multiple}. The function space spanned by MPLs is arguably the simplest one can encounter in the evaluation of multiloop Feynman integrals, and there are now standard automated tools that can be used to evaluate MPLs numerically \cite{Vollinga_2005,Goncharov_2010,buehler2011chaplin,Duhr_2012,Duhr_2012A,Ablinger_2013,Frellesvig_2016,Ablinger_2019,Duhr_2019}. In cases where the $B_{ij}$'s involve differential 1-forms over more complicated geometries (e.g., (punctured) (hyper-)elliptic curves or higher-dimensional Calabi-Yau manifolds), much less is currently known concerning both analytic properties (see, e.g., \cite{brown2011decomposition,Adams_2014,Remiddi_2017,Broedel_2018,Broedel_2018II,Duhr_2020,weinzierl2022feynman,M_ller_2022,Broedel_2018III,Broedel_2019,Wilhelm_2023}) of the associated function spaces and on how to perform systematically high-precision numerical evaluation of the iterated integrals (see, e.g. \cite{Walden:2020odh} and section \ref{bdry}).

\subsection{Elliptic curves, tori and conventions}
\label{subsec_ellipticint}

In this section, we briefly review our conventions and the necessary concepts related to the study of elliptic curves and their relation to complex tori. For a more in-depth review, we refer the reader to \cite{elliptic_calc,weinzierl2022feynman}.

We work with a quartic representation of the elliptic curves as they appear in the maximal cut of the unequal mass $\e$- or $\ee$-sunrise subtopologies of the 5-mass kite \cite{bogner_unequal_2020}
\begin{equation}
\label{elcur}
    y^2_\alpha
    = \prod_{i=1}^4 (x_\alpha-e_i^\alpha)
    \,. 
\end{equation}
Here, $\alpha \in \{\e,\ee\}$ labels the elliptic curve/sunrise subtopology, the roots $e_1^\alpha$ are functions of the kinematic variables (momentum, masses, etc.) and $x_\alpha$ is a scalar depending on the loop momentum (e.g., a Baikov variable \cite{Baikov:1996iu}). 
Explicitly, the roots are:
\begin{subequations}\label{root ordering}
\begin{align}
    \big\{e_i^{\e}\big\} &{=} \Big\{
        {-}\left(\sqrt{X_1}{+}\sqrt{X_2}\right)^2,
        {-}\left(1{+}\sqrt{X_0}\right)^2,
        {-}\left(1{-}\sqrt{X_0}\right)^2,
        {-}\left(\sqrt{X_1}{-}\sqrt{X_2}\right)^2
    \Big\}\,,
    \\
    \big\{e_i^{\ee}\big\} &{=} \Big\{
        {-}\left(\sqrt{X_4}{+}\sqrt{X_5}\right)^2,
        {-}\left(1{+}\sqrt{X_0}\right)^2,
        {-}\left(1{-}\sqrt{X_0}\right)^2,
        {-}\left(\sqrt{X_4}{-}\sqrt{X_5}\right
        )^2
    \Big\}\,,
\end{align}
\end{subequations}
where $i=1,\dots,4$.
Furthermore, we will choose to work in a region of kinematic space such that $e^\alpha_i<e^\alpha_{i+1}$ for both $\alpha=\e$ and $\alpha=\ee$.

For every elliptic curve/complex torus, there are two natural periods, since the (co)homology groups are 2-dimensional. 
Explicitly, these periods are 
\begin{align}
    \psi_1^\alpha 
    =2\int_{e_2^\alpha}^{e_3^\alpha}
        \frac{\rd x_\alpha}{y_\alpha}  
    = \frac{2 K(k^2_\alpha)}{c_4^\alpha}
    \qquad \text{and} \qquad
    \psi_2^\alpha=2\int_{e_4^\alpha}^{e_3^\alpha} \frac{\rd x_\alpha}{y_\alpha} 
    = \frac{2 i K(1-k^2_\alpha)}{c_4^\alpha}
    \,,
\end{align}
where $c_4^\alpha = \frac12 \sqrt{(e_3^\alpha-e_1^\alpha)(e_4^\alpha-e_2^\alpha)}$.\footnote{Here, $c_4$, is an algebraic function that comes from some Jacobians when transforming the quartic elliptic curve to a cubic elliptic curve.}
Here, $K$ is the elliptic integral of first kind and $k$ is the \emph{elliptic modulus}:
\begin{equation}\label{eq:defK}
    K(k^2)=\int_0^1 \dfrac{\text{d}t}{\sqrt{(1-t^2)(1-k^2 t^2)}} 
    \quad \text{with}\quad 
    0\leq k^2=\frac{(e_3-e_2)(e_4-e_1)}{(e_3-e_1)(e_4-e_2)} \leq 1
    \,.
\end{equation}
Explicitly, the elliptic moduli for the elliptic curves are
\begin{subequations}
    \begin{align}
    k^2_\e ={-}\frac{16 \sqrt{X_0} \sqrt{X_1} \sqrt{X_2}}{X_0^2{-}2 \left(X_1{+}X_2{+}1\right) X_0{+}X_1^2{+}\left(X_2{-}1\right){}^2{-}8 \sqrt{X_0} \sqrt{X_1} \sqrt{X_2}{-}2 X_1 \left(X_2{+}1\right)}\,,
    \\
    k^2_\ee = {-}\frac{16 \sqrt{X_0} \sqrt{X_4} \sqrt{X_5}}{X_0^2{-}2 \left(X_4{+}X_5{+}1\right) X_0{+}X_4^2{+}\left(X_5{-}1\right){}^2{-}8 \sqrt{X_0} \sqrt{X_4} \sqrt{X_5}{-}2 X_4 \left(X_5{+}1\right)}\,.
\end{align}
\end{subequations}
In practice, it is convenient to eliminate one degree of freedom and normalize the periods of each elliptic curve such that
\begin{align}\label{def:tau}
   (\psi_1^\alpha,\psi_2^\alpha)
   \mapsto (1,\tau^\alpha) 
   \qquad \text{where} \qquad 
   \tau^\alpha =\frac{\psi_2^\alpha}{\psi_1^\alpha}
    =\dfrac{iK(1-k^2_\alpha)}{K(k^2_\alpha)} 
    \in \mathbbm{H}
   \,.
\end{align}
Before moving on, we pause to highlight the unusual choice for the definition of our periods. 
In most of the literature, the factor of $1/c_4$ is not included in the period.
We have made the choice to absorb this factor in the periods to simplify the presentation of some formulae.

Any given elliptic curve \eqref{elcur} is isomorphic to a complex torus $\mathbbm{T}=\mathbbm{C}/\Lambda_{(1,\tau)}$ with lattice $\Lambda_{(1,\tau)} = \mathbbm{Z}+\mathbbm{Z}\tau$ \cite{Broedel_2018,bogner_unequal_2020}
where we have dropped the label $\alpha$ to keep the notation more readable. 
This isomorphism is realized through \emph{Abel's map}, which maps any point $(x,y)$ on the elliptic curve onto two points $z^\pm$ on the torus (which is, in fact, a double cover of the elliptic curve)
\begin{align}
\label{abelsmap}
    (x,\pm y)\mapsto z^\pm
    &= \pm \frac{1}{\psi_1} \int_{e_1}^x
    \frac{\rd x}{y} 
    \quad\text{mod}\quad\Lambda_{(1,\tau)}
    \nn\\
    &
    = \pm \left[
        e^{i[\arg(x-e_1)-\arg(x-e_2)]}
        \frac{
            F(\sqrt{u_x},k^2)
        }{
            2 K(k^2)
        } 
        + \frac{\tau}{2}
    \right]
    \quad\text{for}\quad x\in\mathbbm{R} 
    \, , 
\end{align}
where 
\begin{align}
\label{ux}
    u_x & = \frac{x-e_2}{x-e_1}\frac{e_1-e_3}{e_2-e_3}\,, 
\end{align}
and $F$ is the incomplete elliptic integral of first kind
\begin{equation}
    \label{ellfirst}
F\left(\arcsin\sqrt{u},k^2\right)=\int_0^{\sqrt{u}} \dfrac{\text{d}t}{\sqrt{(1-t^2)(1-k^2 t^2)}}\, . 
\end{equation}
It is also important to note that one needs to use a branch choice 
\begin{align}
    \frac{y}{|y|}=
    \begin{sqcases}
        -1 & x \leq e_1 \text{ or } x>e_4 \,,
        \\
        -i & e_1 < x_1 \leq e_2 \,,
        \\
        1 & e_2 < x_1 \leq e_3 \,,
        \\
        i & e_3 < x_1 \leq e_4 \,,
    \end{sqcases}
\end{align}
that is compatible with the root ordering $e_i < e_{i+1}$
when comparing the integral definition of Abel's map to the definition in terms of incomplete elliptic integrals.

Now, because $\tau$ is not fixed and there are multiple punctures, the functions and differential forms that appear in this work are actually defined on the \emph{moduli space} of the torus with $n$ marked points, denoted by $\mathcal{M}_{1,n}$. Roughly speaking, each point in $\mathcal{M}_{1,n}$ corresponds to a tuple $(\tau, \mathbf{z})$, where $\tau\in\mathbb{H}$ and $\mathbf{z}\in\mathbb{T}_n$, which parameterizes a genus-1 surface with $n$ marked points. However, the proper definition of this space is quite involved and is not essential to understand the discussion that follows; hence, it is not defined here. Interested readers are referred to \cite[app. F]{weinzierl2022feynman} for more details.

It is often useful to organize functions (and differential forms) on $\mathcal{M}_{1,n}$ by \emph{modular weight}. 
A function $f(z, \tau)$ of modular weight $k+2$ is defined to be one such that
\begin{equation}\label{eq:modW}
    f(z,\tau)\mapsto(c\tau+d)^{k+2}f(z,\tau)\,,
\end{equation}
under the action of $\text{SL}(2, \mathbbm{Z})$ on $\mathbbm{T}$, which takes $\tau \mapsto \frac{a\tau + b}{c\tau + d}$ and $z \mapsto \frac{z}{c\tau + d}$ with $a,b,c,d\in\mathbbm{Z}$ and  $ad - bc = 1$.
In particular, $\d\tau$ and $\d z\vert_{\tau=\text{cte}}$ are modular forms of weight 0 and 1 respectively. 
Quasi-modular forms are a slight generalization of modular forms that will appear in later sections of this text. 
A quasi-modular form of weight $k$ and depth $p$ is a function whose image under a modular ($\text{SL}(2,\mathbbm{Z})$) transformation takes the form
\begin{align}\label{eq:qmf}
    f(z,\tau) 
    \mapsto \sum_{i=0}^p (c\tau+d)^{k+2} 
    \left(\frac{cz}{cz+d}\right)^i f_i(z,\tau)\,,
\end{align}
where the functions $f_i$ are holomorphic on the upper-half plane $\mathbbm{H}$.

As hinted above, it will be beneficial to express the 1-forms in the $\varepsilon$-form differential equation using variables defined on $\mathcal{M}_{1,n}^\e$ and $\mathcal{M}_{1,n}^\ee$. As a first variable on each moduli space, we naturally take the moduli $\tau^\e$ and $\tau^\ee$. 
Similarly, we can choose any number $n$ of punctures $z_i$ on each torus to account for the remaining kinematic degrees of freedom.
Due to the translation invariance on the torus, the punctures should be thought of as differences of points under the image of Abel's map. 
Therefore, we denote these punctures by 
\begin{align}
\label{difference}
    z_i=z_i^+-z_i^- = 2z_i^+\, . 
\end{align}
Furthermore, there may be linear relations between natural candidates for punctures (we will see an example later on) due to this translational invariance.

For the unequal mass sunrise integral, the change of variables between kinematic variables and toric variables is well known \cite{bogner_unequal_2020,giroux_loop-by-loop_2022}. For the $\e$- and $\ee$-sunrise subtopologies of the kite family, these punctures are
\begin{align}
    \label{sunrisepunctures}
    z_i^{{ \textcolor{BrickRed}{(123)}}} =\frac{F\Big(\arcsin\sqrt{u^{{ \textcolor{BrickRed}{(123)}}}_i} ,k_{{ \textcolor{BrickRed}{(123)}}}^2\Big)}{K\Big(k^2_{ \textcolor{BrickRed}{(123)}}\Big)} \qquad \text{ and }\qquad  z_i^{ \textcolor{Blue}{(345)} }=\frac{F\Big(\arcsin\sqrt{u^{{  \textcolor{Blue}{(345)}}}_i} ,k_{{  \textcolor{Blue}{(345)}}}^2\Big)}{K\Big(k^2_{  \textcolor{Blue}{(345)}}\Big)}\,,
\end{align}
where  we have dropped a $\tau = 2 \frac{\tau}{2} \sim 0$ in the difference \eqref{difference} when using \eqref{abelsmap} and
\begin{subequations}\label{sunpunct123}
\begin{align}
    u^{{ \textcolor{BrickRed}{(123)}}}_1
        &= \frac{\left(\sqrt{X_0}+\sqrt{X_1}\right)^2-\left(\sqrt{X_2}-1\right)^{2}}{4\sqrt{X_2}} 
    \, ,\quad
    u^{{ \textcolor{BrickRed}{(123)}}}_2
        = \frac{\left(\sqrt{X_0}+\sqrt{X_2}\right)^2-\left(\sqrt{X_1}-1\right)^{2}}{4\sqrt{X_1}}
    \,,
    \\
    u^{{ \textcolor{Blue}{(345)}}}_4
        &= \frac{\left(\sqrt{X_0}+\sqrt{X_4}\right)^2-\left(\sqrt{X_5}-1\right)^{2}}{4\sqrt{X_5}}  
    \, ,\quad
    u^{{ \textcolor{Blue}{(345)}}}_5
        =\frac{\left(\sqrt{X_0}+\sqrt{X_5}\right)^2-\left(\sqrt{X_4}-1\right)^{2}}{4\sqrt{X_4}} 
    \, .
\end{align}
\end{subequations}
Here, the $u$'s are obtained from $u_\infty$, using (\ref{ux}), and the following mass permutations
\begin{align}
    u_1 = u_\infty\vert_{X_0 \leftrightarrow X_2} 
    \qquad\text{and}\qquad
    u_2 = u_\infty\vert_{X_0 \leftrightarrow X_1}
    \,.
\end{align}
Thus, the 3-dimensional kinematic space of each sunrise is embedded in a complex torus whose independent variables consist of the modulus and two punctures: $\{\tau^\e,z^\e_{i=1,2}\}$ or $\{\tau^\ee,z^\ee_{i=4,5}\}$. There also exists a $z_3$ for both curves
\begin{align}
    u_3^\e = u_1^\e\vert_{m_1\leftrightarrow m_3}
    \,
    \qquad\text{and}\qquad
    u_3^\ee = u_4^\e\vert_{m_4\leftrightarrow m_3}
    \,.
\end{align}
However, these are not independent degrees of freedom since they are linearly related to the other punctures: $z^\e_1+z^\e_2+z^\e_3 = 1 = z^\ee_3+z^\ee_4+z^\ee_5$. 
Thus, a $z_3^\alpha$ will never appear in a final formula. 

Of course, the eyeball and kite integrals have more kinematic degrees of freedom than their sunrise subtopologies. 
Thus, the description of these integrals on the torus will inevitably require additional punctures. In section \ref{epssec}, we extract these new punctures for the eyball and kite integrals in a novel way using only the gauge transformation and the $\varepsilon$-form differential equation.  
An arguably simpler derivation from the maximal cut is described in section \ref{sec_punctureslab}. 
We also rediscover all the sunrise punctures  \eqref{sunrisepunctures} from degenerations of the maximal cut calculations. 
Although their derivation/motivation is provided in sections \ref{epssec} and \ref{sec_punctureslab}, we quote the ``new'' punctures here for the convenience of the reader
\begin{subequations}\label{eq:newz}
\begin{align}
    u_4^\e = u_2^\e
    \frac{
        \big(1 {+} \sqrt{X_1}\big)^2 {-} X_4
    }{
        (\sqrt{X_0} {+} \sqrt{X_2})^2 {-} X_4 
    }
    \,,
    \qquad
    u_5^\e = u_1^\e \frac{(1+\sqrt{X_2})^2-X_5}{(\sqrt{X_0}+\sqrt{X_{1}})^2-X_5}
    \,,
    \\
    u_1^\ee = u_{5}^\ee \frac{(1+\sqrt{X_4})^2-X_1}{(\sqrt{X_0}+\sqrt{X_5})^2-X_1}
    \,,
    \qquad
    u_2^\ee = u_{4}^\ee  \frac{(1+\sqrt{X_5})^2-X_2}{(\sqrt{X_0}+\sqrt{X_4})^2-X_2}
    \,.
\end{align}
\end{subequations}

At this stage, it is important to note that additional branch cuts must be taken into account when using these punctures. A more detailed explanation of this is provided later on around \eqref{eq:z4}. In summary, to fix a branch, we apply the following additional constraints to the kinematic variables $X_i$:
\begin{equation} \label{eq:kinRegion}
\begin{aligned} 
    &X_1{<}(\sqrt{X_0}{-}\sqrt{X_5})^2
    \,,
    \,
    &X_1{<}(1{-}\sqrt{X_4})^2
    \,,
    \quad
    &X_2{<}(\sqrt{X_0}{-}\sqrt{X_4})^2
    \,,
    \,
    &X_2{<}(1{-}\sqrt{X_5})^2
    \,,\\
    &X_4{>}(\sqrt{X_0}{+}\sqrt{X_2})^2
    \,,
    \,
    &X_4{>}(1{+}\sqrt{X_1})^2
    \,,
    \quad
    & X_5{>}(\sqrt{X_0}{+}\sqrt{X_1})^2
    \,,
    \,
    &X_5{>}(1{+}\sqrt{X_2})^2 
    \,.
\end{aligned}
\end{equation}
Only when such conditions are implemented is the numerical evaluation of the various elliptic quantities discussed above and below stable.

\section{Construction of an $\varepsilon$-form basis}
\label{epssec}
In this section, we illustrate a simple algorithm to derive an $\varepsilon$-form basis in the context of the kite integral family. 
For pedagogical clarity, we first analyze the subfamily of integrals with the \textcolor{BrickRed}{($1234$)}-eyeball diagram as its top topology in section \ref{sec_IC}. 
Since this subfamily depends only on one elliptic curve, we can exemplify the key aspects of our algorithm without the complication of two elliptic curves. 
Then, in section \ref{sec_epform}, we generalize this algorithm and obtain the $\varepsilon$-form differential equation for the entire kite integral family.
The gauge transformation to $\varepsilon$-form  constructed in this section, is provided in the ancillary \textsc{Mathematica} notebook \href{https://github.com/StrangeQuark007/kite_ancillary}{\faGithub}. 
\subsection{A warm up: the 4-mass eyeball family}
\label{sec_IC}
\begin{figure}
    \centering
    \includegraphics[align=c, scale=1.3]{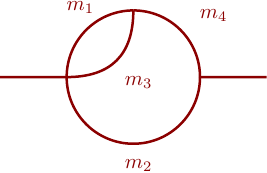}
\caption{The  two-mass eyeball graph of integral \textcolor{BrickRed}{$I_{1,1,1,1,0}$}  with massive propagators $D_{i=1,2,3,4}$.}
\label{fig.3}
\end{figure}

In this section, we construct an $\varepsilon$-form basis for the \textcolor{BrickRed}{($1234$)}-eyeball integral family $I_{\nu_1,\nu_2,\nu_3,\nu_4,0}$ (i.e., the subtopologies of the five-mass kite family with propagator $5$ pinched).
The top-sector of this integral family is spanned by the four-mass eyeball \textcolor{BrickRed}{$I_{1,1,1,1,0}$} (see figure \ref{fig.3}).

This family counts $13$ masters, which we choose to be the following subset of \eqref{eq:5mbasis}
\begin{equation}\label{eq:4mbasis}
    \begin{split}
        \mbf{I}_4^\top
    =(&
    {\color{BrickRed} I_{1,1,0,0,0}}
    ,
    {\color{BrickRed} I_{1,0,1,0,0} }
    ,
    {\color{BrickRed} I_{0,1,1,0,0} }
    ,
    {\color{darkGreen} I_{1,0,0,1,0} }
    , 
    {\color{Blue} I_{0,0,1,1,0} }
    ,
    {\color{BrickRed} I_{1,1,1,0,0} }
    ,
    {\color{BrickRed} I_{2,1,1,0,0} }
    ,
    {\color{BrickRed} I_{1,2,1,0,0} }
    ,
    {\color{BrickRed} I_{1,1,2,0,0} }
    ,
    \\&
    {\color{darkGreen} I_{1,0,1,1,0} }
    ,
    {\color{darkGreen} I_{1,1,0,1,0} }
     {\color{darkGreen}  I_{0,1,1,1,0}}
    ,
    {\color{BrickRed} I_{1,1,1,1,0} }
    )\,.
    \end{split}
\end{equation}
Note that this basis contains all the master integrals that span the {\textcolor{BrickRed}{(123)}-sunrise} family, an element from the {\textcolor{Blue}{(345)}-sunrise} basis, as well as some two- and three-propagator integrals that are not associated with either of the two sunrises. It also satisfies a differential equation like \eqref{eq:5mNCanDEQ}, namely
\begin{align}
\label{eyeballdiffeq}
    \rd \mbf{I}_4= \mbf{A}_4\cdot\mbf{I}_4\, ,
\end{align}
where $\mathbf{A}_4$ is a $13\times13$ submatrix of $\mathbf{A}$. 
Schematically, $\mbf{A}_4$ has the form 
\begin{equation}
\label{smallmat2}
 \mbf{A}_4= \left(\begin{array}{cccccccccccccc}
    \color{darkGreen}{\blacktriangle}& & & \\
    &\color{darkGreen}{\blacktriangle}& & \\
\color{BrickRed}{\blacksquare}& &\color{BrickRed}{123}& &   \\
    \color{darkGreen}{\blacktriangle}&\color{darkGreen}{\blacktriangle}& &\color{darkGreen}{\blacktriangle}&  \\
 & &\color{BrickRed}{\blacksquare}&\color{darkGreen}{\blacktriangle}&\color{BrickRed}{1234}
    \end{array}\right)\,,
\end{equation}
where the $\color{BrickRed}{\blacksquare}$-blocks denote off-diagonal entries associated to the $\e$-elliptic curve (e.g., integrals including the sunrise as a subtopology), while the $\color{darkGreen}{\blacktriangle}$-blocks denote both on- and off-diagonal entries that are \emph{not} associated to this curve.
Furthermore, the diagonal blocks \textcolor{BrickRed}{123} and \textcolor{BrickRed}{1234} denote the top sectors of the \textcolor{BrickRed}{(123)}-sunrise and \textcolor{BrickRed}{(1234)}-eyeball family, respectively. White/empty entries are zero. Since the eyeball is only connected to the \textcolor{BrickRed}{(123)}-elliptic curve, in this subsection, we will often omit the labels referring to the elliptic curve to avoid notational clutter (e.g., $\tau=\tau^{\textcolor{BrickRed}{(123)}}$ here).

Next, we construct an $\varepsilon$-form basis in a systematic and iterative manner by using a series of gauge transformations.

\subsubsection*{The sunrise $\vep$-form transformation}

The first step is to apply the known gauge transformation $\mbf{U}^\e_{4\times4}$ that puts the sunrise block \textcolor{BrickRed}{123} in $\vep$-form \cite{bogner_unequal_2020}. This first transformation matrix has the form 
\begin{equation}\label{eq:U1eb}
    \mbf{U}_4^\rb{1}=
    \text{diag}\big(\vep^2\times\mathbbm{1}_{5\times 5},\mbf{U}^{\e}_{4\times4},
    \vep^2\times \mathbbm{1}_{4\times4}
    \big)\,.
\end{equation}
The change of basis $\mbf{J}_4^\rb{1} = \mbf{U}_4^\rb{1} \cdot \mbf{I}_4$ is the first step towards the $\varepsilon$-basis $\mbf{J}_4$. 
The new basis $\mbf{J}_4^\rb{1}$ satisfies the gauge transformed differential equation $\d \mbf{J}_4^\rb{1} = \mbf{B}_4^\rb{1} \cdot \mbf{J}_4^\rb{1}$ with 
\begin{align}\label{DEQ_sunrise}
    \mbf{B}_4^\rb{1} 
    = \d \mbf{U}_4^\rb{1} 
        \cdot \left(\mbf{U}_4^\rb{1}\right)^{-1} 
    + \mbf{U}_4^\rb{1} \cdot \mbf{A}_4 
        \cdot \left(\mbf{U}_4^\rb{1}\right)^{-1}
    \,.
\end{align}
Although the sunrise block in $\mbf{B}_4^\rb{1}$ is now in $\varepsilon$-form, there are still order $\mathcal{O}(\varepsilon^0)$ entries that we wish to remove: 
\begin{align}
    \left(B_4^\rb{1}\right)_{10,10}\,,
    \left(B_4^\rb{1}\right)_{11,11}\,,
    \left(B_4^\rb{1}\right)_{12,12}\,,
    \left(B_4^\rb{1}\right)_{13,13}
    \,\quad\text{and}\,\quad
    \left(B_4^\rb{1}\right)_{13,6}\,.
\end{align} 
In the next few paragraphs, we explain how to do so systematically.

\subsubsection*{Removing $\mathcal{O}(\varepsilon^0)$ diagonal entries}

The master integrals associated with the four nontrivial $\mathcal{O}(\varepsilon^0)$ diagonal entries mentioned above are all polylogarithmic. Consequently, they can be unambiguously removed by leading singularity renormalizations \cite{Henn:2013pwa}. From the standpoint of the gauge transformation, this amounts to solving the homogeneous equations
\begin{equation}
\label{vii}
\rd \log u_{ii} =-\lim_{\varepsilon\to 0}\big({B}_4^\rb{1}\big)_{ii} \qquad \text{for}~i\in\{10,11,12,13\}\,,
\end{equation}
and multiplying the $i^\text{th}$ basis element by the solution $u_{ii}$. The solutions are 
\begin{subequations}
    \begin{align}
\label{firstvii}
    u_{10,10}&=
    \sqrt{\lambda_{134}}
    \,, 
    \\
    u_{11,11}&=
    u_{12,12}=
    \sqrt{\lambda_{024}}
    \,,
    \\ u_{13,13}&=
    \sqrt{\lambda_{024}}
    \sqrt{\lambda_{134}}
    \,,
    \label{v1313}
\end{align}
\end{subequations}
where $\lambda(a,b,c)=a^2+b^2+c^2-2ab-2ac-2bc$ is the K\"{a}llen function, $\lambda_{abc} = \lambda(X_a,X_b,X_c)$  and $X_3=m_3^2/m_3^2=1$. 
Note that the entries in (\ref{firstvii}-\ref{v1313}) correspond to the inverses of the respective $\vep=0$ maximal cuts. 
For example,
\begin{equation}
    \text{MC}_0(\textcolor{BrickRed}{I_{1,1,1,1,0}})=u_{13,13}^{-1}\, . 
\end{equation}
Thus, the second transformation matrix is
\begin{align}
    \mbf{U}_4^{(2)}&=
    \text{diag}\big(\overbracket[0.4pt]{{1, \dots, 1}}^{\times9},u_{10,10},u_{11,11},u_{12,12},u_{13,13}\big)\,,
    \\
    &= \text{diag}\big(1, \dots, 1,
    \text{MC}_0^{-1}(\textcolor{darkGreen}{I_{1,0,1,1,0}}),
    \text{MC}_0^{-1}(\textcolor{darkGreen}{I_{1,1,0,1,0}}),
    \text{MC}_0^{-1}(\textcolor{darkGreen}{I_{0,1,1,1,0}}),
    \text{MC}_0^{-1}(\textcolor{BrickRed}{I_{1,1,1,1,0}})
    \big)
    \,.
    \nn
\end{align}
As before, we define the next basis in the sequence towards the $\varepsilon$-form by $\mathbf{J}_4^\rb{2} = \mathbf{U}_4^\rb{2} \cdot \mathbf{J}_4^\rb{1}$. 
This basis satisfies a system of differential equations $\d\mbf{J}_4^\rb{2} = \mbf{B}_4^\rb{2}\cdot\mbf{J}_4^\rb{2}$ with connection matrix 
\begin{align}
    \mathbf{B}_4^\rb{2} = 
    \mathrm{d} \mathbf{U}_4^\rb{2} \cdot \left(\mathbf{U}_4^\rb{2}\right)^{-1} 
    + \mathbf{U}_4^\rb{2} \cdot \mathbf{B}_4^\rb{1} \cdot \left(\mathbf{U}_4^\rb{2}\right)^{-1}
    \,.
\end{align} 
Now, the only entry of $\mathbf{B}_4^\rb{2}$ that is not in $\varepsilon$-form is $\left(B_4^2\right)_{13,6}$. 
To remove this entry, we follow the algorithm introduced in \cite{giroux_loop-by-loop_2022}, which is based on the underlying $\text{SL}(2,\mathbb{Z})$ covariance of the differential equation.

\subsubsection*{$\text{SL}(2,\mathbb{Z})$ covariance: removing $\mathcal{O}(\varepsilon^0)$ off-diagonal terms}
\label{subsec_icmod}

First, we make the $\varepsilon$-dependence in the connection explicit:
$\mbf{B}_4^\rb{2}$ has an $\mathcal{O}(\varepsilon^0)$ and $\mathcal{O}(\varepsilon)$ term
\begin{align}
    \mbf{B}_4^{(2)} = \mbf{B}_4^\rb{2,0} 
    + \varepsilon\ \mbf{B}_4^\rb{2,1}
    \,.
\end{align}
In general, the $\mathcal{O}(\varepsilon^0)$ matrix is strictly lower triangular, and, for this example, the only nonvanishing matrix element is $(\mbf{B}_4^\rb{2})_{13,6}$.
This entry can be removed through a third transformation $\mathbf{U}_4^\rb{3}$
that satisfies $\mathbf{U}_4^\rb{3} \cdot \mathbf{B}_4^{(2,0)} + \mathrm{d}\mathbf{U}_4^\rb{3}=0$.
The lower triangular form of $\mathbf{B}_4^{(2,0)}$ suggests the following ansatz
\begin{equation}\label{gagueaway}
    \mbf{U}_4^\rb{3}=\mathbbm{1}_{13\times13}
    + \begin{pmatrix}
        0&\hdots&0&\hdots&0\\
        \vdots&\ddots&\vdots&\ddots&\vdots\\
        0&\hdots&v_{13,6}&\hdots&0
    \end{pmatrix}
    \qquad\text{with}\qquad
    \text{d}u_{13,6} + \Big({B}_4^{(2,0)}\Big)_{13,6} = 0
    \,.
\end{equation}
Note that the inversion of $\text{d}u_{13,6}=-\big({B}_4^{(2,0)}\big)_{13,6}$ is unambiguous only because it is closed. 
When $\mbf{B}_4^\rb{2,0}$ is not closed, an additional gauge transformation is applied that renders it closed before this step.
We will see how to do this systematically when discussing the full kite family.
 
Since the only non-trivial modular objects are the period and its $X_0$-derivative,
we split $\big({B}_4^{(2,0)}\big)_{13,6}$ into terms proportional to $\psi_1$ and ${\partial_0}\psi_1$
\begin{equation}\label{eq:B4136}
    \left({B}_4^{(2,0)}\right)_{13,6}
    = \sigma(X_i,\text{d}X_i)\ \psi_1
    + \rho(X_i,\text{d}X_i)\ {\partial_0} \psi_1
    \,.
\end{equation}
Here, $\sigma$ and $\rho$ are differential 1-forms on $\mathcal{K}$ and are modular invariant under modular transformations within the congruence subgroup $\Gamma(2)$ of $\text{SL}(2,\mathbb{Z})$ \cite{giroux_loop-by-loop_2022}. 

The first step of the ``modular bootstrap'' approach introduced in \cite{giroux_loop-by-loop_2022} is to perform a modular transformation on \eqref{gagueaway}. 
From the right-hand side of \eqref{eq:B4136}, it is clear that $\big(B_4^{(2,0)}\big)_{13,6}$ (and so $\d u_{13,6}$) transforms like the \emph{derivative} of a modular form of weight $1$ (c.f., \eqref{eq:modW}). 
Thus, $u_{13,6}$ must transform as a modular form of weight $1$. 
Applying a modular transformation to \eqref{gagueaway} yields
\begin{align}\label{gagueaway2}
    \underbracket[0.4pt]{{
        (c\tau+d)
        \left[\text{d}u_{13,6} + \Big({B}_4^{(2,0)}\Big)_{13,6} \right]
    }}_{=0}
    +
    c \left( u_{13,6}\ \d\tau + \rho\ {(\partial_0\tau)}\ \psi_1 \right) = 0\,.
\end{align}
Changing the coordinates from the eyeball kinematic variables $(X_0,X_1,X_2,X_4)$ to the toric variables $\boldsymbol{\zeta}\equiv (\tau,z_1,z_2,z_4)$ reveals that $\rho$ depends only on $\text{d}\tau$ and not on punctures $\text{d}z_i$. 
Note that the torus for this eyeball diagram is the $\e$-torus with an additional puncture $z_4$ since the eyeball has one more kinematic degree of freedom than the sunrise
\begin{equation}
    \label{eq:z4}
    z_4=z_4^{\e}
    =\dfrac{\sqrt{(e_1-e_3)(e_2-e_4)}}{2\psi_1}
        \int_{e_2}^{-X_4}\dfrac{\d x}{y}\biggr|_{X_0\leftrightarrow X_1}
    =\dfrac{F\left(\arcsin\sqrt{u_4},k^2\right)}{K(k^2)}\biggr|_{X_0\leftrightarrow X_1}
    \,,
\end{equation}
where
\begin{equation}\label{eq:u4123}
    u_4
    =u_4^{\e}
    = (u_{-X_4})\vert_{X_0\leftrightarrow X_1}
    = u_2^\e
    \frac{
        \big(1 {+} \sqrt{X_1}\big)^2 {-} X_4
    }{
        (\sqrt{X_0} {+} \sqrt{X_2})^2 {-} X_4 
    }
    \,.
\end{equation}
The elliptic functions in \eqref{eq:z4} introduce new branch cuts and to select a branch, we restrict $X_4$ to the region where
\begin{equation}\label{branch z4}
   X_4>(\sqrt{X_0}+\sqrt{X_2})^2\text{   and   } X_4>(1+\sqrt{X_1})^2\, .
\end{equation}
Be aware that, here, $u_4=u_4^{\e}$ should \emph{not} be mistaken for $u_4^{\ee}$ (previously introduced in \eqref{sunrisepunctures}) and the same applies for  $z_4$. 
Although this definition may seem random at first, we explain its origin after giving the solution for $u_{13,6}$ in this section.
A more detailed discussion explaining how the punctures can be systematically obtained from the maximal cut is provided in section \ref{sec_punctureslab}. 

Under the change of variables $(X_0,X_1,X_2,X_4)\mapsto\boldsymbol{\zeta}$, 
the kinematic differentials become
\begin{align}
    \d X_i = \frac{\partial X_i}{\partial \zeta_j} \d\zeta_j\,,
\end{align}
where the partial derivatives are determined from the inverse of the Jacobian matrix $\mathcal{J}_{ij} = \partial\zeta_i/\partial X_j$.
Substituting this into \eqref{gagueaway2} and extracting the $\d\tau$ component yields 
\begin{align}
    u_{13,6} = - \psi_1 \frac{\partial \tau}{\partial X_0} 
        \sum_{i=0,1,2,4} \rho_i \frac{\partial X_i}{\partial \tau}\,,
\end{align}
where $\rho = \sum_{i=0,1,2,4} \rho_i\ \d X_i$ and $\partial X_i /\partial\tau = \mathcal{J}^{-1}_{i\tau}$.
Putting everything together, we obtain an $\varepsilon$-form basis for the eyeball $\mbf{J}_4 = \mbf{U}_4^\rb{3} \cdot \mbf{J}_4^\rb{2}$. 
It satisfies the differential equation
\begin{align}
\label{eyeballep}
\begin{aligned}
    \d\mbf{J}_4 = \varepsilon\ \mbf{B}_4 \cdot \mbf{J}_4\quad \text{where} \quad
    \mbf{B}_4 
    = \mbf{U}_4^\rb{3} \cdot \mbf{B}_4^\rb{2,1} \cdot \left(\mbf{U}_4^\rb{3}\right)^{-1}
    \,,
\end{aligned}
\end{align}
since $        \d \mbf{U}_4^\rb{3} \cdot \left(\mbf{U}_4^\rb{3}\right)^{-1} 
        + \mbf{U}_4^\rb{3} \cdot \mbf{B}_4^\rb{2,0} \cdot \left(\mbf{U}_4^\rb{3}\right)^{-1}
        = 0$
by construction.

Before moving on to the full kite integral family, we provide some motivation for the definition of $z_4$ in \eqref{eq:z4}. Since the sunrise torus only has three scales, we need a \emph{new} elliptic integral that can be interpreted as a puncture to account for the four kinematic scales of the eyeball. The question is \emph{how} to find it without too much guesswork. 

In principle, such a function could be found by directly integrating \eqref{gagueaway}. Of course, this is much harder than the above prescription, which relied only on understanding the modular covariance of the differential equation. Still, by examining a particularly simple piece of $\mbf{B}_4^\rb{2,0}$, we can discover the new puncture. 

Consider now the $\partial_0\psi_1$ term of \eqref{eq:B4136}. 
The $\d X_4$ component of $\rho$ is a function of $X_4$
\begin{align}
    \rho_4 = \frac{f(X_0,X_1,X_2)}{ \sqrt{\lambda(1,X_1,X_4)} \sqrt{\lambda(X_0,X_2,X_4)} }
    \,,
\end{align}
where $f$ is a rational function of its arguments. 
Since $\mbf{B}_4^\rb{2,0}$ is closed, we can obtain $u_{13,6}$ by integrating in $X_4$.
Thus, the gauge transformation must contain the following integral
\begin{align} \label{eq:ellipticFfromMC}
    u_{13,6} \supset (\partial_0\psi_1) f(X_0,X_1,X_2) 
    \int
    \underbracket[0.4pt]{{
         \frac{\d X_4}{ \sqrt{\lambda(1,X_1,X_4)} \sqrt{\lambda(X_0,X_2,X_4)} }
    }}_{
        \d X_4\ \text{MC}({\color{BrickRed} I_{1,1,1,1,0}})
    }
    \,.
\end{align}
This integral is proportional to the ``new'' incomplete elliptic integral $F(u_\phi,k^2_\phi)$ where $u_\phi$ and $k_\phi$ are given in \eqref{eq:kphiEB} and \eqref{eq:uphi}. 
This can be rewritten in terms of $u^\e$ and $k^2_\e$ (as shown in appendix \ref{app:massage}) and leads to the definition \eqref{eq:z4}. 

Recalling that we rescaled our basis by the leading singularities (see equation \eqref{v1313}), it should not be surprising that the integral over the maximal cut appears in the gauge transformation that brings the basis into $\varepsilon$-form. 
This will be explored in more detail and generality in section \ref{sec_punctureslab}. 

\subsection{The 5-mass kite family}
\label{sec_epform}
Using the methodology of section \ref{sec_IC}, we derive an $\vep$-form basis for the full 5-mass kite family \eqref{eq:5mbasis} in this section. The matrix $\mbf{A}$ in the differential equation \eqref{eq:5mNCanDEQ} takes the schematic form 
\begin{equation}
\label{smallmatrix}
  \mbf{A}=\left(\begin{array}{ccccccccccccccccc}
    \color{darkGreen}{\blacktriangle}& & & & & & & & & & &
    \\
    &\color{darkGreen}{\blacktriangle}& & & & & & & & & 
    \\
    \color{BrickRed}{\blacksquare}& &\color{BrickRed}{123}& & & & & & & & 
    \\
    &\color{Blue}{\blacksquare}& &\color{Blue}{345}& & & & & & & 
    \\
    \color{darkGreen}{\blacktriangle}&\color{darkGreen}{\blacktriangle}& & &\color{darkGreen}{\blacktriangle}& & & & & 
    \\
    & &\color{BrickRed}{\blacksquare}& &\color{darkGreen}{\blacktriangle}&\color{BrickRed}{1234}& & & & & 
    \\
    & &\color{BrickRed}{\blacksquare}& &\color{darkGreen}{\blacktriangle}& &\color{BrickRed}{1235}& & & & 
    \\
    & & &\color{Blue}{\blacksquare}&\color{darkGreen}{\blacktriangle}& & &\color{Blue}{2345}& & & 
    \\
    & & &\color{Blue}{\blacksquare}&\color{darkGreen}{\blacktriangle}& & & &\color{Blue}{1345}& & 
    \\
    & & & &\color{darkGreen}{\blacktriangle}& & & & &\color{darkGreen}{\blacktriangle}& 
    \\
    & &\color{BrickRed}{\blacksquare}&\color{Blue}{\blacksquare}&\color{darkGreen}{\blacktriangle}&\color{darkGreen}{\blacktriangle} &\color{darkGreen}{\blacktriangle} &\color{darkGreen}{\blacktriangle} & \color{darkGreen}{\blacktriangle}&\color{darkGreen}{\blacktriangle} &\color{Plum}{\text{12345}}
    \\
    \end{array}\right)\, .
\end{equation}
As before, the blocks $\color{BrickRed}{\blacksquare}$ and $\color{Blue}{\blacksquare}$ represent off-diagonal entries associated with the $\e$ and $\ee$ sunrises while the $\color{darkGreen}{\blacktriangle}$ blocks represent entries that are not associated with either torus. 
Moreover, the diagonal blocks with elliptic subtopologies are denoted by the corresponding propagator labels of the top sector. It is important to note that, with the exception of the top sector (kite), the two elliptic curves are completely decoupled from one another. 

We now proceed with the construction of the $\varepsilon$-form basis. As before, this process can be systematically organized as the construction of a sequence of gauge transformations. 

\subsubsection*{The sunrises $\vep$-form transformations} 

First, we apply the known transformations $\mbf{U}^{\e}_{4\times4}$ and $\mbf{U}^{\ee}_{4\times4}$ for both sunrise subtopologies \cite{bogner_unequal_2020}. Since the corresponding blocks fully decouple, this gauge transformation is block-diagonal and given by
\begin{align}\label{eq:U1}
    \mbf{U}^\rb{1} = \text{diag}\big(\vep^2\times\mathbbm{1}_{8\times8},\mbf{U}^{\e}_{4\times4},\mbf{U}^{\ee}_{4\times4},\vep^2\times\mathbbm{1}_{14\times14}\big)\, .
\end{align}
Then, we define 
\begin{equation}\label{B_for_kite}
    \mbf{J}^\rb{1} = \mbf{U}^\rb{1} \cdot \mbf{I}\,\,\,\,\, \text{such that}\,\,\,\,\,  \mbf{B}^\rb{1} = \d \mbf{U}^\rb{1} \cdot \left(\mbf{U}^\rb{1}\right)^{-1} + \mbf{U}^\rb{1} \cdot \mbf{A} \cdot \left(\mbf{U}^\rb{1}\right)^{-1}\, .
\end{equation}
At this stage, $\mbf{B}^\rb{1}$ has both an $\mathcal{O}(\varepsilon^0)$ and an $\mathcal{O}(\varepsilon)$ part, with the former being lower-triangular.

\subsubsection*{Removing $\mathcal{O}(\varepsilon^0)$ diagonal entries}

To eliminate the $\mathcal{O}(\varepsilon^0)$ terms from the diagonal, we follow the previous method of normalizing the relevant basis elements with the corresponding leading singularities (note that since both sunrise diagonal blocks have been converted to $\varepsilon$-form in the previous step, all these leading singularities are polylogarithmic). Specifically, the second transformation is given by
\begin{align}\label{eq:U2}
    \mbf{U}^\rb{2} = \mathrm{diag}\left(
        \mathbbm{1}_{1\times16},
        \mbf{u}_2
    \right)\,,
\end{align}
where 
\begin{align}
\label{u2max}
    \mbf{u}_2 &= \text{MC}^{-1}_0 \Big( 
    {\color{darkGreen} I_{1,0,1,1,0} }
    ,
    {\color{darkGreen} I_{1,1,0,1,0} }
    ,
    {\color{darkGreen} I_{1,0,0,1,1} }
    ,
    {\color{darkGreen} I_{0,1,0,1,1} }
    ,
    {\color{darkGreen} I_{0,1,1,0,1} }
    ,
    {\color{darkGreen} I_{1,1,0,0,1} }
    ,
    {\color{darkGreen}    I_{1,0,1,0,1}}
    ,    
     {\color{darkGreen}  I_{0,1,1,1,0}}
    ,
    \nn\\&\qquad\qquad\qquad
    {\color{BrickRed} I_{1,1,1,1,0} }
    ,
    {\color{BrickRed} I_{1,1,1,0,1} }
    ,
    {\color{Blue} I_{0,1,1,1,1} }
    ,
    {\color{Blue} I_{1,0,1,1,1} }
    ,
    {\color{darkGreen}   I_{1,1,0,1,1}}
    ,
    {\color{Plum}   I_{1,1,1,1,1} }
    \Big)\,,
    \nn\\
    &= \Big(
       {\color{darkGreen}\sqrt{\lambda_{134}}},
       {\color{darkGreen}\sqrt{\lambda_{024}}},
       {\color{darkGreen}\sqrt{\lambda_{015}}},
       {\color{darkGreen}\sqrt{\lambda_{024}}},
       {\color{darkGreen}\sqrt{\lambda_{235}}},
       {\color{darkGreen}\sqrt{\lambda_{015}}},
       {\color{darkGreen}\sqrt{\lambda_{015}}},
       {\color{darkGreen}\sqrt{\lambda_{024}}},
       \\&\qquad\qquad
       {\color{BrickRed}\sqrt{\lambda_{134}}\sqrt{\lambda_{024}}},
       {\color{BrickRed}\sqrt{\lambda_{235}}\sqrt{\lambda_{015}}},
        {\color{Blue} \sqrt{\lambda_{235}}\sqrt{\lambda_{024}}},
        {\color{Blue} \sqrt{\lambda_{134}}\sqrt{\lambda_{015}}},
       {\color{darkGreen}\sqrt{\lambda_{024}}\sqrt{\lambda_{015}}},
        {\color{Plum}  \lambda_{01245}}
    \Big)\nn\, .
\end{align}
We recall that $\lambda_{ijk}$ is a shorthand for $\lambda(X_i,X_j,X_k)$ defined below \eqref{v1313} and
\begin{equation}\label{eq:lamFull}
\begin{split}
    \lambda_{01245}&=X_1^2 X_2+X_1 X_2^2-X_1 X_2-X_0 X_1 X_2-X_4 X_1 X_2-X_5 X_1 X_2+X_4^2 X_5
    \\&\quad 
    +X_5^2 X_4-X_4 X_5-X_0 X_4 X_5-X_1 X_4 X_5-X_2 X_4 X_5+X_0^2+X_0-X_1 X_0
    \\&\quad 
    -X_2 X_0-X_4 X_0-X_5 X_0+X_1 X_4+X_0 X_1 X_5+X_2 X_5+X_0 X_2 X_4\,.
\end{split}
\end{equation}
As before, we let $\mbf{J}^\rb{2} = \mbf{U}^\rb{2} \cdot \mbf{J}^\rb{1}$ such that $\mbf{B}^\rb{2} = \d \mbf{U}^\rb{2} \cdot \left(\mbf{U}^\rb{2}\right)^{-1} + \mbf{U}^\rb{2} \cdot \mbf{B}^\rb{1} \cdot \left(\mbf{U}^\rb{2}\right)^{-1}$. Thus, $\mbf{B}^\rb{2}$ still has both an $\mathcal{O}(\varepsilon^0)$ and an $\mathcal{O}(\varepsilon)$ part, but the former is now \emph{strictly} lower-triangular. As these two matrices will appear repeatedly in the remaining steps, we name them $\mbf{B}^\rb{2,0}$ and $\mbf{B}^\rb{2,1}$, respectively.

Next, we wish to remove the remaining $\mathcal{O}(\varepsilon^0)$ entries by leveraging the underlying $\text{SL}(2,\mathbb{Z})$ symmetry. However, before proceeding, we must ensure that $\mbf{B}^\rb{2,0}$ is closed under $\d$. Currently, this is not the case since the matrix elements $\big(\mbf{B}^\rb{2,0}\big)_{30,9}$ and $\big(\mbf{B}^\rb{2,0}\big)_{30,13}$ (kite-sunrise components) are not closed. 
Thus, we must perform an $\varepsilon$-independent gauge transformation $\mbf{U}^\rb{3}$ that ensures that the new $\mathcal{O}(\vep^0)$ terms are closed before integrating them out. 
Explicitly, we require a $\mbf{U}^\rb{3}$ such that 
\begin{equation}\label{J3}
    \mbf{J}^\rb{3} = \mbf{U}^\rb{3} \cdot \mbf{J}^\rb{2}\,\,\,\,\, \text{and} \,\,\,\,\,\mbf{B}^\rb{3}=\mbf{U}^\rb{3}\cdot\mbf{B}^\rb{2}\cdot\left(\mbf{U}^\rb{3}\right)^{-1}+\d \mbf{U}^\rb{3}\cdot \left(\mbf{U}^\rb{3}\right)^{-1}\, ,
\end{equation}
with $\d \mbf{B}^\rb{3,0}=0$.

\subsubsection*{$\text{SL}(2,\mathbb{Z})$ covariance: making $\mathcal{O}(\vep^0)$ terms closed} 

To construct $\mbf{U}^\rb{3}$ as described above, we focus on the integrability conditions of $\mbf{B}^\rb{3,0}$
\begin{equation}\label{eq:integrabilityB3}
\d\mbf{B}^\rb{3,0}+\mbf{B}^\rb{3,0}\wedge \mbf{B}^\rb{3,0}=0\,.
\end{equation}
Specifically, if we can find a gauge transformation enforcing $\mbf{B}^\rb{3,0}\wedge \mbf{B}^\rb{3,0}=0$, the closure of $\mbf{B}^\rb{3}$ is guaranteed. Given that $\d \big(\mbf{B}^\rb{2,0}\big)_{30,i}\neq0$ only for $i=9,13$, we consider the following ansatz for $\mbf{U}^\rb{3}$
\begin{equation}\label{eq:U3}
        \mbf{U}^\rb{3} = \mathbbm{1}_{30\times30}
    + \begin{pmatrix}
      \bs{0}_{29\times30}
      \\
      (\mbf{U}^\rb{3})_{1\times30}
    \end{pmatrix} \qquad \text{with} \qquad    \mbf{U}^\rb{3} = \Big(
        \mbf{0}_{1\times24},
        (\tilde{\mbf{u}}_3)_{1\times4}
        ,\mbf{0}_{1\times2}
    \Big)\,,
\end{equation}
where $(\tilde{\mbf{u}}_3)_{1\times4}=(u_{25},u_{26},u_{27},u_{28})$. In this case, it can be shown that one solution to $\mbf{B}^\rb{3,0}\wedge \mbf{B}^\rb{3,0}=0$ is obtained from $u_j$ such that
\begin{equation} \label{eq:intToSolve}
    \d u_j+\big(\mbf{B}^\rb{2,0}\big)_{30,j}=0 \qquad \text{for} ~ j=25,26,27,28\,.
\end{equation}
Integrating the above, the matrix elements of the gauge transformation are 
\begin{equation}
    (\tilde{\mbf{u}}_3)_{1\times4}=\frac{1}{2}\Big(\frac{\Lambda_{01245}}{\sqrt{\lambda_{134}}\sqrt{\lambda_{024}}},\frac{\Lambda_{02154}}{\sqrt{\lambda_{235}}\sqrt{\lambda_{015}}},\frac{\Lambda_{05421}}{\sqrt{\lambda_{235}}\sqrt{\lambda_{015}}},\frac{\Lambda_{04512}}{\sqrt{\lambda_{134}}\sqrt{\lambda_{015}}}\Big)\,,
\end{equation}
where $\Lambda_{abcde}=-X_d^2{-}X_a X_b{+}X_a X_d{+}X_b X_c{+}X_b X_d{+}X_c X_d{-}2X_d X_e{+}X_a{-}X_c{+}X_d$.    
Setting $\mbf{J}^\rb{3} = \mbf{U}^\rb{3} \cdot \mbf{J}^\rb{2}$, we have, as desired, $\mbf{B}^\rb{3} = \d \mbf{U}^\rb{3} \cdot \left(\mbf{U}^\rb{3}\right)^{-1} + \mbf{U}^\rb{3} \cdot \mbf{B}^\rb{2} \cdot \left(\mbf{U}^\rb{3}\right)^{-1}$ such that $\d\mbf{B}^\rb{3,0}=0$. 
Next, we eliminate the remaining off-diagonal terms in $\mbf{B}^\rb{3,0}$.

\subsubsection*{$\text{SL}(2,\mathbb{Z})$ covariance: removing $\mathcal{O}(\varepsilon^0)$ off-diagonal terms}

The only non-zero entries of $\mbf{B}^\rb{3,0}$ are 
\begin{subequations}\label{eq:ODToKill}
    \begin{align}
   \text{$\color{BrickRed}(1234)$- and $\color{BrickRed}(1235)$-eyeball/\textcolor{BrickRed}{$(123)$}-sunrise:}
    \qquad&
    B^\rb{3,0}_{25,9}\,,\
    B^\rb{3,0}_{26,9}\,,\
    \\
    \text{$\color{Blue}(1345)$- and $\color{Blue}(2345)$-eyeball/\textcolor{Blue}{$(345)$}-sunrise:}
    \qquad&
    B^\rb{3,0}_{27,13}\,,\
    B^\rb{3,0}_{28,13}\,,
    \\
    \text{kite/\textcolor{BrickRed}{$(123)$}-sunrise:}
    \qquad&
    B^\rb{3,0}_{30,9}\,,
    \\
    \text{kite/\textcolor{Blue}{$(345)$}-sunrise:}
    \qquad&
    B^\rb{3,0}_{30,13}\,,
    \\
    \text{kite/double-bubble:}
    \qquad&
    B^\rb{3,0}_{30,29}\,.
\end{align}
\end{subequations}
This suggests the following ansatz for the gauge transformation $\mbf{U}^\rb{4}$, whose purpose is to eliminate the nontrivial components in \eqref{eq:ODToKill}. 
We set 
$\mbf{U}^\rb{4} = \mathbbm{1}_{30\times30} + \tilde{\mbf{U}}^\rb{4}$ 
where the only non-zero components of $\tilde{\mbf{U}}^\rb{4}$ are
\begin{align}
    \tilde{U}^\rb{4}_{25,9}\,,
    \tilde{U}^\rb{4}_{26,9}\,,
    \tilde{U}^\rb{4}_{27,13}\,,
    \tilde{U}^\rb{4}_{28,13}\,,
    \tilde{U}^\rb{4}_{30,9}\,,
    \tilde{U}^\rb{4}_{30,13} \quad \text{and} \quad
    \tilde{U}^\rb{4}_{30,29}\,.
\end{align}
Then, requiring that
\begin{align}
\label{diffeqoff}
    \d \tilde{U}^\rb{4}_{i,j} + B^\rb{4,0}_{i,j} = 0\,,
\end{align}
ensures that $\mbf{B}^\rb{4} = \d \mbf{U}^\rb{4} \cdot \left(\mbf{U}^\rb{4}\right)^{-1} + \mbf{U}^\rb{4} \cdot \mbf{B}^\rb{3} \cdot \left(\mbf{U}^\rb{4}\right)^{-1}$ is in $\vep$-form. 

Only the component that is an algebraic function of the $X$'s and does not contain any periods of the elliptic curves is $B^\rb{3,0}_{30,29}$. 
The other non-trivial components of $\mathbf{B}^\rb{4,0}$ have the following uniform structure
\begin{subequations}\label{eq:B4form}
\begin{align}
    B^\rb{3,0}_{i,j} = 
    \psi_{1}^{\textcolor{BrickRed}{(123)}}\ \sigma^{\textcolor{BrickRed}{(123)}}_{i,j} 
    {+} \partial_0\psi_{1}^{\textcolor{BrickRed}{(123)}}\ \rho^{\textcolor{BrickRed}{(123)}}_{i,j}
    &\quad\text{for}\quad 
    (i,j) \in \{(25,9), (26,9), (30,9)\}\,,
    \\
    B^\rb{3,0}_{i,j} = 
    \psi_{1}^{\textcolor{Blue}{(345)}}\ \sigma^{\textcolor{Blue}{(345)}}_{i,j} 
    {+} \partial_0\psi_{1}^{\textcolor{Blue}{(345)}}\ \rho^{\textcolor{Blue}{(345)}}_{i,j}
    &\quad\text{for}\quad 
    (i,j) \in \{ (27,13), (28,13), (30,13)\}\,,
\end{align}
\end{subequations}
where $\sigma^{(a)}_{ij}$ and $ \rho^{(a)}_{ij}$ are differential 1-forms in the kinematic variables $X_a$.
We will integrate the components that contain elliptic functions following the procedure outlined for the 4-mass eyeball in section \ref{subsec_icmod} by leveraging the modular properties of $\mbf{B}^\rb{4}$ in a straightforward systematic manner.

\paragraph{(a) \emph{Kite/double-bubble component}} 
The component $(B^\rb{4,0})_{30,29}$ describes how the differential of the kite couples to the (normalized) double-bubble integral $I_{1,1,0,1,1}$, which fully decouples from both sunrises. Since this component is an algebraic function of the $X$'s, it is simple to integrate equation \eqref{diffeqoff}:
\begin{align}
\scalemath{.98}{
    \tilde{U}^\rb{4}_{30,29}
    = \frac{
        {-} X_0^2 
        {+} X_1 X_0
        {+} X_2 X_0
        {+} X_4 X_0
        {+} X_5 X_0
        {-} 2 X_0
        {+} X_1 X_2
        {-} X_1 X_4
        {-} X_2 X_5
        {+} X_4 X_5
    }{
        2\sqrt{\lambda_{024}}\sqrt{\lambda_{015}}
    }
}
    \,.
\end{align}

\paragraph{(b) \emph{Eyeball/sunrise components}} To remove the kite/eyeball terms at $\mathcal{O}(\vep^0)$ in $\mbf{B}^\rb{4,0}$, we need to solve the differential equations 
\begin{align}
    \d \tilde{U}^\rb{4}_{i,j} + B^\rb{3,0}_{i,j} = 0
    & \quad\text{for}\quad 
    (i,j) = (25,9), (26,9), (27,13), (28,13)\,,
\end{align}  
that involve elliptic functions.
Following the algorithm outlined in section \ref{sec_IC}, the components of the gauge transformation are
\begin{subequations}\label{eq:GTus}
\begin{align}
    \tilde{U}^\rb{4}_{25,9} &= 
        - \psi^\e_1 \frac{\partial \tau^\e}{\partial X_0} 
        \sum_{i} \left(\rho_{25,9}^\e\right)_i\
            \frac{\partial X_i}{\partial \tau^\e}\,,
    \\
    \tilde{U}^\rb{4}_{26,9} &= 
        - \psi^\e_1 \frac{\partial \tau^\e}{\partial X_0} 
        \sum_{i} \left(\rho_{26,9}^\e\right)_i\ 
            \frac{\partial X_i}{\partial \tau^\e}\,,
    \\
    \tilde{U}^\rb{4}_{27,13} &= 
        - \psi^\ee_1 \frac{\partial \tau^\ee}{\partial X_0} 
        \sum_{i} \left(\rho_{27,13}^\ee\right)_i\
            \frac{\partial X_i}{\partial \tau^\ee}\,,
    \\
    \tilde{U}^\rb{4}_{28,13} &= 
        - \psi^\ee_1 \frac{\partial \tau^\ee}{\partial X_0} 
        \sum_{i} \left(\rho_{28,13}^\ee\right)_i\
            \frac{\partial X_i}{\partial \tau^\ee}\,, 
\end{align}
\end{subequations}
where $i=0,1,2,4,5$ in the sums above and $\left(\rho^\alpha_{p,q}\right)_i$ is the $\d X_i$ component of $\rho^\alpha_{p,q}$. The derivatives of the kinematic parameters with respect to the moduli are computed from the inverses of the Jacobians 
\begin{align}
    \mathcal{J}^\e_{ij} = 
        \frac{\partial \zeta^\e_j}{\partial x_i}\,
    \quad \text{and} \quad
    \mathcal{J}^\ee_{ij} = 
        \frac{\partial \zeta^\ee_j}{\partial x_i}\,,
\end{align}
where $\bs{\zeta}^\e = \{z^\e_{i=1,2},z^\e_{i=4,5}\}$, $\bs{\zeta}^\ee = \{z^\ee_{i=1,2},z^\ee_{i=4,5}\}$ and $\bs{x} = \{X_0,X_1,X_2,X_4,X_5\}$. Although we have not yet \emph{derived} all the punctures, we recall that explicit formulas are provided in \cref{sunrisepunctures,sunpunct123,eq:newz}.

As a sanity check, one can show that the expressions in \eqref{eq:GTus} for the gauge transformations are numerically equivalent to the integrals 
\begin{align}
    \int_0^{X_4} \d X_4^\prime \left(B^\rb{3,0}_{i,j}\right)_4
    \quad \text{or} \quad 
    \int_0^{X_5} \d X_5^\prime \left(B^\rb{3,0}_{i,j}\right)_5
    \,,
\end{align}
where $\left(B^\rb{3,0}_{i,j}\right)_a$ is the $\d X_a$ component of $B^\rb{3,0}_{i,j}$.

\paragraph{(c) \emph{Kite/sunrise components}} The final two entries of $\mbf{U}^\rb{4}$ are fixed in exactly the same manner as the eyeball-components of the previous paragraph:
\begin{subequations}
\begin{align}
    \tilde{U}^\rb{4}_{30,9} & = 
    - \psi^\e_1 \frac{\partial \tau^\e}{\partial X_0} 
   \sum_{i} 
   \left(\rho_{30,9}^{\e}\right)_i
   \frac{\partial X_i}{\partial\tau^\e}
    \,,
    \\  
    \tilde{U}^\rb{4}_{30,13} & = 
    - \psi^\ee_1 \frac{\partial \tau^\ee}{\partial X_0}  
    \sum_{i} \left(\rho_{30,13}^{\ee}\right)_i
    \frac{\partial X_i}{\partial\tau^\ee}
    \,.
\end{align}
\end{subequations}
We report that no additional objects or punctures are encountered in the kite components of the gauge transform.\footnote{Note, that this result can still be obtained without knowledge of all punctures by an appropriate splitting of the integration path. For $\tilde{U}^{(4)}_{30,9}$, we can first integrate along $\{X_0,X_1,X_2\}$ keeping $X_4,X_5=0$ fixed and then along $\{X_4,X_5\}$. The latter integral is simple because $B^{(3,0)}_{30,9}$ does not contain objects defined on the $\ee$-torus. The former integral can be performed with the modular bootstrap method described above, but since we keep $X_4,X_5$ constant at $0$, only known punctures $z_1^\e,z_2^\e$ appear. An analogous procedure can be used to obtain $\tilde{U}^{(4)}_{30,13}$. }
The new punctures on each torus obtained from the four eyeball diagrams can be reused for the kite. 

\paragraph{Transformation summary} Finally, we apply all transformations
\begin{align}\label{eq:canonicalU}
 \mbf{U}=   \mbf{U}^\rb{4}\cdot\mbf{U}^\rb{3}\cdot\mbf{U}^\rb{2}\cdot\mbf{U}^\rb{1}\,,
\end{align}
to obtain an $\vep$-form differential equation 
\begin{align}
\label{finaldeq}
    \d\mbf{J}=\vep\,\mbf{B}\cdot\mbf{J} \quad  \text{ with } \quad \mbf{J}=\mbf{U}\cdot\mbf{I} \quad \text{ and } \quad \mbf{B} = \d \mbf{U}\cdot \mbf{U}^{-1} + \mbf{U}\cdot \mbf{A} \cdot \mbf{U}^{-1}\,.
\end{align}
Both $\mbf{U}$ and $\mbf{B}$ are provided in a \textsc{Mathematica} format in the ancillary file accompanying this paper \href{https://github.com/StrangeQuark007/kite_ancillary}{\faGithub}. We emphasize that the outcome $\mbf{B}$ is expressed as a matrix-valued 1-form in kinematic variables (the $X$'s), leading to a highly complicated matrix. 
In section \ref{pullback}, we demonstrate how this complexity is tamed by using the appropriate set of variables and functions. 
This simplified form reveals some of the analytic properties of the Feynman integrals more clearly, such as the simple pole structure and a (partial) connection to Landau singularities. 

To properly set up this discussion, we first provide more details on how to find the new punctures from a maximal cut perspective.

\section{Punctures from maximal cuts}
\label{sec_punctureslab}
As illustrated in \eqref{smallmatrix}, we have a clear decoupling between entries that are naturally expressed on the \textcolor{BrickRed}{(123)}-torus (\textcolor{BrickRed}{$\blacksquare$}), entries that are naturally expressed on the \textcolor{Blue}{(345)}-torus (\textcolor{Blue}{$\blacksquare$}) and entries for which there is no natural choice (\textcolor{darkGreen}{$\blacktriangle$}). Therefore, we are looking for maps from the full kite kinematic space $\mathcal{K}$ onto the moduli space of \emph{either} of these tori. 

One parameter will always be the moduli themselves $\tau^\e$ or $\tau^\ee$. 
The remaining variables are punctures on the corresponding torus $z_i^\e$ or $z_i^\ee$. 
Interestingly, we find that these punctures can be constructed from integrals over the maximal cut of the top sector. 
In section \ref{pullback}, we will see that the choice of punctures obtained in this way leads to a compact differential equation in terms of ``logarithmic'' differential forms on $\mathcal{M}_{1,n}^\e$ or $\mathcal{M}_{1,n}^\ee$.\footnote{Here, $n=3$ for the sunrise family, $n=4$ for an eyeball family and $n=5$ for the kite family.}  
We anticipate that this procedure might also be applicable to higher multiplicity graphs with one or more elliptic subtopologies (e.g., the generic mass cross-box, which additionally depends on the momentum transfer $t$).

The relation between punctures and maximal cuts was already mentioned at the end of section \ref{sec_IC}.  
There, we found that a new elliptic function is needed in the gauge transformation to eliminate the $\mathcal{O}(\varepsilon^0)$ terms of the differential equation (similar observations were recently made for various diagrams in \cite{G_rges_2023}, although the setup is somewhat different). 
Of course, these new functions also appear in the differential equation at $\mathcal{O}(\varepsilon)$.
In our case, we find it useful to interpret them as punctures on one of the two tori. 

By direct calculation, we observe that the new punctures are simply integrals over maximal cuts. This makes sense physically, because the cut integrals satisfy the same differential equation as the uncut integrals and should in some sense ``know'' about these punctures. Therefore, we propose that this maximal cut perspective provides a simple way to obtain these punctures \emph{without} relying on information from the differential equation and gauge transformation. We motivate and further explain this procedure for both the eyeball and the kite in sections \ref{subsec_icecone} and \ref{fivepunc}.

\subsection{Four punctures for an eyeball}
\label{subsec_icecone}

We begin by elaborating on how and why the punctures introduced in section \ref{subsec_ellipticint} are obtained from the maximal cut of the eyeball diagrams. To explain this, we recall the steps that led to the $\varepsilon$-form in \eqref{eyeballep}. Without loss of generality, we work on the sunrise cut such that only the sunrise and eyeball diagrams remain. In what follows, we refer back to section \ref{sec_IC} as much as possible to ensure that the connection between notation in both sections remains clear.

The starting point is the cut version of the differential equation \eqref{DEQ_sunrise}
\begin{equation}
\text{d}\slashed{\mbf{J}}_4^{(1)}=\slashed{\mbf{B}}_4^{(1)}\cdot\slashed{\mbf{J}}_4^{(1)}\, ,
\end{equation}
where the \emph{slashed} notation indicates we work on the sunrise cut and throw away all but the eyeball and sunrise topologies. 
Here, $\slashed{\mbf{B}}_4^{(1)}$ takes the form
\begin{align}
\label{schemA4}
    \slashed{\mbf{B}}_4^{(1)}=\begin{pmatrix}
0 & 0  \\
\mbf{OD}^{(0)} & \mbf{EB}^{(0)}
\end{pmatrix}+\varepsilon \begin{pmatrix}
\mbf{SR}^{(1)} & 0  \\
\mbf{OD}^{(1)} & \mbf{EB}^{(1)}
\end{pmatrix}\, ,
\end{align}
where $\mathbf{SR}$ denotes the ($4\times4$) sunrise block, $\mathbf{EB}$ the ($1\times 1$) eyeball block, and $\mathbf{OD}$ the leftover ($1\times 4$) off-diagonal block and 
the $\varepsilon$ dependence is tracked through the superscripts (e.g., $\mbf{EB}=\mbf{EB}^{(0)}+\varepsilon\mbf{EB}^{(1)}$). 

At this stage, only the gauge transformation \eqref{eq:U1eb} has been applied to the original basis. 
Therefore, only the sunrise sector is in $\varepsilon$-form and $\mbf{OD}$ involves elliptic functions of $X_0$, $X_1$ and $X_2$, but \emph{not} of $X_4$.
Elliptic functions of $X_4$ appear after gauging away the reaming $\mathcal{O}(\varepsilon^0)$ part of the differential equation. 
As explicitly shown in section \ref{sec_IC}, this is done in two steps. We first remove $\mbf{EB}^{(0)}$ (the diagonal part of \eqref{schemA4}'s first term) and then remove the remaining off-diagonal term $\mbf{OD}^{(0)}$. 

Concretely, the first step is accomplished by the ($\varepsilon$-free) gauge transformation 
\begin{equation}
    \slashed{\mbf{U}}_4^{(2)}
    = \cancel{\eqref{v1313}}
    = \begin{pmatrix}
        \mathbbm{1} & 0
        \\
        0 & u_{13,13}
    \end{pmatrix}
    = \begin{pmatrix}
        \mathbbm{1}_{4\times4} & 0  
        \\
        0 & \text{MC}^{-1}_0
    \end{pmatrix}
    = \begin{pmatrix}
        \mathbbm{1}_{4\times4} & 0  
        \\
        0 & \sqrt{\lambda_{134}}\sqrt{\lambda_{024}}
    \end{pmatrix}
    \,,
\end{equation}
where $\cancel{\eqref{v1313}}$ contains only the components of $\eqref{v1313}$ that survive the sunrise cut, $u_{13,13}=\text{MC}^{-1}_0$ is the inverse of the  maximal cut of the eyeball diagram in 2-dimensions ($\vep=0$) and 
\begin{align}
    \d \slashed{\mbf{U}}_4^\rb{2} \cdot \big(\slashed{\mbf{U}}_4^\rb{2}\big)^{-1} = \begin{pmatrix}
        0 & 0  
        \\
        0 & -\mbf{EB}^{(0)}
    \end{pmatrix}
    \,.
\end{align}
The resulting differential equation is 
\begin{equation}
    \text{d}\slashed{\mbf{J}}_4^{(2)} 
    = \slashed{\mbf{B}}_4^{(2)} \cdot \slashed{\mbf{J}}_4^{(2)}\,,
\end{equation}
with 
\begin{equation}
    \slashed{\mbf{B}}_4^{(2)} 
    = \d \slashed{\mbf{U}}_4^\rb{2} 
        \cdot \big(\slashed{\mbf{U}}_4^\rb{2}\big)^{-1}
    + \slashed{\mbf{U}}_4^\rb{2} 
        \cdot \slashed{\mbf{B}}_4^{(2)}  
        \cdot \big(\slashed{\mbf{U}}_4^\rb{2}\big)^{-1}
    \qquad \text{and} \qquad
    \slashed{\mbf{J}}_4^{(2)} 
    = \slashed{\mbf{U}}_4^\rb{2}
        \cdot \slashed{\mbf{J}}_4^{(2)}
    \,.
\end{equation}
Here, only the off-diagonal block of $\slashed{\mbf{B}}_4^{(2)}$ is not in $\varepsilon$-form 
\begin{equation}
\begin{split}
    \slashed{\mbf{B}}_4^{(2)}
    =\slashed{\mbf{B}}_4^{(2,0)}
    + \varepsilon\ \slashed{\mbf{B}}_4^{(2,1)}
    =\begin{pmatrix}
    0 & 0 
    \\
    \text{MC}_0^{-1}\ \mbf{{OD}}^{(0)} & 0 
\end{pmatrix}+\varepsilon\begin{pmatrix}
    \mbf{SR}^{(1)} & 0
    \\
    \text{MC}_0^{-1}\  \mbf{{OD}}^{(1)} 
    & \mbf{EB}^{(1)}
\end{pmatrix}\,.
\end{split}
\end{equation}
To eliminate the remaining non-$\varepsilon$-form block $ \text{MC}_0^{-1}\ \mbf{{OD}}^{(0)}$, we introduce the ($\varepsilon$-free) gauge transformation $\slashed{\mbf{U}}_4^{(3)}$ (c.f., \eqref{gagueaway}) 
\begin{align}
\slashed{\mbf{U}}_4^{(3)}=\begin{pmatrix} \mathbbm{1}&0\\ \hat{\mbf{U}} &\mathbbm{1}\end{pmatrix} \quad \text{satisfying} \quad \slashed{\mbf{U}}_4^{(3)}\cdot\slashed{\mbf{B}}_4^{(2,0)}+\text{d}\slashed{\mbf{U}}_4^{(3)}=0\,.
\end{align}
This differential equation is formally solved by
\begin{equation}\label{elliptic_origin}
    \hat{\mbf{U}}
    = -\int \text{MC}_0^{-1}\ \mbf{OD}^{(0)} 
    \sim -\int (\cdots)\ \text{MC}_0
    \,.
\end{equation}
since, from \eqref{schemA4}, we know that $\mbf{OD}^\rb{0} \sim \text{MC}^2_0$.
Although $\hat{\mbf{U}}$ has only one nonvanishing entry, its general form is
\begin{equation}
\begin{split}
    \hat{\mbf{U}}
    =&-\int \sum_{j=0}^2 \text{d} X_j\  \mbf{g}_j
    -\int \text{d}X_4\ \dfrac{\mbf{f}_0 +\mbf{f}_1\ X_4 + \mbf{f}_2\ X_4^2}{\sqrt{\lambda_{134}}\sqrt{\lambda_{024}}}\, ,
\end{split}
\end{equation}
where the $\mbf{f}_j$ are vector-valued rational functions of $X_0$, $X_1$ and $X_2$, and, the $\mbf{g}_j$ are vector-valued algebraic functions of $X_0$, $X_1$, $X_2$ and $X_4$. 
Since the $\d X_4$-component of the integrand of $\hat{\mbf{U}}$ contains a square root of a degree four 
polynomial in $X_4$ (c.f., \eqref{v1313}), we expect the integrated expression of $\hat{\mbf{U}}$ to contain a \emph{new} elliptic function (c.f., \eqref{eq:z4}). 

Crucially, we observe that we could \emph{alternatively} have found \eqref{eq:z4} just by looking at indefinite integrals over the eyeball's maximal cut. 
This is the perspective we advocate to conceptualize the origin of the new punctures without referring to an $\vep$-form differential equation and a gauge transform. 
Not only is this prescription related to physical quantities, it also provides a universal definition.
Since there is inherently a lot of freedom in defining the punctures without interfering with the gauge transformation algorithm of section \ref{sec_epform}, obtaining them from maximal cuts is likely as close to a canonical choice as we can make.\footnote{Even with the punctures defined as indefinite integrals of the maximal cut there is freedom in choosing the branch of the integrated expression. For example, naively integrating the maximal cut of the $\color{BrickRed}(1235)$-eyeball with \textsc{Mathematica}'s \texttt{Integrate[]} sometimes yields the expected puncture $z_5^\e = z_4^\e\vert_{y_4\leftrightarrow y_5}$ shifted by a half-period $-\frac{1}{2}\tau^\e$. For this reason, we prefer to define the remaining punctures by acting with the discrete symmetries of the kite topology on $z_4^\e$.}
How to obtain punctures related to the sunrises from this perspective is also discussed below.

Integrating $\text{MC}_0$ with respect to $X_4$ results in an incomplete elliptic integral of the first kind and takes the form of a puncture \eqref{abelsmap}. 
We motivate the choice of integration variable from the observation that $\mathbb{T}^{\e}$ does not ``know'' about $m_4$ and, consequently, $X_4$ should be integrated out. Explicitly, the integral reads
\begin{equation}
\label{ellc4}
\int \rd X_4~\text{MC}_0({\color{BrickRed}I_{1,1,1,1,0}}) =\int \dfrac{\text{d}X_4}{\sqrt{\lambda_{134}\lambda_{024}}}=\dfrac{1}{2c_4^\prime}F\left( \arcsin\sqrt{u_\phi},k_\phi^2\right)\,,
\end{equation}
with
\begin{subequations}
    \label{eq:c4Prime-uphi-kphi}
    \begin{align}
        c_4^\prime&=-\frac{\sqrt{(+++)(+--)(-+-)(--+)}}{4}\,,
        \\
        k_\phi^2 &= \frac{(++-)(-++)(+-+)(---)}{(+++)(-+-)(--+)(+--)}\,,\label{eq:kphiEB}\\
        u_\phi&=  \frac{(+++)(-+-)}{(+-+)(---)}\, \, \frac{1-2\sqrt{X_1}+X_1-X_4}{1+2\sqrt{X_1}+X_1-X_4}\,.\label{eq:uphi}
    \end{align}
\end{subequations}
where $(\pm\pm\pm)=1\pm\sqrt{X_0}\pm\sqrt{X_1}\pm\sqrt{X_2}$. The choice of branch on the right-hand side of \eqref{ellc4} is valid for $0<k_\phi^2<1$ and $0<X_4<1$. Moreover, since $k^2_\phi$ is a cross-ratio, it is manifestly invariant under any permutation of the kinematic variables. 
Here, $c_4^\prime = i c_4/k_\phi$; up to an overall phase, $c_4^\prime$ is the Galois-conjugate ($\pm\mapsto \mp$ in $(\pm,\pm,\pm)$) of $c_4$.

It is not yet clear how \eqref{ellc4} is related to \eqref{eq:z4}. To make such a relation manifest, we first need to understand how $k_\phi^2$ relates to $k^2_\e$ (c.f., \eqref{eq:defK}) and how $u_\phi$ is connected to $u_4^{\e}$ (c.f., \eqref{eq:u4123}). A simple calculation reveals that
\begin{equation}\label{eq:kuEB}
    k_\phi^2= \frac{1}{1-k_{\e}^2} \quad \text{and} \quad u_4^{\e}=\frac{1}{1-u_\phi}\,.
\end{equation}
From there, using the results in appendix \ref{app:massage}, we obtain
\begin{equation}\label{eq:maxCutToEC}
    \eqref{ellc4}
    = \frac{1}{2}\psi_1
    \Bigg[\
        {\underbracket[0.4pt]{        
            \dfrac{F\left(\arcsin\sqrt{u_4},k_{\e}^2\right)}{K\left(k_{\e}^2\right)}
        }_{z_4^\e}}
        - 1
    \Bigg]
    \,.
\end{equation}
Thus, in light of \eqref{abelsmap}, it seems quite natural to define the extra eyeball puncture as the term in brackets in \eqref{eq:maxCutToEC}. 
Note that the additional term of ``$-1$'' inside the above brackets does not affect the definition because $z_4^\e$ and $z_4^\e-1$ are equivalent modulo the lattice $\Lambda_{(1,\tau^\e)}$.

Although we previously assumed that the two punctures from the sunrise subtopology were known from \cite{bogner_unequal_2020}, these can also be obtained systematically from a maximal cut argument. In fact, we obtain them through specific kinematic limits of $u_4$ in \eqref{eq:kuEB}. In particular, the $X_4\rightarrow \infty$ limit (corresponding to an infinitely heavy $m_4$ effectively pinching the $\color{BrickRed}(1234)$-eyeball into the $\e$-sunrise) gives (c.f., \eqref{sunpunct123})
\begin{equation}\label{ulim}
    \lim_{X_4\to \infty}u_4^{\e}=u_2^{\e}=u_1^{\e}|_{X_1\leftrightarrow X_2}\, ,
\end{equation}
where $X_1\leftrightarrow X_2$ is a symmetry of the sunrise graph.
The discrete symmetries of the $\e$-sunrise graph correspond to the $3!$ permutations of the masses $m_1,m_2$ and $m_3$. 
As observed in \eqref{ulim}, $u_1^\e$ and $u_2^\e$ are related by a discrete symmetry. 
The remaining permutations only yield $u_1, u_2 $ and $u_3$. 
Once plugged into (\ref{sunrisepunctures}) (note that $k^2_{\e}$ is invariant under all these symmetries) these correspond to the punctures in \cite{bogner_unequal_2020} subject to the relation $z_1+z_2+z_3=1$ that originates from the translation invariance of the torus.\footnote{Naively, this symmetry argument can also be applied to the eyeball graph which is symmetric under the exchange $m_1\leftrightarrow m_3$. However, $u_4^{\e}$ remains invariant under this operation. This is expected because the eyeball should only need one new independent puncture.}
The key takeaway from this example is that one can identify \emph{all} the necessary punctures to define the embedding of $\mathcal{K}_{\text{$\color{BrickRed}(1234)$-eyeball}}$ into $\mathcal{M}_{1,4}^{\e}$ simply by analyzing the integral of the corresponding maximal cut!

In the same way, we can identify another new puncture on the $\e$-torus from the $X_5$ integral of the maximal cut of {\textcolor{BrickRed}{$I_{1,1,1,0,1}$}}:
\begin{align} \label{eq:z15}
    z_5^\e = \dfrac{F\left(\arcsin\sqrt{u_5^\e},k_{\e}^2\right)}{K\left(k_{\e}^2\right)}
    \quad\text{with}\quad
    u_5^\e = u_1^\e \frac{(1+\sqrt{X_2})^2-X_5}{(\sqrt{X_0}+\sqrt{X_1})^2-X_5}
    \,.
\end{align}
As expected, $z_5^\e = z_4^\e\vert_{\substack{X_1 \leftrightarrow X_2\\X_5 \leftrightarrow X_4}}$ and in the heavy $m_5$ limit this reduces to one of the $\e$-sunrise punctures.
Likewise, two new punctures, $z_1^{\textcolor{Blue}{(345)}}$ and $z_2^{\textcolor{Blue}{(345)}}$, on the $\ee$-torus can be obtained from the maximal cuts of {\textcolor{Blue}{$I_{1,0,1,1,1}$}} and {\textcolor{Blue}{$I_{0,1,1,1,1}$}} respectively:
\begin{subequations}\label{eq:z21}
    \begin{align}
    z_1^\ee &= \dfrac{F\left(\arcsin\sqrt{u_1^\ee},k_{\ee}^2\right)}{K\left(k_{\ee}^2\right)}
    \quad\text{with}\quad
    u_1^\ee = u_{5}^\ee \frac{(1+\sqrt{X_4})^2-X_1}{(\sqrt{X_0}+\sqrt{X_5})^2-X_1}
    \,,
    \\
    z_2^\ee &= \dfrac{F\left(\arcsin\sqrt{u_2^\ee},k_{\ee}^2\right)}{K\left(k_{\ee}^2\right)}
    \quad\text{with}\quad
    u_2^\ee = u_{4}^\ee \frac{(1+\sqrt{X_5})^2-X_2}{(\sqrt{X_0}+\sqrt{X_4})^2-X_2}
    \,.
\end{align}
\end{subequations}
When combined with the sunrise punctures of each torus
(which can also be derived through similar limits as in \eqref{ulim}),
there is a total of four linearly independent punctures $\{z_1,z_2,z_4,z_5\}^{\alpha}$ on each torus $\alpha=\e,\ee$ (recall that we use transitional invariance to remove $z_3^\alpha$ on each torus). 
Consequently, we can establish a complete embedding of $\mathcal{K}$ within $\mathcal{M}_{1,5}^\e$ or $\mathcal{M}_{1,5}^\ee$ simply by analyzing the maximal cuts of all eyeball subtopologies. 
The complete set of punctures closes under various mass permutations, which have been collected in appendix \ref{puncturesappendix}. 
In the following section, we will also demonstrate that these punctures are consistent with the double integral of the maximal cut of the kite.

\subsection{Punctures from the kite maximal cut}
\label{fivepunc}

A natural question to ask is what happens when we integrate the maximal cut of the kite twice? Do we recover the punctures of section \ref{subsec_icecone}? 

To answer these questions, the most natural thing to do is to integrate the maximal cut of the kite with respect to the variables \emph{exclusive} to one elliptic curve
\begin{align}\label{eq:extPunct}
    \int\frac{\rd X_i~\rd X_j}{\lambda_{01245}} \quad \text{where $i,j\in\{1,2\}$ or $i,j\in\{4,5\}$ with $i\neq j$}\,.
\end{align}
 For example, suppose that we take $i=4$ and $j=5$. Then, 
\begin{equation}
\begin{aligned}
    \int\frac{\rd X_4 \rd X_5}{\lambda_{01245}} 
    &=  \int \d X_4\ 
    \frac{2\ \text{MC}_0({\color{BrickRed}I_{1,1,1,1,0}})}{X_0}
    \text{arctanh}\Big( Q\ \text{MC}_0({\color{BrickRed}I_{1,1,1,1,0}}) \Big)
    \,,
    \\
    &=  \int \d X_4\ 
    \frac{i\ \text{MC}_0({\color{BrickRed}I_{1,1,1,1,0}})}{X_0}
    \log \frac{
        Q 
        + i \text{MC}_0^{-1}({\color{BrickRed}I_{1,1,1,1,0}})
    }{
        Q 
        - i \text{MC}_0^{-1}({\color{BrickRed}I_{1,1,1,1,0}})
    }
    \,,
\end{aligned}
\end{equation}
where $\text{MC}_0^{-1}({\color{BrickRed}I_{1,1,1,1,0}}) = \sqrt{\lambda_{314}
}\sqrt{\lambda_{024}}$ and $Q = X_0 (1-X_1+X_4) + X_2(X_1+X_4-1) + X_4(X_1-X_4-2X_5+1)$. Unfortunately, this integral takes the form of the Feynman parametric representation of the sunrise integral in $2$-dimensions \cite{Wilhelm:2022wow}.
Since the 2-dimensional sunrise integral is especially complicated, performing the $X_4$ integration is not useful.  However, as expected, the $X_5 \to \infty$ limit of the above integral sends $\log(\cdots) \to i \pi$ and we recover the integral \eqref{ellc4} up to an overall factor. 

It is also tempting to define the punctures of the kite by studying the image of the singular points of the logarithm in \eqref{eq:extPunct} under Abel's map for the $\text{MC}_0^{-1}({\color{BrickRed}I_{1,1,1,1,0}})$ elliptic curve. 
It would be interesting to see if these provide more natural/useful choices than those we have made in this work.

\section{Differential equation's pullback onto the tori's moduli space}
\label{pullback}
As pointed out in section \ref{epssec}, the entries of $\mbf{B}$ in \eqref{finaldeq} depend on the following quantities
\begin{align}
  \Big\{  X_i,\psi_1^{{\e}}, \psi_1^{{\ee}},\dfrac{\partial  \psi^{{\e}}_1}{\partial X_0},\dfrac{\partial  \psi^{{\ee}}_1}{\partial X_0},\dfrac{\partial X_i}{\partial \tau^{{\e}}},\dfrac{\partial X_i}{\partial \tau^{{\ee}}}\Big\}\,. 
\end{align} 
Recall that the periods $\psi_1$, their derivatives, and the derivatives with respect to $\tau$ were introduced earlier from the transformation to $\varepsilon$-form in \eqref{eq:canonicalU}. We now wish to rewrite the entries of $\mathbf{B}$ in terms of the variables ($\tau$ and $z_i$) defined on either of the two tori.

The rule for choosing which torus to work on for a given entry is simple: if the entry couples to any of the sunrises (i.e., if it belongs to either the $\textcolor{BrickRed}{\blacksquare}$- or $\textcolor{Blue}{\blacksquare}$-blocks in \eqref{smallmatrix}), then we consider its pullback on $\mathcal{M}_{1,5}^\e$ or $\mathcal{M}_{1,5}^\ee$ respectively. In contrast, if the entry falls outside this classification (i.e., if it belongs to a $\textcolor{darkGreen}{\blacktriangle}$-block or any of the remaining diagonal entries in \eqref{smallmatrix}), it is a dlog form, and either torus is a valid choice. Note that this choice may break the overall $\e\leftrightarrow\ee$ symmetry of the differential equation pulled back on the tori. This may also have a nontrivial effect on whether or not the quantities defined on $\mathcal{M}_{1,5}^\e$ and $\mathcal{M}_{1,5}^\ee$ couple in the differential equation solution at a given order in $\vep$. In section \ref{bdry}, we will briefly discuss a naive organization strategy to use in order to prevent this from happening.

Once we determine on which torus a given entry is defined, we can pull back the corresponding kinematic 1-form onto it. The target space of differential forms is universally spanned by modular forms and Kronecker-Eisenstein forms, which we review below (see section \ref{omegas}).

As we shall see, once we determine into which of the above categories an entry of $\mathbf{B}$ falls, information about the location of its poles and modular properties is enough to fix its pullback from a natural class of ansatzes: linear combinations of modular and Kronecker-Eisenstein forms over $\mathbbm{Q}+i\mathbbm{Q}$. In particular, the pole structure will be used to determine the $z$-dependence of Kronecker-Eisenstein forms (see section \ref{poles}), while the modular properties will be used to establish the modular weight of the differential forms that are allowed to appear (the modular behavior of each matrix element is summarized in \eqref{bigmatrix}; the matrix contains dlog forms of modular weight $0$ and (quasi-)modular forms of weights $\{0,1,2,3,4\}$). The coefficients in $\mathbbm{Q}+i\mathbbm{Q}$ are always fixed numerically from a PSLQ fit \emph{a posteriori}.\footnote{This can be done, e.g., using \textsc{Mathematica}'s \texttt{FindIntegerNullVector[]}.}

For the PSLQ reconstruction to work, it is important that a branch is chosen consistently throughout the whole calculation. To be more precise, the root ordering in (\ref{root ordering}) will pick a branch for $\tau^{\e}$, $z_1^{\e}$, $z_2^{\e}$ and $\tau^{\textcolor{Blue}{(345)}}$, $z_4^{\textcolor{Blue}{(345)}}$, $z_5^{\textcolor{Blue}{(345)}}$.
Additionally, we enforce \eqref{branch z4} to fix the branch for $z_4^{\e}$ and do similarly for $z_5^{\e}$ by requiring
\begin{equation}
      X_5>(\sqrt{X_0}+\sqrt{X_1})^2\text{   and   } X_5>(1+\sqrt{X_2})^2\,.
\end{equation}
Finally, we pick a branch for $z_1^{\textcolor{Blue}{(345)}}$ and $z_2^{\textcolor{Blue}{(345)}}$ by adding the constraints
\begin{equation}
\begin{split}
      &X_1<(\sqrt{X_0}-\sqrt{X_5})^2\text{   and   } X_1<(1-\sqrt{X_4})^2\, ,\\
      &X_2<(\sqrt{X_0}-\sqrt{X_4})^2\text{   and   } X_2<(1-\sqrt{X_5})^2\, .
\end{split}
\end{equation}
We checked the stability of the PSLQ reconstruction and the integrability condition of the resulting matrix over a random sample of $\mathcal{O}(10^{6})$ numerical points in $\mathcal{K}$ satisfying these conditions. 

In what follows, we discuss the pullback procedure in detail and provide the complete matrix $\mathbf{B}$ in terms of differential forms on the tori (for the eyeball and kite in sections \ref{subsPullbackIc} and \ref{pullbackkite}, respectively). The result of the full pullback is available in a \textsc{Mathematica} format in the ancillary file attached to this article \href{https://github.com/StrangeQuark007/kite_ancillary}{\faGithub}.

\subsection{Review on differential forms on the torus}
\label{omegas}

In this section, we outline the definitions and additional conventions essential for the discussion that follows. We start with the aforementioned \emph{Kronecker-Eisenstein form} \cite{elliptic_calc,weinzierl2022feynman} 
\begin{equation}
\omega_k(z,\tau)=(2\pi
)^{2-k}\Big(g^{(k-1)}(z,\tau)\text{d}z+(k-1)g^{(k)}(z,\tau)\frac{\text{d}\tau}{2\pi i}\Big)\,,
\end{equation}
where the $g$-kernels are defined from the generating \emph{Kronecker-Eisenstein series} \cite{weil1976elliptic,Brown:2011wfj,Broedel:2014vla,Broedel:2018qkq} (following the $\pi$-prescription of \cite[eq. (13.190)]{weinzierl2022feynman})
\begin{equation}\label{kronecker-eisenstein}
   F(z,\eta,q)=\pi \dfrac{\theta_1'(0,\tau) \theta_1(\pi(z+\eta),\tau)}{\theta_1(\pi z,\tau)\theta_1(\pi \eta,\tau)}=\sum_{\alpha=0}^\infty \eta^{\alpha-1} g^{(\alpha)}(z,\tau)\,, 
\end{equation}
where $\theta_1'(z,\tau)=\partial_z \theta_1(z,\tau)$ and $\theta_1(z,\tau)$ is the odd Jacobi theta function.
The $g^\rb{k}(z,\tau)$'s are the genus-1 analogous of the rational function $1/z$. 
In particular, the $g$-kernels have a simple pole at each lattice point $z=a+b\tau$ where $a,b\in\mathbbm{Z}$ and $b\neq0$. 
For $k=1$, $g^\rb{1}(z,\tau)$ has a simple pole on all lattice points, including those with $b=0$.

Importantly, the $g$-kernels transform under $\text{SL}(2,\mathbbm{Z})$ as \emph{quasi-modular forms} (c.f., \eqref{eq:qmf}) 
\begin{equation}
    g^{(k)}\Big(\dfrac{z}{c\tau+d},\dfrac{a\tau+b}{c\tau+d}\Big)=(c\tau+d)^k\sum^k_{j=0}\dfrac{(2\pi i~c~z)^j}{j!(c\tau+d)^j} g^{(k-j)}(z,\tau)\,.
\end{equation}
It is not hard to see that this implies that $\omega_k(z,\tau)$ is a quasi-modular form of weight $k$ (according to our definition of modular weight in \eqref{eq:modW}). For $k=0,...,4$ we have the following useful expansions around $q=e^{i\pi \tau}=0$
\begin{equation}
\begin{split}
    g^\rb{0}(z,\tau) &= 1
    \\
    g^{(1)}(z,\tau)&=\pi\cot(\pi z)+4\pi \sin(2\pi z)q^2 +\mathcal{O}(q^4)\,,\\
    g^{(2)}(z,\tau)&=-\dfrac{\pi^2}{3}+8\pi^2\cos(2\pi z)q^2+\mathcal{O}(q^4)\,,
    \\
    g^{(3)}(z,\tau)&=-8\pi^3\sin(2\pi z)q^2+\mathcal{O}(q^4)\,,
      \\
    g^{(4)}(z,\tau)&=-\frac{\pi^4}{45}-\frac{16 \pi^4}{3}\cos(2\pi z)q^2+\mathcal{O}(q^4)\,.
\end{split}
\end{equation}
Clearly, for $-\infty<\text{Im} z<\infty$, only $g^{(1)}(z,\tau)$ has a pole (at the origin), while $g^{(2,3,4)}(z,\tau)$ do not. 
In fact, one can show that $g^{(k>1)}(z,\tau)$ are regular at the origin \cite{bogner_unequal_2020}.

Finally, for some entries of $\mbf{B}$, we also need \emph{modular forms}, which are puncture independent. Such forms appear whenever the $\d z$ dependence of the ansatz is fixed, but the $\d \tau$ dependence is not. In that regard, we introduce the modular forms 
\begin{subequations}
\begin{align}
    \eta_2(\tau)&=[e_2(\tau)-2e_2(2\tau)]\frac{\textnormal{d}\tau}{2\pi i}\,,\label{eq:EAT2}\\
          \eta_4(\tau)&=e_4(\tau)\frac{\textnormal{d}\tau}{(2\pi i)^3}\,,
\end{align}
\end{subequations}
where the $e_\bullet$'s are \emph{Eisenstein series} (as defined in \cite[eq. (13.126)]{weinzierl2022feynman}). In terms of more tractable elliptic functions, these are given by
\begin{equation}
    e_2(\tau)=\frac{2\pi^2}{6}\frac{\theta_1^{'''}(0,\tau)}{\theta_1^{'}(0,\tau)} \quad \textnormal{and} \quad  e_4(\tau)=\frac{\pi^4}{90}\big({\theta_{2}^{8}\!\big(0,\tau\big) + \theta_{3}^{8}\!\big(0,\tau\big) + \theta_{4}^{8}\!\big(0,\tau\big)}\big)\,.
\end{equation}
It is not difficult to show that $e_2(\tau) - 2e_2(2\tau) \in \mathcal{M}_2(\Gamma_0(2))$ and $e_4(\tau) \in \mathcal{M}_4(\textnormal{SL}(2,\mathbbm{Z}))$, where $\mathcal{M}_{k}(\Gamma)$ represents the set of modular forms of weight $k$ in the congruence subgroup $\Gamma$ of $\textnormal{SL}(2,\mathbbm{Z})$. For more on modular forms and congruence subgroups, see \cite{Adams:2017ejb}.

\subsection{All poles from the diagonal dlog terms and connection to Landau}\label{poles}

In order to pull $\mathbf{B}$ back to the torus, we need to construct an ansatz of differential 1-forms and know where the entries of $\mathbf{B}$ can diverge on the torus. 
Our strategy is to identify a (optimally \emph{minimal}) set of matrix elements $B_{ij}$ that contains all possible poles in a way that is easily mapped to the torus. 
In our case, we find that this is a property satisfied by the diagonal dlog terms of $\mathbf{B}$.\footnote{This observation is motivated by the naive expectation that the diagonal should ``know'' about all off-diagonal singularities, which could, in principle, be made more precise using the language of spectral sequences.}

The dlog terms are modular invariant objects that often pull back to specific linear combinations of $\omega_2(z,\tau)$. An important example is
\begin{equation}\label{eq:om2}
\Omega_2(z,\tau)=\omega_2(z,\tau)-2\omega_2(z,2\tau)\,.
\end{equation}
Note that the $g$-kernels with the argument $2\tau$ (hidden in $\omega_2(z,2\tau)$) are required to make the above expressions modular invariant and do not change the location of any singular points on the torus. 
Although $g^\rb{k}(z,2\tau) = \frac12 \left[ g^\rb{k}(\frac{z}{2}, \tau) + g^\rb{k}(\frac{z+1}{2}, \tau) \right]$ (see \cite[\S 13]{weinzierl2022feynman}; alternatively, this can be extracted from the formula described in appendix \ref{sec_newrel}) these $g$-kernels are singular only on a subset of the points where $g$-kernels with argument $\tau$ are singular.

To determine what $z$ is, given a dlog entry of $\mathbf{B}$, it is instructive to compare the $q$-expansion of $\omega_2(z, \tau)$ with that of the entry. On one hand, we have
\begin{equation}
    \omega_2(z,\tau)=\text{d}\log \dfrac{\theta_1(\pi z,\tau)}{\eta(\tau)}=\text{d}\log \sin(\pi z) +\mathcal{O}(q^{2})\,,
\end{equation}
where $\eta(\tau)$ is the Dedekind eta function. This allows us to rewrite any linear combination of $\omega_2$'s at leading order in $q$ as
\begin{equation}\label{omega leading q}
    \sum_j c_j \, \omega_2(f_j(\boldsymbol{z}),\tau)=\text{d}\log \Big( \prod_j \sin^{c_j}\pi f_j(\boldsymbol{z})\Big)+\mathcal{O}(q^2)\,,
\end{equation}
where $c_j\in \mathbbm{Q}+i\mathbbm{Q}$. On the other hand, the kinematic dlog entries we wish to pull back factorize as
\begin{equation}
\text{d}\log \Big(\dfrac{ A_{i_1}\hdots A_{i_n}}{A_{j_1}\hdots A_{j_n}}\Big)=\text{d}\log A_{i_1}+\hdots +\text{d}\log A_{i_n}-\text{d}\log A_{j_1}-\hdots -\text{d}\log A_{j_n}\,.
\end{equation}
Here, each algebraic function $A_i$ can be pulled back separately onto the torus (i.e., expressed as 0-forms in terms of $\tau$ and the $z$ variables) before being expanded in $q$. Then, expressing the leading-order terms of these expansions as the dlog of a product of sines gives an efficient way of reading off the $Z$-arguments $f_j(\boldsymbol{z})$ appearing in \eqref{omega leading q}.

Crucially, we observe that applying this procedure to all diagonal dlog terms is \emph{sufficient} to find all the arguments needed to express the matrix $\mathbf{B}$ completely in terms of modular and Kronecker-Eisenstein forms.

\paragraph{Landau analysis of the $\e$-sunrise subtopology} 

For an integral family such as the sunrise, all of the elliptic structure appears in the top sector. 
When the kinematics is localized to the Landau variety of the top sector,
the elliptic geometry necessarily degenerates, meaning that $\tau^\alpha \to 0$ or $i\infty$ (with $\alpha=\e,\ee$). 
In fact, it is fairly easy to show that the elliptic modulus $k^2_\alpha$ and its complement $1-k^2_\alpha$ are rational functions of the polynomials that define the Landau variety for the corresponding sunrise diagram; when the kinematics is restricted to a subvariety in $\mathcal{K}$ defined by any of these polynomials,
$k^2_\alpha \to 0$ or $1$ and $\tau^\alpha \to i\infty$ or $0$.
In the following, we focus (without loss of generality) on $\alpha=\e$ and suppress this label on the variables.

Near any ($\e$-sunrise) Landau singularity where $\tau\to i\infty$, the polynomials in the original kinematic variables $X_i$ that describe this Landau singularity are given by the leading term in the corresponding $q$-series expansion. 
For example, we have 
\begin{subequations}
\begin{align}
    X_0 &= 16 q^2 \sin^2(\pi z_1) \sin^2(\pi z_2) 
        \to 0
    \,,
    \\
    X_1 &= \sin^2(\pi z_1)  \sin^{-2}\big(\pi (z_1+z_2) \big)
    \,,\label{eq:X1land}
    \\
    X_2 &= \sin^2(\pi z_2) \sin^{-2}\big(\pi (z_1+z_2) \big)
    \label{eq:X2land}
    \,,
\end{align}
\end{subequations}
when $\tau \to i \infty$ (or equivalently 
$q\to0$). 
Recalling that $z_3 = 1 - z_1-z_2 \simeq z_1+z_2 \text{ mod } \Lambda_{(1,\tau)}$, we see that the subleading Landau equations $X_{i=1,2}=0,\infty$ are satisfied whenever the $\e$-punctures vanish in the $\tau \to i\infty$ limit.
In summary, we find that the subleading Landau singularities $X_{i=1,2}=0,\infty$ of the sunrise integral are associated to the points $(\tau, z_{i=1,2}) \to (i\infty,0)$ in $\mathcal{M}_{1,3}^\e$. 

The above analysis of the $\tau \to i\infty$ limit is the first step towards constructing an ansatz for the pullback of a kinematic quantity to $\mathcal{M}_{1,3}^\e$. 
For example, from \eqref{eq:X1land}, we know that the form $\d\log(X_1)$ should have simple poles whenever $z_1$ and $z_1+z_2$ vanish. 
Therefore, we conclude that $z_1$ and $z_1+z_2$ should appear in the arguments of the $g$-kernels in our ansatz.
Indeed, one finds  
\begin{align}
    \d\log(X_1) 
    = 2\Omega_2 (z_1+z_2,\tau) 
    - 2\Omega_2 (z_1,\tau) 
    \,, 
\end{align}
where each $\Omega_2$ is modular invariant. 

In contrast to the subleading Landau singularities, the (pseudo-)normal threshold singularities $(\pm\pm\pm)=1\pm\sqrt{X_0}\pm\sqrt{X_1}\pm\sqrt{X_2}=0$ (c.f., \eqref{eq:c4Prime-uphi-kphi}) arise in the limit $\tau\to 0$ (or equivalently $q\to1$) \cite{giroux_loop-by-loop_2022}.
Since this limit defines the boundary of the analyticity domain of $\theta_1(z,q)$, studying the (pseudo-)normal thresholds on $\mathcal{M}_{1,3}^\e$ is, strictly speaking, only possible after analytic continuation of $\theta_1(z,q)$ via a modular $S$-transformation ($\tau \mapsto -1/\tau$).\footnote{Since the (psudo-)normal thresholds correspond to $k^2\to\infty$, the punctures must also be analytically continued to the region with $k^2>1$ before taking the (psudo-)normal threshold limit.}

However, even if we cannot access these singularities without doing a $S$-transformation, the $\tau\to i\infty$ limit can still be used to find the corresponding singular locus on $\mathcal{M}_{1,3}^\e$ since $\d\log(\text{theshold singularity})$ is modular invariant. 
For example, consider the complementary elliptic modulus $1-k^2$. 
It is a rational function of all the polynomials defining the normal and pseudo-normal thresholds:
\begin{align}\label{eq:compK}
    1-k^2 = 
    \frac{
        (--+) (-+-) (+--) (+++)
    }{
        (---) (-++) (+-+) (++-)
    }\,.
\end{align}
Multiplying the numerator and the denominator of \eqref{eq:compK} yields the polynomial
\begin{equation}\label{eq:Tdef}
    T = (--+) (-+-) (+--) (+++) (---) (-++) (+-+) (++-)\,,
\end{equation}
which vanishes at all normal and pseudo-normal thresholds of the $\e$-sunrise. 
By pulling back $\d\log T$ and looking at its poles, we learn how the zeroes of $T$ are embedded into $\mathcal{M}^\e_{1,3}$. 
As outlined earlier, to pull back this differential form, we start by analyzing the $\tau\to i \infty$ limit:
\begin{align}\label{eq:limitT}
    \lim_{\tau\to i\infty} T = \left[
    2 
    \sin(\pi  z_1) 
    \sin(\pi  z_2) 
    \sin^{-1}\big(\pi  (z_1+z_2) \big)
    \right]^4.
\end{align}
This suggests that we should use the $Z$-arguments $z_1$, $z_2$ and $z_1+z_2$ in our ansatz.
Fixing the coefficients using the PSLQ algorithm, we find
\begin{align}\label{eq:pbT}
    \d\log T
    =4\Big[\Omega_2(z_1 + z_2,\tau)
    - \Omega_2(z_1,\tau)
    - \Omega_2(z_2,\tau)
    - 3 \eta_2(\tau)\Big]
    \,,
\end{align}
where $\eta_2$ is a form that only depends on $\tau$ (c.f., \eqref{eq:EAT2}) and is added \emph{after} fixing the $\d z$-part. 
It is interesting that the $Z$-arguments for the $T=0$ Landau singularities are the same as the combined $Z$-arguments of the $X_1=0$  and $X_2=0$ Landau singularities (c.f., \cref{eq:X1land,eq:X2land}). 
In the end, the $Z$-arguments associated with the full Landau variety in $\mathcal{M}^\e_{1,3}$ are fully determined by the location of the subleading Landau singularities!
The major difference between $\d\log T$ and $\d\log X_1$ is the pure $\d\tau$ term in $\d\log T$. 
Indeed, the subleading Landau singularities occur at $\tau\to i\infty$ while the (pseudo-)normal thresholds occur at $\tau\to 0^+$ in our parameterization.

\paragraph{Landau analysis of the {\color{BrickRed}(1234)}-eyeball (and kite)} 

For integral families such as the eyeball and kite diagrams, the elliptic structure is inherited from a subtopology. 
Thus, it is no longer obvious that the pullback of the entire Landau variety corresponds to the degenerate limits $\tau^\alpha \to i \infty$ or $\tau^\alpha \to 0$; components not associated with the elliptic subtopology could be missing. 
This opens up the interesting possibility that part of the Landau variety pulls back to \emph{finite} (nonzero) $\tau^\alpha$, which we explore further below. 

While we will focus on the Landau singularities of the {\color{BrickRed}(1234)}-eyeball here, the discussion carries over straightforwardly to the kite. 
Once again, we suppress the label $\e$ for readability. 

Even if there are some components of the Landau variety that are not associated with the $\tau^\alpha\to i\infty$ limit, the $q^\alpha$-series is still extremely useful for determining the location of Landau singularities as illustrated in \eqref{eq:limitT}. 
For example, using the $\tau\to i \infty$ limit of $X_4$ in \eqref{dlogsin}, we find that the subleading Landau equation $X_4=0$ is associated with the vanishing of $z_1$, $z_1+z_2\pm z_4$ and $z_1 \pm z_2$ 
\begin{align} \label{eq:dlogX4pullback}
    \hspace{-0.4cm}\d\log X_4 &= 
    2 \omega_2\left(z_1,2 \tau\right)
    {+} 2 \omega_2\left(z_2,2 \tau\right)
    {-} 2 \omega_2\left(z_1,\tau\right)
    {+} \omega_2\big(z_1{+}\tfrac{z_2+z_4}{2},\tau\big)
    {+} \omega_2\left(z_1{+}\tfrac{z_2-z_4}{2},\tau\right)
    \nn\\&\quad
    - \omega_2\left(\tfrac{z_2-z_4}{2},\tau\right)
    - \omega_2\left(\tfrac{z_2+z_4}{2},\tau\right)
    - 2 \omega_2\left(z_1+z_2,2 \tau\right)
    - 2 \eta_2\left(\tau\right)
    \,.
\end{align}
Note that the expression above is manifestly modular invariant on the left-hand side, even though it is not expressed in terms of $\Omega_2$'s.

At this stage, one may wonder why we pulled back the complicated polynomial $T$ in \eqref{eq:Tdef} instead of one of its simpler components $(\pm\pm\pm)$. 
Although we can use PSLQ to match the $\d z$-part of our ansatz with the $\d z$-part of $\d\log(\pm\pm\pm)$, it seems that a more robust ansatz beyond $\eta_2$ is needed to match the $\d\tau$-parts.
We report that naively adding Kronecker-Eisenstein forms evaluated at rational points $\omega_2(\mathbbm{Q}+i\mathbbm{Q},N \tau)$ to the PSLQ ansatz does not seem to solve this issue. 
In fact, it might be impossible to pull back the $\d\tau$-part of $\d\log(\pm\pm\pm)$ using only rational coefficients in the ansatz. 
Therefore, it currently appears that only the dlog of \emph{special} polynomials can be easily pulled back to $\mathcal{M}^\e_{1,3}$. 

Interestingly, we can also pull back functions or 0-forms. 
For example, using \eqref{eq:dlogX4pullback}, we find that 
\begin{align} \label{eq:invX4pullback}
    \frac{1}{X_4} &= \frac{\partial z_4}{\partial X_4}
    \bigg[
        g^\rb{1}(z_2{+}z_4,\tau)
        {-} g^\rb{1}(z_2{-}z_4,\tau)
        {+} g^\rb{1}(z_1{+}\tfrac{z_2{+}z_4}{2},\tau)
        {-} g^\rb{1}(z_1{+}\tfrac12(z_2{-}z_4),\tau)
    \bigg]
    \,.
\end{align}
Without knowing how to pull back a dlog-form, the coefficient $\partial z_4/\partial X_4$ could be predicted simply by requiring modular invariance; one would find that the coefficients of a $g$-kernel ansatz needs to transform like the derivative of a puncture. 
Furthermore, note that while $z_1$ and $z_2$ appear as $Z$-arguments in \eqref{eq:dlogX4pullback}, they do not appear as $Z$-arguments in \eqref{eq:invX4pullback}. 
This suggests that either much of the singularity structure of the $X_4^{-1}$ pullback is hidden inside the partial derivative $\partial z_4/\partial X_4$ or that there are some nontrivial cancellations in the pullback of $\d\log X_4$. 

Moreover, the pullback of 0-forms in terms of $g$-kernels (e.g., \eqref{eq:invX4pullback}) seems to be new. 
That is, we are unaware of other explicit results that convert rational functions of kinematics $X_i$ into linear combinations of $g$-kernels. 
It would be interesting if such expressions could help sharpen statements about the pullback of the Landau variety.

In principle, the list of $Z$-arguments associated with the Landau variety could be constructed without knowledge of the differential equation; only the punctures and the Landau variety in $\mathcal{K}$ are needed. 
This is essentially what we have done since the singularities appearing in the pullback of $\mbf{B}$ are just the Landau singularities in disguise. 
However, there is an important advantage to the way the Landau singularities are organized in the dlog-forms on the diagonal of $\mathbf{B}$: their arguments are correctly factorized. 
Identifying and factorizing the square-root singularities of a Feynman integral is a well-known and difficult problem that is the subject of ongoing research (see \cite{Drummond:2019cxm, Mago:2020kmp, Mago:2021luw, Ren:2021ztg, Yang:2022gko,  Henke:2021ity} for some recent references).

For example, the singularity $\sqrt{\lambda_{134}}=0$ appears in a very specific way. 
One instance of this comes from the matrix element $B_{25,25}$ ($\e$-eyeball---$\e$-eyeball component of $\mathbf{B}$):
\begin{align}
    \d\log
        \frac{\sqrt{\lambda_{134}}+X_1-X_4+1}{\sqrt{\lambda_{134}}-X_1+X_4-1}
    = \frac{\left(X_1+X_4-1\right) \d X_1 - 2 X_1 \d X_4}{X_1 \sqrt{\lambda_{134}}}
    \,.
\end{align}
The above dlog form is easy to map to the torus and 
consequently, the coefficients of $\d X_i$ (rational functions of $X_i$) are also easy to pull back. 
For example, from the $\d X_4$-component, we find
\begin{align} \label{eq:invlambda}
    \frac{-2}{\sqrt{\lambda_{134}}} =
    \frac{\partial z_4}{\partial X_4}
    \bigg[
        \mathcal{G}^\rb{1}_2\Big(\tfrac12(z_2+z_4)\Big)
        + \mathcal{G}^\rb{1}_2\Big(\tfrac12(z_2-z_4)\Big)
    \bigg]\,,
\end{align}
where $\mathcal{G}^\rb{k}_{2}(z) = g^\rb{k}(z,\tau) -  2k\ g^\rb{k}(z,2\tau)$ are simply the linear combinations of $g$-kernels that appear in the components of the modular invariant differential form $\Omega_2(z,\tau)$ in \eqref{eq:om2}.
Interestingly, we find that it is much harder to derive the pullback of $1/\lambda_{134}$ in terms of $g$-kernels.
Of course, we can simply square both sides of \cref{eq:invlambda} but this would correspond to something with double poles on the torus, which is unexpected because $1/\lambda_{134}$ has only simple poles in the roots of $\lambda_{134}$.

On the other hand, it is not too difficult to construct the pullback of $\d\log\lambda_{134}$:
\begin{align} \label{eq:dloglambda}
    \d\log(\lambda_{134})
    &= 2\Omega_2(z_1,\tau)
    + \Omega_2(z_2,\tau)
    -2\Omega_2(z_1{+}z_2,\tau)
    + \Omega_2(z_4,\tau)
    -\omega_2(z_2,\tau)
    \nn\\&\qquad
    -\omega_2(z_4,\tau)
    +2\omega_2(\tfrac{z_2{+}z_4}{2},\tau)
    +2\omega_2(\tfrac{z_2{-}z_4}{2},\tau)
    -4 \eta_2(\tau)
    \,.
\end{align}
Once again we note the absence of $Z$-arguments in \eqref{eq:invlambda} when compared to \eqref{eq:dloglambda} and ask whether there are singularities secretly hiding in the partial derivative $\partial z_4/\partial X_4$ or if there are nontrivial cancellations in \eqref{eq:dloglambda}.

\subsection{A warm up: the eyeball pullback}
\label{subsPullbackIc}

As a warm-up, let's first exemplify this approach from the perspective of the \textcolor{BrickRed}{$I_{1,1,1,1,0}$} eyeball integral. This integral is naturally expressed on $\mathcal{M}^\e_{1,4}$. To avoid clutter, within this section we will suppress the $\e$-label on the variables, since the label $\ee$ is never involved.

\paragraph{Building an ansatz}

We begin by listing all the distinct, linearly independent combinations of dlog forms appearing on the diagonal of the eyeball's row. We find the following 
\begin{equation}\label{eq:poleEB}
    \text{d}\log\{
     X_0,\,  X_4 ,\,  \lambda_{024},\,\lambda_{314}
    \}\,,
\end{equation}
where $\lambda_{ijk}=\lambda(X_i,X_j,X_k)$ is defined below \eqref{v1313}. Next, we change variables from $X_0$, $X_1$, $X_2$, and $X_4$ to $\tau$, $z_1$, $z_2$, and $z_4$ (as previously defined in sections \cref{subsec_ellipticint,sec_punctureslab}). Then, we retain only the leading order term of their $q$-expansion and factorize the $\mathrm{d}\log$ argument into $\sin$ functions. This yields 
    \begin{subequations}\label{dlogsin}
        \begin{align}
        \text{d}\log\Big( X_0|_{q^2}\Big)&= \text{d}\log\Big(\sin^2 \Big(\pi z_1\Big)\sin^2 \Big(\pi z_2\Big)\Big)\,, 
        \\
        \text{d}\log \Big(X_4|_{q^0}\Big)&=\text{d}\log\dfrac{\sin \Big(\dfrac{\pi}{2}( 2z_1{+}z_2{+}z_4)\Big)\sin \Big(\dfrac{\pi}{2}( 2z_1{+}z_2{-}z_4)\Big)\sin^2 \Big(\pi z_2\Big)}{\sin^2\Big(\pi(z_1{+}z_2)\Big)\sin\Big(\dfrac{\pi}{2}(z_2{+}z_4)\Big)\sin\Big(\dfrac{\pi}{2}(z_2{-}z_4)\Big)}\,,\\
        \text{d}\log \Big(\lambda_{024}|_{q^0}\Big)&=\text{d}\log\dfrac{\sin^2 \Big(\pi z_1\Big)\sin^4 \Big(\pi z_2\Big)}{\sin^2\Big(\pi(z_1{+}z_2)\Big)\sin^2\Big(\dfrac{\pi}{2}(z_2{+}z_4)\Big)\sin^2\Big(\dfrac{\pi}{2}(z_2{-}z_4)\Big)}\,,\\
        \text{d}\log \Big(\lambda_{134}|_{q^0}\Big)&=\text{d}\log\dfrac{\sin^2 \Big(\pi z_1\Big)\sin^2 \Big(\pi z_2\Big)\sin^2 \Big(\pi z_4\Big)}{\sin^2\Big(\pi(z_1{+}z_2)\Big)\sin^2\Big(\dfrac{\pi}{2}(z_2{+}z_4)\Big)\sin^2\Big(\dfrac{\pi}{2}(z_2{-}z_4)\Big)}\, . 
\end{align}
    \end{subequations}
From \eqref{dlogsin}, we can deduce all the building blocks necessary to fix the ansatz \eqref{omega leading q} for all terms in the eyeball row of $\mathbf{B}$. In particular, the spanning set of forms contains only $\omega_k(\mathcal{L}_i^{\e}(z_1, z_2, z_4), N\tau)$ and $\eta_k(\tau)$ where $N=1,2$, $k=1,2,3$ is the modular weight and 
\begin{equation}\label{eq:ebLs}
    \big\{\mathcal{L}_i^{\e}\big\}_{\substack{i=1\\i\neq4}}^{9}=\Big\{z_1,z_2,z_4,z_1+z_2,\frac{z_2+z_4}{2},\frac{z_2-z_4}{2},z_1+\frac{z_2+z_4}{2},z_1+\frac{z_2-z_4}{2}\Big\}\,, 
\end{equation}
are the $Z$-arguments of the Kronecker-Eisenstein forms. 
Note that the case $i=4$ is skipped since it is reserved for $\mathcal{L}_4^{\e}=z_5$, which will only appear in the full kite; this way, all single term $Z$-arguments are indexed first. 
Furthermore, recall that $\eta_k(\tau)$ is, \emph{a priori}, included in this basis as a means to account for \emph{pure} $\mathrm{d}\tau$ corrections in the Kronecker-Eisenstein form ansatz \eqref{omega leading q}.

\paragraph{The eyeball pullback} Below, we present the restults for the pullback of all entries that are part of the {\color{BrickRed}(1234)}-eyeball line of $\mathbf{B}$. The presentation is organized by modular weight.

\paragraph{\emph{Modular weight 1}}

\begin{equation}\label{eq:EBw1}
\begin{split}
\textcolor{BrickRed}{B_{25,12}}&=2\omega_1\Big(\mathcal{L} ^{{\e}}_3,\tau\Big)\,.
\end{split}
\end{equation}

\paragraph{\emph{Modular weight 2}}

\begin{subequations}\label{eq:EBw2}
\begin{align}
   \textcolor{BrickRed}{ B_{25,10}}&=\dfrac{1}{2}\left[\omega_2\Big(\mathcal{L} ^{{\e}}_6,\tau\Big)-\omega_2\Big(\mathcal{L} ^{{\e}}_7,\tau\Big)+\omega_2\Big(\mathcal{L} ^{{\e}}_8,\tau\Big)-\omega_2\Big(\mathcal{L} ^{{\e}}_9,\tau\Big)\right]\,,\\
    \textcolor{BrickRed}{B_{25,11}}&=\frac{1}{2}\left[3\omega_2\Big(\mathcal{L} ^{{\e}}_7,\tau\Big)-3\omega_2\Big(\mathcal{L} ^{{\e}}_6,\tau\Big)+\omega_2\Big(\mathcal{L} ^{{\e}}_8,\tau\Big)-\omega_2\Big(\mathcal{L} ^{{\e}}_9,\tau\Big)\right]\,,\\
\textcolor{BrickRed}{B_{25,17}}&=\Omega_2\Big(\mathcal{L} ^{{\e}}_6,\tau\Big)+
\Omega_2\Big(\mathcal{L} ^{{\e}}_7,\tau\Big)+\Omega_2\Big(\mathcal{L} ^{{\e}}_8,\tau\Big)+\Omega_2\Big(\mathcal{L} ^{{\e}}_9,\tau\Big)\,,\\
   \textcolor{BrickRed}{ B_{25,18}}&=\Omega_2\Big(\mathcal{L} ^{{\e}}_7,\tau\Big)-\Omega_2\Big(\mathcal{L} ^{{\e}}_6,\tau\Big)-\Omega_2\Big(\mathcal{L} ^{{\e}}_8,\tau\Big)+\Omega_2\Big(\mathcal{L} ^{{\e}}_9,\tau\Big)\,,\\
    \textcolor{BrickRed}{B_{25,24}}&=-\Omega_2\Big(\mathcal{L} ^{{\e}}_6,\tau\Big)+\Omega_2\Big(\mathcal{L} ^{{\e}}_7,\tau\Big)+\Omega_2\Big(\mathcal{L} ^{{\e}}_8,\tau\Big)-\Omega_2\Big(\mathcal{L} ^{{\e}}_9,\tau\Big)\,,\\
\textcolor{BrickRed}{B_{25,25}}&=-2\omega_2(\mathcal{L} ^{{\e}}_1,2\tau)-2\omega_2(\mathcal{L} ^{{\e}}_2,2\tau)-2\omega_2(\mathcal{L} ^{{\e}}_3,\tau)-4\omega_2(\mathcal{L} ^{{\e}}_5,\tau)\notag\\
&\quad+6\omega_2(\mathcal{L} ^{{\e}}_5,2\tau)+3\omega_2\Big(\mathcal{L} ^{{\e}}_6,\tau\Big)+3\omega_2\Big(\mathcal{L} ^{{\e}}_7,\tau\Big)+\omega_2\Big(\mathcal{L} ^{{\e}}_8,\tau\Big)\\
&\quad+\omega_2\Big(\mathcal{L} ^{{\e}}_9,\tau\Big)+6\eta_2(\tau)\,.\notag
\end{align}
\end{subequations}
The mass-permutation symmetry of the eyeball graph requires $B_{25,18}=B_{25,24}\bigr|_{m_1\leftrightarrow m_3}$. 
As discussed in appendix \ref{puncturesappendix}, we find that $z_2$ and $z_4$ are left invariant under the exchange of $m_1$ and $m_3$, while $z_1\mapsto 1-z_1-z_2$ under the same permutation. 

\paragraph{\emph{Modular weight 3}}

\begin{equation}\label{eq:EBw3}
\begin{split}
\hspace{-0.2cm}
\frac{\textcolor{BrickRed}{B_{25,9}}}{4}&= \omega_3\Big(\mathcal{L}^{{\e}}_3,\tau\Big){-}3\omega_3\Big(\mathcal{L} ^{{\e}}_6,\tau\Big){+}3\omega_3\Big(\mathcal{L} ^{{\e}}_7,\tau\Big){-}\omega_3\Big(\mathcal{L} ^{{\e}}_8,\tau\Big) {+}\omega_3\Big(\mathcal{L} ^{{\e}}_9,\tau\Big)\,.
\end{split}
\end{equation}

\subsection{The kite pullback}
\label{pullbackkite}

Next, we turn to the full kite integral, which is naturally expressed on either $\mathcal{M}_{1,5}^{\alpha=\e,\ee}$.

\paragraph{Building an ansatz}

Just as with the eyeball, we now want to determine which $Z$-arguments to feed into the pullback ansatz for the kite row. Once again, we should list the minimal set of linearly independent dlog forms, denoted $\mathfrak{S}_{\e} \cup \mathfrak{S}_{\ee}$. Here, the entries of $\mathfrak{S}_{\e}$ can naturally be expressed on the $\mathcal{M}_{1,5}^\e$, while the entries in $\mathfrak{S}_{\ee}$ are naturally expressed on $\mathcal{M}_{1,5}^\ee$. The list is as follows
\begin{equation}
   \mathfrak{S}_{{\e}}\cup \mathfrak{S}_{{\ee}}
   = \text{d}\log\{ X_0,\,  X_1,\,  X_2,\, X_4,\,  X_5,\,\lambda_{024},\,  \lambda_{314}, \lambda_{315},\, \lambda_{325},\,\lambda_{01245}\}\cup\{\substack{1\leftrightarrow 4\\2\leftrightarrow 5}\}\,,
\end{equation}
where $\lambda_{ijk}$ and $\lambda_{01245}$ are defined below \eqref{v1313} and in \eqref{eq:lamFull}, respectively. Besides the function $\lambda_{01245}$, which we shall discuss soon, all the above dlog forms may be expressed (up to mass permutations; see appendix \ref{puncturesappendix}) as Kronecker-Eisenstein and modular forms using the results we found for the eyeball (i.e., \eqref{dlogsin}).

Following the procedure detailed in section \ref{poles}, we find that $\text{d}\log \lambda_{01245}$ introduces \emph{four}
new $Z$-arguments: $\mathcal{L}_{10}$, $\mathcal{L}_{11}$, $\mathcal{L}_{12}$, and $\mathcal{L}_{13}$ (the analogues of equation \eqref{dlogsin} in this case are too lengthy to be quoted here). These $Z$-arguments can be naturally expressed on the moduli space of either torus. Depending on which torus we wish to work on, we use $\omega_k(\mathcal{L}_i(z_1,z_2,z_4,z_5),n\tau)$ with $n$, $k$ as before and either 
\begin{subequations}\label{eq:L123}
    \begin{align}
       \hspace{-0.54cm}\big\{\mathcal{L}_i^{\e}\big\}_{i=1}^{17}&{=}\Big\{z_1,z_2,z_4,z_5,z_1{+}z_2,\frac{z_2{+}z_4}{2},\frac{z_2{-}z_4}{2},z_1{+}\frac{z_2{+}z_4}{2},z_1{+}\frac{z_2{-}z_4}{2},\frac{z_1{+}z_5}{2},\frac{z_1{-}z_5}{2},\\&\hspace{-1.7cm}z_2{+}\frac{z_1{+}z_5}{2},z_2{+}\frac{z_1{-}z_5}{2},\underbracket[0.4pt]{{\frac{z_1{{+}}z_2{{+}}z_4{{+}}z_5}{2}}, {\frac{z_1{{+}}z_2{{+}}z_4{-}z_5}{2}},{\frac{z_1{{+}}z_2{-}z_4{{+}}z_5}{2}},{\frac{z_1{{+}}z_2{-}z_4{-}z_5}{2}}}_{\text{new $Z$-arguments on 
       $\mathcal{M}_{1,5}^{\e}$}}\Big\}\,,\notag\\
               \hspace{-0.54cm}\big\{\mathcal{L}_i^{\ee}\big\}_{i=1}^{17}&{=}\Big\{z_4,z_5,z_1,z_2,z_4{+}z_5,\frac{z_5{+}z_1}{2},\frac{z_5{-}z_1}{2},z_4{+}\frac{z_5{+}z_1}{2},z_4{+}\frac{z_5{-}z_1}{2},\frac{z_4{+}z_2}{2},\frac{z_4{-}z_2}{2},\\&\hspace{-1.7cm}z_2+\frac{z_4{+}z_2}{2},z_2+\frac{z_4{-}z_2}{2},\underbracket[0.4pt]{{\frac{z_4{+}z_5{+}z_1{+}z_2}{2}},{\frac{z_4{+}z_5{+}z_1{-}z_2}{2}}, {\frac{z_4{+}z_5{-}z_1{+}z_2}{2}},{\frac{z_4{+}z_5{-}z_1{-}z_2}{2}}}_{\text{new $Z$-arguments on $\mathcal{M}_{1,5}^{\ee}$}}\Big\}
               .\notag
\end{align}
\end{subequations}
with $(\tau,z_i)=(\tau^{\e},z_i^{{\e}})$ and $(\tau,z_i)=(\tau^{\ee},z_i^{{\ee}})$, respectively. 

\paragraph{The kite on the torus}\label{kite_pulled_back}

The entries in the kite row exhibit the following modular structure
\begin{equation}
\Big(\begin{array}{cccccccccccccccccccccccccccccc}
    -&-&-&-&-&-&-&-&\color{BrickRed}{3}&\color{BrickRed}{2}&\color{BrickRed}{2}&\color{BrickRed}{1}&\color{Blue}{3}&\color{Blue}{2}&\color{Blue}{2}&\color{Blue}{1}&\textcolor{BrickRed}{\d\mathrm{l}}&\textcolor{BrickRed}{\d\mathrm{l}}&\textcolor{BrickRed}{\d\mathrm{l}}&\textcolor{BrickRed}{\d\mathrm{l}}&\textcolor{BrickRed}{\d\mathrm{l}}&\textcolor{BrickRed}{\d\mathrm{l}}&\textcolor{BrickRed}{\d\mathrm{l}}&\textcolor{BrickRed}{\d\mathrm{l}}&\textcolor{BrickRed}{\d\mathrm{l}}&\textcolor{BrickRed}{\d\mathrm{l}}&\textcolor{BrickRed}{\d\mathrm{l}}&\textcolor{Blue}{\d\mathrm{l}}&\textcolor{Blue}{\d\mathrm{l}}&\textcolor{BrickRed}{\d\mathrm{l}}
    \end{array}\Big)\,.
\end{equation}
    Here, entries marked with ``$-$'' vanish, those labeled as ``dl'' represent dlog forms, and finally, entries denoted with $1, 2, 3$ are quasi-modular forms of the corresponding modular weight. Additionally, the color coding used here is as before: the entries in \textcolor{BrickRed}{red} are pulled back onto $\mathcal{M}_{1,5}^\e$, while those in \textcolor{Blue}{blue} are pulled back onto $\mathcal{M}_{1,5}^\ee$. The dlog terms could be pulled back onto either $\mathcal{M}_{1,5}^{\alpha=\e,\ee}$; the predominance of \textcolor{BrickRed}{red} for most dlog entries is simply \emph{a choice}.\footnote{When integrating the differential equation in terms of iterated integrals, this choice may have important effects as it can lead to mixing of the elliptic curves; we will get back to this issue later in section \ref{bdry}.} The few \textcolor{Blue}{blue} dlog entries are specifically associated with the eyeballs that include the $\ee$-sunrise as a subtopology. 

All entries in this row are provided below explicitly. Note that we use the labels ${\e}$ or ${\ee}$ on the $Z$-arguments $\mathcal{L}_i^{(ijk)}$ to indicate which torus the entry is pulled back onto, and we simply write $\tau$ for the respective moduli, omitting the label. As before, we organize the presentation by modular weight and use the shorthand $\Tilde{B}_{ij}=B_{ij}\vert_{\mathcal{L}_i^{{\e}}\rightarrow \mathcal{L}_i^{{\ee}}, \tau^{{\e}}\rightarrow \tau^{{\ee}}}$. 

\paragraph{\emph{Modular weight 1}}
\begin{subequations}\label{eq:KTw1}
    \begin{align}
\textcolor{BrickRed}{B_{30,12}}&=\omega_1\Big(\mathcal{L}^{{\e}}_1,\tau\Big)+\omega_1\Big(\mathcal{L}^{{\e}}_2,\tau\Big)\,,\\
\textcolor{Blue}{B_{30,16}}&=\Tilde{B}_{30,12}=\omega_1\Big(\mathcal{L}^{{\ee}}_1,\tau\Big)+\omega_1\Big(\mathcal{L}^{{\ee}}_2,\tau\Big)\,.
\end{align}
\end{subequations}

\paragraph{\emph{Modular weight 2}}

\begin{subequations}\label{eq:KTw2}
\begin{align}
\textcolor{BrickRed}{B_{30,10}}&=\frac{1}{4} \Big[\sum_{k={8,9,12,13}}\omega_2\Big(\mathcal{L}_k^{{\e}},\tau\Big)-\sum_{\ell={6,7,10,11}}\omega_2\Big(\mathcal{L}_\ell^{{\e}},\tau\Big)\Big]\,,\\
\textcolor{Blue}{B_{30,14}}&=\Tilde{B}_{30,10}\,,\\
\textcolor{BrickRed}{B_{30,11}}&=
\frac{1}{4} \Big[\sum_{k={8,9,10,11}}\omega_2\Big(\mathcal{L}_k^{{\e}},\tau\Big)-\sum_{\ell={6,7,12,13}}\omega_2\Big(\mathcal{L}_\ell^{{\e}},\tau\Big)\Big]\,,\\
\textcolor{Blue}{B_{30,15}}&=\Tilde{B}_{30,11}\,,\\
\textcolor{BrickRed}{B_{30,17}}&=\frac{1}{2}\Big[ \Omega_2\Big(\mathcal{L}_6^{{\e}},\tau\Big)- \Omega_2\Big(\mathcal{L}_7^{{\e}},\tau\Big)
+ \Omega_2\Big(\mathcal{L}_{8}^{{\e}},\tau\Big)- \Omega_2\Big(\mathcal{L}_9^{{\e}},\tau\Big)\Big]\,,\\
\textcolor{BrickRed}{B_{30,18}}&=-\frac{1}{2}\Big[ \Omega_2\Big(\mathcal{L}_6^{{\e}},\tau\Big)+ \Omega_2\Big(\mathcal{L}_7^{{\e}},\tau\Big)
+ \Omega_2\Big(\mathcal{L}_{8}^{{\e}},\tau\Big)+ \Omega_2\Big(\mathcal{L}_9^{{\e}},\tau\Big)\Big]\,,\\
\textcolor{BrickRed}{B_{30,19}}&=\Omega_2\Big(\mathcal{L}_{10}^{{\e}},\tau\Big)+\Omega_2\Big(\mathcal{L}_{11}^{{\e}},\tau\Big)\,,\\
\textcolor{BrickRed}{B_{30,20}}&=\Omega_2\Big(\mathcal{L}_6^{{\e}},\tau\Big)+\Omega_2\Big(\mathcal{L}_7^{{\e}},\tau\Big)\,,\\
\textcolor{BrickRed}{B_{30,21}}&=\frac{1}{2}\Big[ \Omega_2\Big(\mathcal{L}_{10}^{{\e}},\tau\Big)- \Omega_2\Big(\mathcal{L}_{11}^{{\e}},\tau\Big)+ \Omega_2\Big(\mathcal{L}_{12}^{{\e}},\tau\Big)- \Omega_2\Big(\mathcal{L}_{13}^{{\e}},\tau\Big)\Big]\,,\\
\textcolor{BrickRed}{B_{30,22}}&=-\frac{1}{2}\Big[ \Omega_2\Big(\mathcal{L}_{10}^{{\e}},\tau\Big)+ \Omega_2\Big(\mathcal{L}_{11}^{{\e}},\tau\Big)+ \Omega_2\Big(\mathcal{L}_{12}^{{\e}},\tau\Big)+ \Omega_2\Big(\mathcal{L}_{13}^{{\e}},\tau\Big)\Big]\,,\\
\textcolor{BrickRed}{B_{30,23}}&=\frac{1}{2}\Big[\Omega_2\Big(\mathcal{L}_{12}^{{\e}},\tau\Big)+ \Omega_2\Big(\mathcal{L}_{13}^{{\e}},\tau\Big)- \Omega_2\Big(\mathcal{L}_{10}^{{\e}},\tau\Big)- \Omega_2\Big(\mathcal{L}_{11}^{{\e}},\tau\Big)\Big]\,,\\
\textcolor{BrickRed}{B_{30,24}}&=\frac{1}{2}\Big[\Omega_2\Big(\mathcal{L}_{8}^{{\e}},\tau\Big)+\Omega_2\Big(\mathcal{L}_9^{{\e}},\tau\Big)- \Omega_2\Big(\mathcal{L}_6^{{\e}},\tau\Big)-\Omega_2\Big(\mathcal{L}_7^{{\e}},\tau\Big)\Big]\,,\\
\textcolor{BrickRed}{B_{30,25}}&=\frac{1}{2}\Big[\sum_{k={6,8,16,17}}\omega_2\Big(\mathcal{L}_k^{{\e}},\tau\Big)-\sum_{\ell={7,9,14,15}}\omega_2\Big(\mathcal{L}_\ell^{{\e}},\tau\Big)\Big]\,,\\
\textcolor{BrickRed}{B_{30,26}}&=\frac{1}{2}\Big[\sum_{k={10,12,15,17}}\omega_2\Big(\mathcal{L}_k^{{\e}},\tau\Big)-\sum_{\ell={11,13,14,16}}\omega_2\Big(\mathcal{L}_\ell^{{\e}},\tau\Big)\Big]\,,\\
    \textcolor{BrickRed}{B_{30,27}}&=\frac{1}{2}\Big[\Omega_2\Big(\mathcal{L}_{14}^{\e},\tau\Big)-\Omega_2\Big(\mathcal{L}_{15}^{\e},\tau\Big)+\Omega_2\Big(\mathcal{L}_{16}^{\e},\tau\Big)-\Omega_2\Big(\mathcal{L}_{17}^{\e},\tau\Big)\Big]\,,\\
\textcolor{BrickRed}{B_{30,28}}&=\frac{1}{2}\Big[\Omega_2\Big(\mathcal{L}_{14}^{\e},\tau\Big)+\Omega_2\Big(\mathcal{L}_{15}^{\e},\tau\Big)-\Omega_2\Big(\mathcal{L}_{16}^{\e},\tau\Big)-\Omega_2\Big(\mathcal{L}_{17}^{\e},\tau\Big)\Big]\,,\\
\textcolor{BrickRed}{B_{30,29}}&= -\frac{1}{2}\Big[\Omega_2\Big(\mathcal{L}_{14}^{{\e}},\tau\Big)-\Omega_2\Big(\mathcal{L}_{15}^{{\e}},\tau\Big)-\Omega_2\Big(\mathcal{L}_{16}^{{\e}},\tau\Big)-\Omega_2\Big(\mathcal{L}_{17}^{{\e}},\tau\Big)\Big]\,,\\
\textcolor{BrickRed}{B_{30,30}}&=-2 \omega_2\Big(\mathcal{L}_1^{{\e}},2 \tau\Big)-2 \omega_2\Big(\mathcal{L}_2^{{\e}},2 \tau\Big)+6 \omega_2\Big(\mathcal{L}_5^{{\e}},2 \tau\Big)-2 \omega_2\Big(\mathcal{L}_5^{{\e}},\tau\Big)\notag\\
   &+2
   \omega_2\Big(\mathcal{L}_6^{{\e}},\tau\Big)+2 \omega_2\Big(\mathcal{L}_7^{{\e}},\tau\Big)+2 \omega_2\Big(\mathcal{L}_{10}^{{\e}},\tau\Big)+2 \omega_2\Big(\mathcal{L}_{11}^{{\e}},\tau\Big)\\
   &{-}\omega_2\Big(\mathcal{L}_{14}^{{\e}},\tau\Big){-}\omega_2\Big(\mathcal{L}_{15}^{{\e}},\tau\Big){-}\omega_2\Big(\mathcal{L}_{16}^{{\e}},\tau\Big){-}\omega_2\Big(\mathcal{L}_{17}^{{\e}},\tau\Big){+}6 \eta_2(\tau)\,.\notag
\end{align}
\end{subequations}

\paragraph{\emph{Modular weight 3}}
\begin{subequations}\label{eq:KTw3}
    \begin{align}
\textcolor{BrickRed}{B_{30,9}}&=2\Big[\sum_{\ell=14}^{17}\omega_3\Big(\mathcal{L}_{\ell}^{{\e}},\tau
   \Big)-\sum_{\ell=5}^{13}\omega_3\Big(\mathcal{L}_{\ell}^{{\e}},\tau
   \Big)\Big]\,,\\
\textcolor{Blue}{B_{30,13}}&=  \Tilde{B}_{30,9}\,.
\end{align}
\end{subequations}

\section{Boundary value and integration}
\label{bdry}
In this section, we provide a boundary value for the kite integral basis in \eqref{finaldeq}. We also detail the (numerical) integration technique, highlighting certain technical challenges. Subsequently, we elaborate on the stress tests conducted on the integrated results within a simplified region of the kinematic space and conclude with an outlook on numerical integration improvements.

\subsection{Boundary value} We compute the boundary value in a region of the kinematic space where all the master integrals are finite and simple.  A suitable choice for this is the kinematic point $\mbf{X}_0\in\mathcal{K}$, located in the neighborhood
\begin{equation}\label{eq:p0}
m_i=m>0 \quad\text{and}\quad p^2\to 0^+ \qquad \forall~i \,.
\end{equation}
At $\mbf{X}_0$, the master basis in \eqref{finaldeq} becomes significantly simpler, primarily due to the identities
\begin{equation}
  \frac{\partial X_i}{\partial \tau^\alpha}=\left.\frac{\partial X_0}{\partial \tau^\alpha}\right\vert_{\mbf{X}_0}=0 \qquad \text{for}\quad i\neq 0 \quad \text{and} \ \alpha=\e,\ee\,,
\end{equation}
which is easily checked numerically. Analytically, these identities follow from the observation that, in the neighborhood of $\mbf{X}_0$, we have $X_i\to1 $ 
for all $i\neq 0$ and approach
\begin{align}\label{eq:p0onT}
            \tau^{\e}&=\tau^{\ee}=i\infty\,,\notag\\ z_{1}^\e&=z_{2}^\e=z_{4}^\ee=z_{5}^\ee= \frac{1}{3}\,, \\z_{4}^\e&=z_{5}^\e=z_{1}^\ee=z_{2}^\ee=\frac{1}{2}-i\infty\,.\notag
\end{align}
Thus, the only nonzero components of the basis at $\mbf{X}_0$ are
\begin{equation}
\label{incond63}
\begin{split}
\mathbf{J}_0=\varepsilon^{2}\Big(\underbracket[0.4pt]{J_1}_{\times 8},\frac{\sqrt{3} J_2}{2}, \underbracket[0.4pt]{0}_{\times2},-\frac{\sqrt{3} J_2}{4},\frac{\sqrt{3} J_2}{2},\underbracket[0.4pt]{0}_{\times2}, -\frac{\sqrt{3}
   J_2}{4},i \sqrt{3} J_2,\underbracket[0.4pt]{0}_{\times3},i \sqrt{3} J_2,\underbracket[0.4pt]{0}_{\times9}\Big)\,,
\end{split}
\end{equation}
where $J_1=I_{1,1,0,0,0}$ and $J_2=I_{1,1,1,0,0}$. In this case, we see that there are only two\footnote{This counting takes into account IBP relations between the integrals at $\mbf{X}_0$. It also accounts for the fact that $I_{0,1,1,0,1}$ converges to the sunrise $I_{1,1,1,0,0}$ in a sufficiently small neighborhood of $\mbf{X}_0$. This can be directly observed through an analysis by regions and by noting that both integrals satisfy the same Symanzik polynomials in the contributing region.} out of the 30 integrals in \eqref{eq:5mFam} to evaluate explicitly in the neighborhood of $\mbf{X}_0$.

The double-tadpole $I_{1,1,0,0,0}$ is independent of $p^2$ and is easily evaluated in $d=2-2\varepsilon$ dimensions using Feynman parameters. The result is
\begin{equation}\label{eq:icTad}
    J_1=I_{1,1,0,0,0}(2-2\varepsilon;\mbf{X}_0)=e^{2\gamma_{\text{EM}}\varepsilon}\Gamma^2(\varepsilon)\,.
\end{equation}

Next, we evaluate the sunrise integral around $\mbf{X}_0$ using the method of regions \cite{Beneke:1997zp,Jantzen:2011nz,Jantzen:2012mw}. We find that there is only one region contributing to the integral. Consequently, the parametric representation of the integrand can be expanded at leading order in $p^2$ before loop integration. Moreover, performing the change of variables in \cite[eq. (74)]{Adams:2013nia} on the expanded integrand yields a fully off-shell 1-loop triangle integrand with massless internal states in $4+2\varepsilon$ dimensions, which is given in \cite[App.~V]{Bern:1993kr}. Putting everything together, we obtain the exact result
\begin{equation}\label{eq:icSun}
\hspace{-0.1cm} J_2=I_{1,1,1,0,0}(2{-}2\varepsilon;\mbf{X}_0)=\mathcal{N}\Big[\left(-e^{\frac{2 i \pi }{3}}\right)^{-\varepsilon} F_{\varepsilon}\left({\tfrac{2 i \pi }{3}}\right){-}\left(-e^{-\frac{2 i \pi }{3}}\right)^{-\varepsilon} F_{\varepsilon}\left({-\tfrac{2 i \pi }{3}}\right){+}\frac{\pi}{\varepsilon}\Big]\,,
\end{equation}
where $\mathcal{N}=\tfrac{e^{2 \gamma_\text{EM}\varepsilon} \Gamma (1{+}2 \varepsilon)}{(-1)^{1+2\varepsilon}3^{\frac{1}{2}+\varepsilon}}$ and $F_{\varepsilon}(z)=\frac{3 i \Gamma^2(\varepsilon+1)}{2 \varepsilon^2 \Gamma (2 \varepsilon+1)}\, _2F_1(-2 \varepsilon,-\varepsilon,1-\varepsilon,e^{z})$. The $\varepsilon$-expansion of both \eqref{eq:icTad} and \eqref{eq:icSun} can be easily performed using the identity \cite{Adams:2015ydq}
\begin{align}
     {}_2F_1\left(-2\varepsilon,-\varepsilon;1-\varepsilon; x \right)
 & =
 1 + 2 \varepsilon^2 \mathrm{Li}_2\left(x\right)
 + \varepsilon^3 \left[ 2 \mathrm{Li}_3\left(x\right) - 4 \mathrm{Li}_{2,1}\left(x,1\right) \right]
 \nonumber \\
 &
 + \varepsilon^4 \left[ 2 \mathrm{Li}_4\left(x\right) - 4 \mathrm{Li}_{3,1}\left(x,1\right) + 8 \mathrm{Li}_{2,1,1}\left(x,1,1\right) \right]
 + {\mathcal O}\left(\varepsilon^5\right)\,,
\end{align}
where $\mathrm{Li}_{n_1,n_2,...,n_k}$ denotes the classical multiple polylogarithm. Explicit expressions for the expansion can be found in the ancillary file \href{https://github.com/StrangeQuark007/kite_ancillary}{\faGithub} and were verified numerically up to weight three using \texttt{PySecDec} \cite{Borowka:2017idc,Heinrich:2023til}.

\subsection{Discussion: numerical integration, checks and outlook} 
An important motivation for deriving the differential equation in \eqref{finaldeq} and the boundary condition in \eqref{incond63} is to systematize the numerical evaluation of the master integrals to which they correspond. The starting point is the formal solution in \eqref{eq:ittSol}. 

For elliptic Feynman integrals, one strategy is to express the iterated integrals that show up in the solution in terms of \emph{elliptic multiple polylogarithms} \cite{Broedel:2014vla, Broedel:2017kkb,Walden:2020odh,weinzierl2022feynman}
\begin{equation}
\label{eq:gamT}
    \tilde{\Gamma}\left(\left.\substack{n_1\\ w_1}\substack{n_2\\ w_2}\substack{...\\ ...}\substack{n_k\\ w_k};w\right|\tau\right)=
    \int_{0}^{w} \textnormal{d} w'\,g^{(n_{1})}(w'{-}w_{1},\tau)\tilde{\Gamma}\left(\left.\substack{n_2\\ w_2}\substack{...\\ ...}\substack{n_k\\ w_k};w'\right|\tau\right)\,,
\end{equation}
if we integrate along a path in $z$-space or in terms of \emph{iterated integrals over modular forms} $\{f_i\}$ and ELi\emph{-series} \cite{Adams:2014vja,Adams:2015gva,Adams:2015ydq,bogner_unequal_2020,Weinzierl:2020fyx,Walden:2020odh,weinzierl2022feynman} if we integrate along a path in $\tau$-space
\begin{subequations}\label{eq:cteZints}
    \begin{align}
    \mathfrak{M}(f_1,f_2,...,f_k\vert \tau)&=\displaystyle\int\limits_{i \infty}^{\tau}
 f_1\left(t_1\right)
 \displaystyle\int\limits_{i \infty}^{t_1} 
 f_2\left(t_2\right)
 ...
\displaystyle\int\limits_{i \infty}^{t_{k-1}}
 f_k\left(t_k\right)\,,\\
      \mathrm{ELi}_{n_1,...,n_\ell;m_1,...,m_\ell;\sigma_1,...,\sigma_{\ell-1}}^{x_1,...,x_\ell;y_1,...,y_\ell;\bar{q}}
 &=\prod_{\alpha=1}^\ell \sum_{\substack{j_\alpha=1\\k_\alpha=1}}^\infty\frac{x_\alpha^{j_\alpha}}{j_\alpha^{n_\alpha}}\frac{y_\alpha^{k_\alpha}}{k_\alpha^{m_\alpha}}\bar{q}^{j_\alpha k_\alpha}\prod\limits_{i=1}^{\ell-1} \left(j_i k_i{+} ... {+} j_\ell k_\ell \right)^{-\sigma_i/2}\,,\label{eq:ELI}
\end{align}
\end{subequations}
where $\Bar{q}=e^{2\pi i \tau}$. Their convergence properties are summarized in \cite{Walden:2020odh}. Of course, the high-precision numerical evaluation of \eqref{eq:gamT} and \eqref{eq:cteZints} comes with few obvious challenges, which we enumerate below.

\paragraph{Technical challenge I: \emph{elliptic curves mixing}} In cases where many elliptic curves are involved, the function space spanned by \eqref{eq:gamT} and \eqref{eq:cteZints} is sufficient \emph{only} if the quantities defined on $\mathcal{M}_{1,5}^\e$ and $\mathcal{M}_{1,5}^\ee$ fully decouple. 

As for the kite family, the two sources of elliptic functions in the differential equation, namely the sunrises \emph{rectangular} (shaded) blocks in
\begin{equation}
\label{bigmatrix}
\mbf{B}= 
\adjustbox{valign=c}{\includegraphics[scale=1]{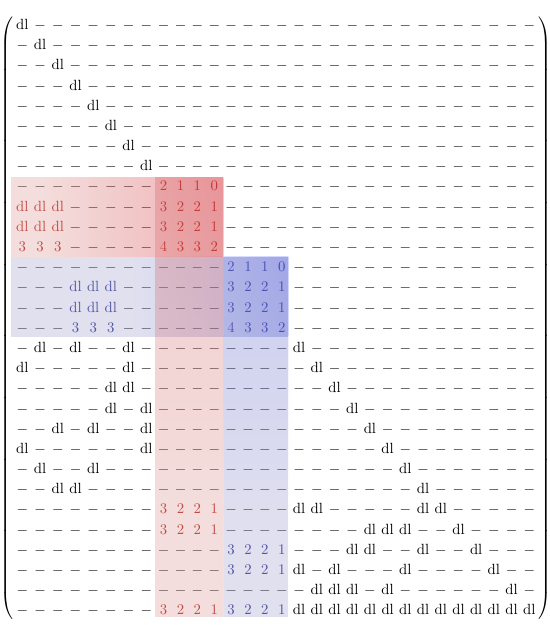}}\,,
\end{equation}
completely decouple. Here, as before, ``$\d\mathrm{l}$'' indicates that the entry is a dlog form in kinematic variables, while the ``$-$'' entries vanish. Otherwise, the number gives the (quasi-)modular weight of the entry. Below, it is assumed that these non-dlog entries have \emph{already} been pulled back to the moduli space of their respective torus: $\mathcal{M}_{1,5}^\e$ for entries overlapping with the \textcolor{BrickRed}{red}-shaded region and $\mathcal{M}_{1,5}^\ee$ for those overlapping with the \textcolor{Blue}{blue}-shaded region.

Now, from \eqref{bigmatrix}, one can show to arbitrarily high order in the solution's Laurent expansion that iterated integrals, which can potentially depend simultaneously on both $\e$ and $\ee$, appear \emph{only} through the interaction of dlog entries with any entries in either the blue or red shaded rectangular blocks. From this perspective, a simple strategy to prevent mixing of $\e$ and $\ee$ becomes quite clear: we always have the freedom to pullback the dlog form on the torus so that no mixing occurs. 

Although conceptually clear, this way of organizing the calculation requires keeping track of \emph{local} choices for the pullback of the dlog forms at the level of each iterated integral order by order in the $\varepsilon$-expansion. Of course, from a technical point of view, this quickly becomes inefficient. To us, this serves as a first indication that the approach of performing the numerical evaluation of the master basis through that of \eqref{eq:gamT} and \eqref{eq:cteZints} might not be optimal when \emph{multiple} tori are involved. A potential alternative direction is mentioned towards the end of this section. 

\paragraph{Technical challenge II: \emph{convergence}} Once we have ensured that $\e$ and $\ee$ decouple in the iterated integrals, there is yet another technical challenge to face when numerically evaluating the master integrals from \eqref{eq:gamT} and \eqref{eq:cteZints}. Similarly to Feynman integrals that evaluate to MPLs, we often want to evaluate these integrals at points near or beyond the boundary of their region of convergence (for methods addressing these issues more formally, see, for example, \cite{Passarino:2016zcd,Duhr:2019rrs}).

For the kite master, this occurs, for example, when the differential equation is solved to finite $\tau$'s starting from \eqref{eq:p0onT}. Because $z_{4}^\e=z_{5}^\e=z_{1}^\ee=z_{2}^\ee=\frac{1}{2}-i\infty$ remain fixed, the leading-order solution 
\begin{align}
          \hspace{-0.38cm} J_{30}^{(2)}&{=}
\sum_{j=6}^9{\Delta_{6,7}^{j}}\Big(\left[\Omega _{2,j,1}^{\e},\Omega _{2,1,1}^{\e}\right]{-}2 \left[\Omega _{2,j,1}^{\e},\Omega _{2,1,2}^{\e}\right]{-}\left[\Omega _{2,j,1}^{\e},\Omega _{2,5,1}^{\e}\right]{+}2
   \left[\Omega _{2,j,1}^{\e},\Omega _{2,5,2}^{\e}\right]\Big)\\&
  \hspace{-0.2cm} {+}\sum_{j=10}^{13}{\Delta_{10,11}^{j}}\Big(\left[\Omega _{2,j,1}^{\e},\Omega _{2,2,1}^{\e}\right]{-}2 \left[\Omega _{2,j,1}^{\e},\Omega _{2,2,2}^{\e}\right]{-}\left[\Omega _{2,j,1}^{\e},\Omega _{2,5,1}^{\e}\right]{+}2
   \left[\Omega _{2,j,1}^{\e},\Omega _{2,5,2}^{\e}\right]\notag\Big)\\&
   \hspace{-0.28cm}+(\e\leftrightarrow\ee)\,,\notag
\end{align}
where $\Delta_{n,m}^{j}=(-1)^{\delta_{j,n}{+}\delta_{j,m}}$, $\Omega_{a,b,c}^d=\omega_a(\mathcal{L}_b^d,c~\tau_d)$ and\footnote{Note that, in practice, the lower bound of the integration domain ($\tau=i\infty$ or $q=0$) may introduce (spurious) poles. These singularities are typically dealt with through the use of tangential base-point regularization scheme \cite{deligne1989groupe,Brown:2014pnb,Panzer:2015ida,Bonisch:2021yfw}.} 
\begin{equation}\label{eq:itIntTau}
    [w_1,w_2,...,w_k]
 =
\displaystyle\int\limits_{i \infty}^{\tau}
 w_1\left(t_1\right)
 \displaystyle\int\limits_{i \infty}^{t_1} 
 w_2\left(t_2\right)
 ...
\displaystyle\int\limits_{i \infty}^{t_{k-1}}
 w_k\left(t_k\right)\,,
\end{equation}
once expressed, for example, through ELi-series (c.f., \eqref{eq:ELI}), does not converge \emph{numerically}. Naturally, this divergence arises because the solution involves a series representation that extends beyond its radius of convergence; in this particular case, the use of the $g$-kernels quasiperiodicity in $\tau$ resolves the issue (see \cite[eq. (13.200)]{weinzierl2022feynman}). Furthermore, such divergences are absent from subtopologies involving three or fewer propagators, corresponding to master integrals independent of the ``extra'' punctures $z_{4}^\e=z_{5}^\e=z_{1}^\ee=z_{2}^\ee=\frac{1}{2}-i\infty$. In particular, we find that the sunrise components after the $\tau$ integrations match \cite[eq. (106)]{bogner_unequal_2020}, as expected.

\paragraph{Numerical checks in the soft ($p^2=0^+$) regime} For the kite family, a simple alternative to avoid the need of continuing the integration path beyond the region of convergence of \eqref{eq:gamT} and \eqref{eq:cteZints} is to \emph{first} integrate the differential equation from \eqref{eq:p0onT} to finite values of $z_{4}^\e, z_{5}^\e, z_{1}^\ee$ and $z_{2}^\ee$, for example, setting $z_{4}^\e = z_{5}^\e = z_{1}^\ee = z_{2}^\ee = \frac{1}{2}$, while keeping everything else constant. 

A technical benefit of performing the integration at fixed $\tau^\e = \tau^\ee = i\infty$ is the significant simplification that occurs at the level of the pullback differential equation \eqref{bigmatrix}. In fact, along the path $\gamma_0$ parameterized by the straight line $z_{4}^\e = z_{5}^\e = z_{1}^\ee = z_{2}^\ee = t$, where $t$ varies from $\frac{1}{2} - i\infty$ to $\frac{1}{2}$, the matrix elements are purely trigonometric and the resulting iterated integrals combine \emph{nontrivially} into finite quantities. The result can then be efficiently computed using \textsc{Mathematica}'s \texttt{NIntegrate[]} function, as exemplified in the ancillary file \href{https://github.com/StrangeQuark007/kite_ancillary}{\faGithub}.

At this point, let us emphasize that the map\footnote{To be more precise, the relevant map is obtained by solving simultaneously the equations
\begin{equation*}
    \text{sn}^2\left(  \theta _3^2\left(0,e^{i \pi  \tau_\alpha }\right)\frac{\pi z_i^\alpha}{2},\frac{\theta _2^4\big(0,e^{i \pi  \tau_\alpha
   }\big)}{\theta _3^4\left(0,e^{i \pi  \tau_\alpha }\right)}\right)=u_i^\alpha(\boldsymbol{x}) \qquad \forall~ i=1,2,4,5~\text{and}~ \alpha=\e,\ee\,,
\end{equation*}
for $\boldsymbol{x}=\{X_0,X_1,X_2,X_4,X_5\}$. Here, $\text{sn}(x,k)$ denotes the Jacobi sn-function, and the $u$'s are given in \cref{sunpunct123,eq:u4123,eq:z15,eq:z21}. The above equation is nothing but the formal inverse of, e.g., \eqref{sunrisepunctures} and is easily checked numerically.} between the space $\mathfrak{M}=\mathcal{M}_{1,5}^\e\times\mathcal{M}_{1,5}^\ee$ (parametrized by the moduli spaces of the two tori) and the kinematic space $ \mathcal{K} $ is \emph{not} invertible everywhere. For example, the map fails to be surjective because, while $\mathfrak{M}$ is 10-dimensional, $ \mathcal{K} $ is only 5-dimensional. (This phenomenon is intrinsically related to the fact that we are considering a multiscale problem with \emph{ multiple} tori. In particular, it does not occur for the unequal mass sunrise discussed in \cite{bogner_unequal_2020} because the number of kinematic degrees of freedom happens to match that on the torus, i.e., the map is an isomorphism.) For this reason, it turns out that the point 
\begin{equation}\label{eq:p0onT-alt}
\begin{split}
        \tau^{\e}&=\tau^{\ee}=i\infty\,,\\ z_{1}^\e&=z_{2}^\e=z_{4}^\ee=z_{5}^\ee= \frac{1}{3}\,, \\z_{4}^\e&=z_{5}^\e=z_{1}^\ee=z_{2}^\ee=\frac{1}{2}\,,
\end{split}
\end{equation}
we land on after integrating along $\gamma_0$ is \emph{not} the image of any (complex) point in $\mathcal{K}$.\footnote{This can be verified using, e.g., \textsc{Mathematica}'s \texttt{FindInstances[]}.} This implies that the numerical result does not coincide with the master integrals evaluated over $\mathcal{K}$, but instead corresponds to their extensions to $\mathfrak{M}$. Practically, this makes direct comparisons with software like \texttt{PySecDec} \cite{Heinrich:2023til} unfeasible for some coordinate values in $\mathfrak{M}$ such as \eqref{eq:p0onT-alt}. 

Nonetheless, we can validate our results by moving from \eqref{eq:p0onT-alt} to points in the $\mathfrak{M}$ space that have support in $\mathcal{K}$. By restricting ourselves to the submanifold of $\mathfrak{M}$ where $\tau^\e=\tau^\ee=i\infty$, we can efficiently obtain numbers using the same method as outlined earlier (integrating successively along straight lines in $\mathfrak{M}$ with \texttt{NIntegrate[]}).\footnote{In doing so, one must ensure that the differential equation on $\mathfrak{M}$ (depending on 10 variables) remains integrable along each integration path.} For instance, the master integrals computed at the particular point
\begin{equation}
\begin{split}
    \tau^\e&=\tau^\ee=i\infty\,,\quad
    z_1^\e=z_4^\ee=1/4\,,\quad
    z_2^\e=z_5^\ee=1/2\,,\\
    z_4^\e=&z_1^\ee=-\frac{2i\log(i\sqrt{2+\sqrt{3}})}{\pi}\,,\quad z_5^\e=z_2^\ee=\frac{i\log(8+3\sqrt{7})}{2\pi}\,,
\end{split}
\end{equation}
(corresponding to $\mbf{X}=\{X_0,X_1,X_2,X_4,X_5\}=\{0^+,1,2,1,2\}$ in $\mathcal{K}$) perfectly agrees with the prediction of \texttt{PySecDec} at leading and subleading order in $\varepsilon$. Similar checks were carried out successfully at various other points within the soft region.

\paragraph{Outlook: \emph{improving numerics}} We conclude this section by noting that, outside the \emph{simplifying} soft limit $\tau^\e=\tau^\ee=i\infty$, we found it impractical to apply the \texttt{NIntegrate[]}-based method (used above) for high-precision numerical evaluation of iterated integrals. The primary hindrance being the slow convergence caused by the integration kernels being elliptic ($g$-kernels) rather than trigonometric.

Although numerical values can, in principle, be obtained by directly evaluating iterated integrals in the form of \eqref{eq:gamT} and \eqref{eq:cteZints}, we report that this approach converges too slowly using the available tools (e.g., \cite{Walden:2020odh}) to make \emph{satisfactory} comparisons of our results with those found in the literature \cite{Broadhurst:2022bkw} or through \texttt{PySecDec} within \emph{generic} kinematic regions. 

This is a situation that we wish to improve in the near future. In particular, we believe that there are more efficient and generalizable approaches to the problem of numerical integration of differential equations defined on moduli spaces. Prime candidates are generalizations of the series methods employed, e.g., in \cite{Caron-Huot:2020vlo, Abreu:2020jxa, Hidding:2020ytt,Armadillo:2022ugh} for the high-precision evaluation of polylogarithmic Feynman integrals. For all of these approaches to work, the only inputs required are a differential equation, a boundary value, and a path of integration, thus bypassing the need for an explicit evaluation of MPLs. In this regard, potential generalizations to \emph{elliptic} Feynman integrals are exciting because they are meant to be agnostic of elliptic curve mixing and of explicit evaluation of eMPLs in generic kinematic regions.

\section{Conclusion}
In this paper, we present the first analytic results for the 5-mass kite family of Feynman integrals near two space-time dimensions using the method of differential equations. 

Our approach involved investigating the relationship between two elliptic curves associated with the sunrise subsectors of the 5-mass kite family of Feynman integrals. In particular, we introduced a method to construct an $\varepsilon$-form basis, leveraging the underlying $\text{SL}(2,\mathbbm{Z})$ covariance of these integrals. Moreover, we identified a relevant parameterization of the kinematic space in terms of the moduli and location of punctures on a torus. The puncture locations were physically motivated and take the form of integrals over maximal cuts. From there, we described how to systematically pull back the $\varepsilon$-form differential equation from kinematic space to a much more compact form involving both modular and KE forms on the moduli space of the tori. Specifically, we pointed out how to identify the relevant set of $Z$-arguments (combinations of punctures to feed the ansatz of KE forms) through a simple analysis of the $q$-expansions of the \emph{diagonal} entries of the differential equation. The output of this procedure is one of the main results of this paper: a schematic representation of the differential equation is quoted in \eqref{bigmatrix} and two of its rows are explicitly recorded in \cref{eq:EBw1,eq:EBw2,eq:EBw3,eq:KTw1,eq:KTw2,eq:KTw3}. The full matrix in can be found in the ancillary files: \href{https://github.com/StrangeQuark007/kite_ancillary}{\faGithub}. 
We further leveraged this analysis to sharpen the connection between Landau singularities in the kinematic space $\mathcal{K}$ with those on the moduli space. 
This was examplified by deriving explicit expressions for various polynomials that define the Landau variety and their associated dlog-forms (c.f., \cref{eq:pbT,eq:dlogX4pullback,eq:invX4pullback,eq:invlambda,eq:dloglambda}) in terms of $g$-kernels and modular forms.
To our knowledge, this is the first time that such relations between 0-forms in $\mathcal{K}$ and $g$-kernels have been explicitly written down. 
Lastly, we derived a boundary value for the kite family of master integrals (see \eqref{incond63}) that facilitates its numeric evaluation and a cross-check of our results in a particular simplifying limit (unequal mass soft limit). 
Our hope is that these developments lead to a number of new directions, some of which are briefly discussed below. 

We anticipate that the methods introduced in this paper will also be applicable to more general multiscale 2-loop examples involving one or more elliptic curves. Now, since the kite is the most general 2-point function (in cubic theories), natural candidates for future work are the 3- and 4-point analogues: the 4-mass parachute (see \cite[fig. 1 (h)]{Fevola:2023fzn}) and the 5-mass crossed-box/acnode graph (see \cite[fig. 3]{Mizera:2021ujs}), respectively. In particular, the latter still involves two elliptic curves and five masses, but also depends on an additional variable: the momentum transfer $t$. 

Furthermore, it would be interesting to explore how the ideas introduced in this paper could be applied to examples more directly related to Standard Model processes. A prime example is the 3-parameter double box, which was recently discussed in \cite{G_rges_2023}. As highlighted in this reference, this graph is associated with an elliptic curve and contributes to the planar corrections of H+jet production through a loop of massive (top) quarks ``cut in half'' vertically by a gluon. At LHC energies ($\sqrt{s}\approx 13$ TeV), 
the cross section for H+jet production is about half of the inclusive Higgs production rate \cite[fig. 4.62]{Campbell:2017hsr}.
Although amplitudes for this process have been obtained at NNLO \cite{Harlander:2002wh,Anastasiou:2002yz,Ravindran:2003um,Bonciani:2019jyb}, most of these calculations have been performed in the large top-quark mass effective theory, where the top quark is integrated out and the Higgs boson directly couples to gluons through higher-dimensional operators (for lower order results in the full theory see, e.g., \cite{Ellis:1987xu,Neumann:2018bsx,Jones:2018hbb,Bonciani:2022jmb}). However, these effective calculations are expected to break down for jet energy \emph{greater} than $\sim1.6$ times the Higgs mass $m_H=125~\text{GeV}$ \cite{Grazzini:2013mca,Neumann:2014nha,Dawson:2014ora}.
In view of the forthcoming high-luminosity phase of the LHC and higher-energy muon colliders, it seems sensible to direct future efforts towards lower-order perturbative calculations within the \emph{full} theory. At present, a complete NNLO calculation is only conceivable numerically, although an $\varepsilon$-form differential equation is known \cite{G_rges_2023}. 

While systematic, the $\vep$-form algorithm of \cite{G_rges_2023} introduces integrals over a rational function times the period of the elliptic curve. However, the relationship between these integrals and the class of functions discussed earlier in \eqref{eq:gamT} and \eqref{eq:cteZints} is rather obscure. It is natural to ask to what extent these functions are related to $g$-kernels and modular forms.

It would also be interesting to further solidify our understanding of the interplay between the $Z$-arguments on the tori's moduli space and Landau singularities in kinematic space. Although a partial analysis for the kite integral was initiated in section \ref{pullback}, investigating whether $Z$-arguments can, in some way, be derived more directly from Landau analysis is an exciting avenue for future research. This is particularly relevant for the Symbol Prime bootstrap \cite{Wilhelm:2022wow, Morales:2022csr, McLeod:2023qdf,Cao:2023tpx,He:2023umf}.
Since it is possible to extract the symbol of the physical non $\vep$-form kite integral $I_{1,1,1,1,1}$ from the boundary condition and differential equation provided in this work (the gauge transformation to $\vep$-form must be undone), it would be interesting to compare our symbol with the result from a bootstrap.

As discussed in section \ref{bdry}, verifying our results involved a challenging high-precision numerical evaluation of the basis of master integrals by solving the pulled back differential equation in \emph{generic} kinematic regions. In particular, using publicly available tools (see \cite{Walden:2020odh}), we found that the numerical evaluation of the eMPLs entering the evaluation of the kite master integral converged very slowly. 
This made the comparison with the literature \cite{Broadhurst:2022bkw} and \texttt{PySecDec} (public numeric evaluation software) in generic kinematic regions particularly difficult and, in our view, unsatisfactory. A pressing problem is therefore to clarify how to efficiently extract high-precision numerical values for multiscale elliptic Feynman integrals given a differential equation written in KE and modular forms. A promising direction involves the development of a series method that extends those established in \cite{Caron-Huot:2020vlo, Abreu:2020jxa, Hidding:2020ytt} to the context of elliptic Feynman integrals.

Lastly, looking beyond the elliptic world, it would also be interesting to see if the method introduced in this paper can be generalized to more complicated higher-genus geometries. For example, given a Feynman diagram involving a hyperelliptic curve (see \cite{Marzucca:2023gto} for examples), is there a systematic way to define punctures when needed? (Note that, while the generalization of modular and KE forms to higher-genus surfaces is still under extensive investigation \cite{DHoker:2023vax,DHoker:2023khh}, we find it legitimate to ask this question even at this early stage). Of course, once an appropriate set of differential forms is established, extending our method to higher-dimensional geometries, such as Calabi-Yau spaces, also presents a natural avenue for future research.

\section*{Acknowledgements}
We thank Claude Duhr for the suggestion of this project, as well as for many fruitful discussions. 
We also thank Babis Anastasiou, Simon Caron-Huot, Miguel Correia, Christoph Nega, Oliver Schlotterer, Marcus Spradlin, Lorenzo Tancredi, Anastasia Volovich, Fabian Wagner, Stefan Weinzierl and Matthias Wilhelm for helpful discussions. A special thanks goes to Sebastian Mizera for sharing results on the Landau singularities of the kite integral before publication of the \texttt{PLD.jl} package.
We also thank Simon Caron-Huot, Claude Duhr, Oliver Schlotterer and Matthias Wilhelm for comments on the manuscript.
FP and AP thank Uppsala University for hospitality during parts of this work.  
The authors acknowledge the hospitality of ETH Zürich and the University of Zürich during the workshop and associated school \emph{Elliptics 2023}.  MG's work is supported by the National Science and
Engineering Council of Canada (NSERC) and the Canada Research Chair program. FP's work was co-funded by the European Union (ERC Consolidator Grant LoCoMotive 101043686). Views and opinions expressed are those of the author(s) only and do not necessarily reflect those of the European Union or the European Research Council. Neither the European Union nor the granting authority can be held responsible for them. 
AP is supported by the US Department of Energy under contract DESC0010010 Task F.
YS acknowledges support from the Centre for Interdisciplinary Mathematics at Uppsala University and partial support by the European Research Council under ERC- STG-804286 UNISCAMP.

\newpage

\begin{appendix}

\section{Punctures and $Z$-arguments under mass permutations}
\label{puncturesappendix}

In this appendix, we provide some examples of how the discrete symmetries of the sunrise, eyeball, and kite topologies act on the punctures and $Z$-arguments.

The $\e$- and $\ee$-sunrises are invariant under the action of the respective symmetric groups $S_3(\{m_1,m_2,m_3\})$ and $S_3(\{m_3,m_4,m_5\})$, 
while the {\color{BrickRed}(1234)}-, {\color{BrickRed}(1235)}-, {\color{Blue}(2345)}-, and {\color{Blue}(1345)}-eyeball diagrams are only invariant under the respective exchanges $m_1 \leftrightarrow m_3$, $m_2 \leftrightarrow m_3$, $m_2 \leftrightarrow m_5$ and $m_3 \leftrightarrow m_4$.
The {\color{Plum}kite} has two discrete symmetries, namely 
$m_1 \leftrightarrow m_4$,  $m_2 \leftrightarrow m_5$ and 
$m_1 \leftrightarrow m_5$,  $m_2 \leftrightarrow m_4$. 
In \cref{tab:permTable}, we provide a partial list of how these symmetries affect punctures and $Z$-arguments.

\begin{table}[H]
    \centering
    \begin{tabular}{||c| c| c||}
    \hline
    Mass permutations & Puncture permutations & $Z$-argument permutations\\
 \hline\hline
  $m_1\leftrightarrow m_2$  & $z_1^{\textcolor{BrickRed}{(123)}}\leftrightarrow z_2^{\textcolor{BrickRed}{(123)}}$& $\mathcal{L}_1^{\textcolor{BrickRed}{(123)}}\leftrightarrow \mathcal{L}_2^{\textcolor{BrickRed}{(123)}}$\\\hline
  $m_1\leftrightarrow m_3$  & $z_1^{\textcolor{BrickRed}{(123)}}\leftrightarrow 1-z_1^{\textcolor{BrickRed}{(123)}}-z_2^{\textcolor{BrickRed}{(123)}}$& $\mathcal{L}_1^{\textcolor{BrickRed}{(123)}}\leftrightarrow -\mathcal{L}_5^{\textcolor{BrickRed}{(123)}}$\\\hline
   $m_2\leftrightarrow m_3$  & $z_2^{\textcolor{BrickRed}{(123)}}\leftrightarrow 1-z_1^{\textcolor{BrickRed}{(123)}}- z_2^{\textcolor{BrickRed}{(123)}}$& $\mathcal{L}_2^{\textcolor{BrickRed}{(123)}}\leftrightarrow -\mathcal{L}_5^{\textcolor{BrickRed}{(123)}}$
   \\\hline
\end{tabular}
    
    \vspace{5mm} 

   \begin{tabular}{||c| c| c||}
    \hline
    Mass permutations & Puncture permutations & $Z$-argument permutations\\
 \hline\hline
  $m_1\leftrightarrow m_3$  & $z_1^{\textcolor{BrickRed}{(123)}}\leftrightarrow 1-z_1^{\textcolor{BrickRed}{(123)}}-z_2^{\textcolor{BrickRed}{(123)}}$& $\mathcal{L}_1^{\textcolor{BrickRed}{(123)}}\leftrightarrow -\mathcal{L}_5^{\textcolor{BrickRed}{(123)}},\, \mathcal{L}_8^{\textcolor{BrickRed}{(123)}}\leftrightarrow -\mathcal{L}_9^{\textcolor{BrickRed}{(123)}}$\\\hline
\end{tabular}

    \vspace{5mm} 
 
 \begin{tabular}{||c| c| c||}
    \hline
    Mass permutations & Puncture permutations & $Z$-argument permutations\\
 \hline\hline
  $m_1\leftrightarrow m_2,\, m_4\leftrightarrow m_5$  & $z_1^{\textcolor{BrickRed}{(123)}}\leftrightarrow z_2^{\textcolor{BrickRed}{\textcolor{BrickRed}{(123)}}},\, z_4^{\textcolor{BrickRed}{(123)}}\leftrightarrow z_5^{\textcolor{BrickRed}{(123)}}$&\begin{tabular}{@{}c@{}}$\mathcal{L}_1^{\textcolor{BrickRed}{(123)}}\leftrightarrow \mathcal{L}_2^{\textcolor{BrickRed}{(123)}},\, \mathcal{L}_4^{\textcolor{BrickRed}{(123)}}\leftrightarrow \mathcal{L}_5^{\textcolor{BrickRed}{(123)}},$ \\ $ \mathcal{L}_6^{\textcolor{BrickRed}{(123)}}\leftrightarrow \mathcal{L}_{10}^{\textcolor{BrickRed}{(123)}} ,\, \mathcal{L}_7^{\textcolor{BrickRed}{(123)}}\leftrightarrow \mathcal{L}_{11}^{\textcolor{BrickRed}{(123)}} ,$\\ $\mathcal{L}_8^{\textcolor{BrickRed}{(123)}}\leftrightarrow \mathcal{L}_{12}^{\textcolor{BrickRed}{(123)}},\,\mathcal{L}_9^{\textcolor{BrickRed}{(123)}}\leftrightarrow \mathcal{L}_{13}^{\textcolor{BrickRed}{(123)}},$\\
  $\mathcal{L}_{15}^{\textcolor{BrickRed}{(123)}}\leftrightarrow \mathcal{L}_{16}^{\textcolor{BrickRed}{(123)}},\,\mathcal{L}_{17}^{\textcolor{BrickRed}{(123)}}\leftrightarrow -\mathcal{L}_{17}^{\textcolor{BrickRed}{(123)}}$\end{tabular} \\\hline
  $m_1\leftrightarrow m_4,\,m_2\leftrightarrow m_5$&\begin{tabular}{@{}c@{}}$\tau^{\textcolor{BrickRed}{(123)}}\leftrightarrow \tau^{\textcolor{Blue}{(345)}},$\\
  $z_1^{\textcolor{BrickRed}{(123)}}\leftrightarrow z_4^{\textcolor{Blue}{(345)}},\, z_2^{\textcolor{BrickRed}{(123)}}\leftrightarrow z_5^{\textcolor{Blue}{(345)}},$\\
  $z_4^{\textcolor{BrickRed}{(123)}}\leftrightarrow z_1^{\textcolor{Blue}{(345)}},\,z_5^{\textcolor{BrickRed}{(123)}}\leftrightarrow z_2^{\textcolor{Blue}{(345)}}$\end{tabular}&$\mathcal{L}^{\textcolor{BrickRed}{(123)}}_i\leftrightarrow \mathcal{L}^{\textcolor{Blue}{(345)}}_i$\\\hline
   $m_1\leftrightarrow m_5,\,m_2\leftrightarrow m_4$&\begin{tabular}{@{}c@{}}$\tau^{\textcolor{BrickRed}{(123)}}\leftrightarrow \tau^{\textcolor{Blue}{(345)}},$\\$z_1^{\textcolor{BrickRed}{(123)}}\leftrightarrow z_5^{\textcolor{Blue}{(345)}},\, z_2^{\textcolor{BrickRed}{(123)}}\leftrightarrow z_4^{\textcolor{Blue}{(345)}},$\\
  $z_4^{\textcolor{BrickRed}{(123)}}\leftrightarrow z_2^{\textcolor{Blue}{(345)}},\,z_5^{\textcolor{BrickRed}{(123)}}\leftrightarrow z_1^{\textcolor{Blue}{(345)}}$\end{tabular}&\begin{tabular}{@{}c@{}}$\mathcal{L}_1^{\textcolor{BrickRed}{(123)}}\leftrightarrow \mathcal{L}_2^{\textcolor{Blue}{(345)}},\, \mathcal{L}_4^{\textcolor{BrickRed}{(123)}}\leftrightarrow \mathcal{L}_5^{\textcolor{Blue}{(345)}},$ \\ $ \mathcal{L}_6^{\textcolor{BrickRed}{(123)}}\leftrightarrow \mathcal{L}_{10}^{\textcolor{Blue}{(345)}} ,\, \mathcal{L}_7^{\textcolor{BrickRed}{(123)}}\leftrightarrow \mathcal{L}_{11}^{\textcolor{Blue}{(345)}} ,$\\ $\mathcal{L}_8^{\textcolor{BrickRed}{(123)}}\leftrightarrow \mathcal{L}_{12}^{\textcolor{Blue}{(345)}},\,\mathcal{L}_9^{\textcolor{BrickRed}{(123)}}\leftrightarrow \mathcal{L}_{13}^{\textcolor{Blue}{(345)}},$\\
  $\mathcal{L}_{15}^{\textcolor{BrickRed}{(123)}}\leftrightarrow \mathcal{L}_{16}^{\textcolor{Blue}{(345)}},\,\mathcal{L}_{17}^{\textcolor{BrickRed}{(123)}}\leftrightarrow -\mathcal{L}_{17}^{\textcolor{Blue}{(345)}}$\\
  $\mathcal{L}_3^{\textcolor{BrickRed}{(123)}}\leftrightarrow \mathcal{L}_3^{\textcolor{Blue}{(345)}},\,\mathcal{L}_{14}^{\textcolor{BrickRed}{(123)}}\leftrightarrow\mathcal{L}_{14}^{\textcolor{Blue}{(345)}}$\end{tabular}\\\hline
\end{tabular}

    \caption{\textbf{(Top)} Symmetries of the sunrise $\color{BrickRed} I_{1,1,1,0,0}$ and its implications on punctures and $Z$-arguments. \textbf{(Middle)} Symmetries of the eyeball $\color{BrickRed} I_{1,1,1,1,0}$ and its implications on punctures and $Z$-arguments. \textbf{(Bottom)} Symmetries of the kite $\color{Plum} I_{1,1,1,1,1}$ and its implications on punctures and $Z$-arguments.}
    \label{tab:permTable}
\end{table}

\section{Relating $F(u_\phi,k_\phi^2)$ to $F(u_4,k^2_\e)$ }
\label{app:massage}

In \eqref{ellc4} we found that integrating the eyeball's maximal cut gave rise to incomplete elliptic integrals of the first kind, which we interpreted as additional punctures. However, when naively performing such indefinite integrals in, e.g., \textsc{Mathematica}, we obtain incomplete elliptic integrals of the first kind with a \emph{different} modulus squared $k^2$.
Below, we show how certain relations between elliptic integrals can be applied to relate \eqref{ellc4} with indefinite integration over the eyeball's maximal cut.

\paragraph{Summary of the relations used}
For $0\leq k^2\leq 1$ and $0\leq \phi \leq \pi/2$, we have 
\begin{enumerate}[(i)]
    \item $F(-\phi,k^2)=-F(\phi,k^2)$\,,
    \item $F(\phi,k^2)+F(\psi,k^2)=\pm K(k^2)\,\,\,$   for $\,\,\,\psi=\pm \arccos\left(\pm \sin(\phi)\sqrt{1-k^2}/\sqrt{1-k^2\sin^2(\phi)}\right)$\,,
    \item $F\left(\phi,k^2\right)+F\left(\psi,k^2\right)=iK(1-k^2)\,\,\,$   for $\,\,\,\psi=\arcsin\left(-1/(k\sin(\phi))\right)$\,,
    \item $F\left(\phi,1/k^2\right)=k F\left(\psi,k^2\right)\,\,\,$ for $\,\,\,\psi=\arcsin\left(\sin(\phi)/k\right)$\,,
    \item $F(\phi,1-k^2)=i F(\psi,k^2)-i K(k^2)\,\,\,$ for $\,\,\,\psi=\arcsin\left(1/\sqrt{1-(1-k^2)\sin^2(\phi)}\right)$\,.
\end{enumerate}
 Interestingly, we find numerically that identities (i), (ii) and (iv) are valid for a larger domain, namely $0\leq k^2\leq 1$ and $0\leq \text{Re}(\phi) \leq \pi/2$ without restriction on $\text{Im}(\phi)$.

\paragraph{Modular transformations and shifts}
The moduli $\tau$ and the puncture $z$ were defined (see \cref{def:tau,abelsmap}, respectively) such that
\begin{equation}
\tau=\dfrac{iK(1-k^2)}{K(k^2)} \qquad \text{and} \qquad z=\dfrac{F(\phi,k^2)}{K(k^2)}\,.
\end{equation}
We can apply the formulas introduced above to establish useful transformation rules for punctures on a torus. Of particular interest are the transformation rules of $z$ under the generators $T$ and $S$ of $\text{SL}(2,\mathbbm{Z})$ and under the discrete shifts along the $A$- and $B$-cycles. Under $T$, $z$ remains fixed, while under $S$, $z\mapsto z' = \frac{z}{\tau}$. For shifts along the $A$- and $B$-cycles we have, respectively, $z\mapsto z'=z+1$ and $z\mapsto z'=z+\tau$.

For the $S$ transformation, we have
\begin{equation}
\begin{split}
    z'=\dfrac{z}{\tau}&= \dfrac{F(\phi,k^2)}{i K(1-k^2)}=\frac{F(\psi_1,1-k^2)-K(1-k^2)}{K(1-k^2)}=\dfrac{F(\psi_2,1-k^2)}{K(1-k^2)}\,, 
\end{split}
\end{equation}
where $\psi_1=\arcsin\left(1/\sqrt{1-k^2 \sin^2(\phi)}\right)$ and $\psi_2=-\text{arccot}\left(k\tan(\psi_1)\right)$.
Similarly, we have for the $A$-cycle shift
\begin{equation}
   z'=z+1=\dfrac{F(\phi,k^2)}{K(k^2)}+\dfrac{ K(k^2)}{K(k^2)}=\dfrac{F\left(\psi,k^2\right)}{K(k^2)}\, ,
\end{equation}
where $\psi=\arccos\left(-\sin(\phi)\sqrt{1-k^2}/\sqrt{1-k^2\sin^2(\phi)}\right)$.
The $B$-cycle shift can be used to cast a given elliptic integral back into the range $0\leq \phi\leq \pi/2$ by writing
\begin{equation}
   z'=z+\tau=\dfrac{F(\psi,k^2)}{K(k^2)}+\dfrac{i K(1-k^2)}{K(k^2)}=\dfrac{F\left(\phi,k^2\right)}{K(k^2)}\, ,
\end{equation}
where $\psi=\arcsin\left(1/\left(k\sin(\phi)\right)\right)$.

\paragraph{Example: extracting $z^{\e}_4$ from the maximal cut of the eyeball}
We have seen in \eqref{ellc4} that integrating the maximal cut of the eyeball yields
\begin{equation}
\int \rd X_4~\text{MC}_0({\color{BrickRed}I_{1,1,1,1,0}}) =\dfrac{1}{2c_4^\prime}F\Big( \arcsin\sqrt{u_\phi},k_\phi^2\Big)\,,
\end{equation}
with $k_{\e}^2= 1-1/k_\phi^2$ (c.f., \eqref{eq:kuEB}).
Using successively relation (iv) and (v), we find
\begin{equation}
\begin{split}
\dfrac{1}{2c_4'}F(\arcsin\sqrt{u_\phi},k^2_\phi)=&\dfrac{1}{2c_4'k_\phi}\, F\Big(\arcsin\left(k_\phi\sqrt{u_\phi}\right),\frac{1}{k_\phi^2}\Big)\\
=&\dfrac{i}{2c_4'k_\phi}\Big[ F\Big(\arcsin\Big(\frac{1}{\sqrt{1-u_\phi}}\Big),1-\frac{1}{k_\phi^2}\Big)-K\Big(1-\frac{1}{k_\phi^2}\Big)\Big]\,.
\end{split}
\end{equation}
Consequently, we see that $u_{\e}=1/(1-u_\phi)$ and $c_4=-i c_4' k_\phi$ as previously quoted in \eqref{eq:kuEB}.
Finally, plugging $\psi_1=2K(k^2_{\e})/c_4$, we find \eqref{eq:maxCutToEC}:
\begin{equation}
\int \rd X_4~\text{MC}_0({\color{BrickRed}I_{1,1,1,1,0}})=\dfrac{1}{2}\psi_1\Bigl[\dfrac{F\left(\arcsin\sqrt{u_4},k^2_{\e}\right)}{K\left(k^2_{\e}\right)}-1\Bigr]\, .
\end{equation}

\section{New relations between $g$-kernels}\label{sec_newrel}
In this appendix, we present and derive a new relation between the $g$-kernels (defined earlier in \eqref{kronecker-eisenstein}), namely\footnote{We would like to thank Claude Duhr and Einan Gardi for sharing the formula in the case $N=2$.}
\begin{equation}\label{newrel}
g^{(k)}(Nz,\tau)=\sum_{n=0}^{k} \dfrac{(2\pi i)^n}{n!} N^{(k-2-n)} \sum_{l=0}^{N-1} l^n \sum_{m=0}^{N-1} g^{(k-n)}\left(\dfrac{l\tau}{N}+\dfrac{m}{N}+z,\tau\right)\,,
\end{equation}
for $N\in \mathbb{N}$ and with the convention that $0^0=1$. 
In practice, this identity can be used to cast iterated integrals of the form
\begin{equation}
    \int_0^z g^{(k_1)}(N_1 z_1,\tau)\text{d}z_1\int_0^{z_1}g^{(k_1)}(N_2 z_2,\tau)\text{d}z_2...\int_0^{z_{n-1}}g^{(k_n)}(N_n z_n,\tau)\text{d}z_n \qquad (N_i\in\mathbbm{N})\,,
\end{equation}
into a linear combinations of eMPLs (as defined in \eqref{eq:gamT}).

To streamline the derivation of \eqref{newrel}, we first introduce and prove an auxiliary identity for the functions $\Omega(z,\eta,\tau)$
(doubly periodic completion of the Kronecker-Eisenstein functions $F(z,\eta,\tau)$ defined in equation \eqref{kronecker-eisenstein})\footnote{We would like to thank Oliver Schlotterer for invaluable discussions on the proof.}
\begin{equation}\label{O to F}
\Omega(z,\eta,\tau)=\exp\left(2\pi i \eta \dfrac{\text{Im}z}{\text{Im}\tau}\right) F(z,\eta,\tau)\,.
\end{equation} 
The needed auxiliary identity is
\begin{equation}\label{toprove}
\Omega(Nz,\eta,\tau)=\dfrac{1}{N}\sum_{l=0}^{N-1}\sum_{m=0}^{N-1} \Omega\left(\dfrac{l\tau}{N}+\dfrac{m}{N}+z,N\eta,\tau\right)\, . 
\end{equation}
To prove this identity, we start by introducing the lattice sum representation of $\Omega(z,\eta,\tau)$ in comoving coordinates ($z=u\tau+v$ with $u,v\in\mathbbm{R}$) \cite{elliptic_calc, Gerken:2018jrq}
\begin{equation}\label{lattice_sum}
\Omega(z,\eta,\tau)=\sum_{q,r\in \mathbbm{Z}} \dfrac{e^{2\pi i (qv-ru)}}{q\tau+r+\eta}\, .
\end{equation}
Next, we plug (\ref{lattice_sum}) into the right-hand side (RHS) of (\ref{toprove}) and pull out a factor of $1/N$ 
\begin{equation}
\text{RHS of\,\,}\eqref{toprove}=\dfrac{1}{N^2}\sum_{q,r\in \mathbbm{Z}} \dfrac{e^{2\pi i (qv-ru)}}{\frac{q}{N}\tau+\frac{r}{N}+\eta}\sum_{m=0}^{N-1}e^{2\pi i q m/N}\sum_{l=0}^{N-1}e^{-2\pi i r l/N}\, .
\end{equation}
Following this, we split the outer sum into two sums over the sets 
\begin{equation}
    I=N\mathbbm{Z}^2 \quad \text{and} \quad \tilde{I}=\mathbbm{Z}^2\setminus N \mathbbm{Z}^2,
\end{equation}
to find
\begin{equation}\label{intermediate_step}
\begin{split}
\text{RHS of\,\,}\eqref{toprove}&=\dfrac{1}{N^2}\sum_{(q,r)\in I} \dfrac{e^{2\pi i (qv-ru)}}{\frac{q}{N}\tau+\frac{r}{N}+\eta}\sum_{m=0}^{N-1}e^{2\pi i q m/N}\sum_{l=0}^{N-1}e^{-2\pi i r l/N}\\
&+\dfrac{1}{N^2}\sum_{(q,r)\in\tilde{I}} \dfrac{e^{2\pi i (qv-ru)}}{\frac{q}{N}\tau+\frac{r}{N}+\eta}\sum_{m=0}^{N-1}e^{2\pi i q m/N}\sum_{l=0}^{N-1}e^{-2\pi i r l/N}\, .
\end{split}
\end{equation}
For the first and last term on the left-hand side (LHS) of \eqref{intermediate_step}, we respectively have 
\begin{equation}\label{eq:sums}
\sum_{m=0}^{N-1}e^{2\pi i q m/N}\sum_{l=0}^{N-1}e^{-2\pi i r l/N}=\begin{sqcases}
       N^2 & \text{for} \quad (q,r)\in I\,, \\
       0 & \text{for}\quad (q,r)\in \tilde{I}\,.
\end{sqcases}
\end{equation}
where $\sum_{k=0}^wx^k=\dfrac{x^{1+w}-1}{x-1}$ is used to show that the second pair of sums in the LHS of \eqref{intermediate_step} vanishes. 
Substituting \eqref{eq:sums} into \eqref{intermediate_step} yields the LHS of \eqref{toprove}:
\begin{equation}
\begin{split}
\text{RHS of\,\,}\eqref{toprove}&=\sum_{q,r\in N\mathbbm{Z}} \dfrac{e^{2\pi i (qv-ru)}}{\frac{q}{N}\tau+\frac{r}{N}+\eta}+0\\
&=\sum_{q,r\in \mathbbm{Z}} \dfrac{e^{2\pi i N(qv-ru)}}{q\tau+r+\eta}\\
&=\Omega(Nz,\eta,\tau)\\
&=\text{LHS of\,\,}\eqref{toprove}\, .
\end{split}
\end{equation}
In the second step above we have relabeled the summation indices, while in the third step we have identified the result with the definition for the lattice sum representation \eqref{lattice_sum}. 
This completes the proof of the auxiliary identity \eqref{toprove}.

Using \eqref{toprove}, it is then easy to prove \eqref{newrel} in two steps. 
First, we insert \eqref{toprove} into \eqref{O to F} and find
\begin{equation}\label{eq:Fid}
F(Nz,\eta,\tau)=\dfrac{1}{N}\sum_{l=0}^{N-1}\sum_{m=0}^{N-1} \exp\left(2\pi i l\eta\right)F\left(\dfrac{l\tau}{N}+\dfrac{m}{N}+z,N\eta,\tau\right)\,.
\end{equation}
Subsequently, \eqref{newrel} follows directly from the series expansion of \eqref{eq:Fid} in $\eta$. Let us also mention that the series expansion of \eqref{toprove}  in $\eta$ leads to a relation for the so-called $f$-kernels (doubly periodic and non-meromorphic analogue of $g$), which are of particular relevance in string theory \cite{Broedel:2014vla, Dolan:2007eh, Broedel:2017jdo}
\begin{equation}
f^{(k)}(Nz,\tau)=N^{k-2}\sum_{l=0}^{N-1}\sum_{m=0}^{N-1}  f^{(k)}\left(\dfrac{l\tau}{N}+\dfrac{m}{N}+z,\tau\right)\, . 
\end{equation}

\end{appendix}

\bibliographystyle{JHEP}
\bibliography{refs.bib}
\end{document}